\documentclass[rmp,aps,nofootinbib,twocolumn]{revtex4}
\usepackage{graphics}
\usepackage{epsfig}

\begin{document}
\title{Universality classes in nonequilibrium lattice systems
\footnote{Scheduled for publication in Reviews of Modern Physics}}

\author{G\'eza \'Odor}
\email{odor@mfa.kfki.hu}
\affiliation{Research Institute for Technical Physics and Materials Science, 
\\ H-1525 Budapest, P.O.Box 49, Hungary}

\begin{abstract}
This work is designed to overview our present knowledge about
universality classes occurring in nonequilibrium systems defined 
on regular lattices.
In the first section I summarize the most important critical exponents,
relations and the field theoretical formalism used in the text.
In the second section I briefly address the question of scaling behavior 
at first order phase transitions. 
In section three I review dynamical extensions of basic static classes,
show the effect of mixing dynamics and the percolation behavior.
The main body of this work is given in section four where genuine,
dynamical universality classes specific to nonequilibrium systems are 
introduced. 
In section five I continue overviewing such nonequilibrium classes but 
in coupled, multi-component systems.
Most of the known nonequilibrium transition classes are explored in low 
dimensions between active and absorbing states of reaction-diffusion 
type of systems. However by mapping they can be related to universal 
behavior of interface growth models, which I overview in section six. 
Finally in section seven I summarize families of absorbing state system
classes, mean-field classes and give an outlook for further directions of
research.
\end{abstract}
\maketitle

\tableofcontents

\section {Introduction} \label{introsect}

Universal scaling behavior is an attractive feature in statistical physics 
because a wide range of models can be classified purely in terms of their 
collective behavior. Scaling phenomenon has been observed in many branches
of physics, chemistry, biology, economy ... etc., most frequently by critical 
phase transitions. Nonequilibrium phase transitions may appear in models of 
population \cite{Alb}, epidemics \cite{Ligget,Mollison}, catalysis \cite{ZGB}, 
cooperative transport \cite{Hav,Chowd}, enzyme biology \cite{Berry} 
and markets \cite{Bou} for example. 

Dynamical extensions of static universality classes --- established in 
equilibrium --- are the simplest nonequilibrium models systems, 
but beyond that critical phenomena, with new classes have been 
explored so far \cite{DickMar,GrasWu,Hayeof}.
While the theory of phase transitions is quite well understood
in thermodynamic equilibrium its research in nonequilibrium
is rather new. In general phase transitions, scaling and universality 
retain most of the fundamental concepts of equilibrium models. 
The basic ingredients affecting universality classes are again the
collective behavior of systems, the symmetries, the conservation laws 
and the spatial dimensions as described by renormalization group theory.
Besides them several new factors have also been identified recently. 
Low dimensional systems are of primary interest 
because the fluctuation effects are relevant, hence the mean-field 
type of description is not valid.
In the past decades this field of research was growing very quickly
and now we are faced with a zoo of models, universality classes, 
strange notations and abbreviations. This article aims to help
newcomers as well as researchers to navigate in the literature by 
systematically reviewing most of the explored universality classes.
I define models by their field theory (when it is available),
show their symmetries or other important features and list the 
critical exponents and scaling relations.
 
Nonequilibrium systems can be classified into two categories:
(a) Systems which do have a hermitian Hamiltonian and whose stationary 
states are given by the proper Gibbs-Boltzmann distribution. 
However, they are prepared in an initial condition which is far from 
the stationary state and sometimes, in the thermodynamic limit, 
the system may never reach the true equilibrium.
These nonequilibrium systems include, for example, phase ordering systems,
spin glasses, glasses etc. I begin the review of classes by showing the
scaling behavior of the simplest prototypes of such systems in section 
\ref{eqintro}
These are defined by adding simple dynamics to static models.
(b) Systems without a hermitian Hamiltonian defined by transition rates,
which do not satisfy the detailed balance condition 
(the local time reversal symmetry is broken). 
They may or may not have a steady state and even if they have one, it is
not a Gibbs state.
Such models can be created by combining different dynamics or 
by generating currents in them externally.
The critical phenomena of these systems are referred here as 
``Out of equilibrium classes'' and discussed in section \ref{eqintro}.
There are also systems, which are not related to equilibrium models,
in the simplest case these are lattice Markov processes of interacting
particle systems \cite{Ligget}. These are referred here as 
``Genuinely non-equilibrium systems'' and are discussed in the rest of
the work. The discussion of the latter type of systems is splitted to 
three parts:
In section \ref{genchap} phase transition classes of simplest
such models is presented. These are usually reaction-diffusion (RD) type of 
models exhibiting phase transition to absorbing states.
In section \ref{multichap} I list the known classes which occur by
combinations of basic genuine class processes. These models are
coupled multi-component RD systems. 
While the former two sections are related to critical phenomena
near extinction in section \ref{growth} I discuss universality classes
in systems, where site variables are non-vanishing, in surface growth models.
The bosonic field theoretical description is applicable for them as well.
I point out mapping between growth and RD systems when it is possible.
In section \ref{1storder} I briefly touch the point of discontinuous
nonequilibrium phase transitions especially because dynamical scaling
may occur at such points. 

I define a critical universality class by the complete set of exponents
at the phase transition. Therefore different dynamics split up the basic
static classes of homogeneous systems. I emphasize the role of
symmetries, boundary conditions which affect these classes.
I also point out very recent evidence according to which in low
dimensional systems symmetries are not necessarily the most relevant
factors of universality classes.
Although the systems covered here might prove to be artificial to
experimentalists or to application oriented people they constitute the 
fundamental blocks of understanding of nonequilibrium critical phenomena.
Note that the understanding of even so simple models runs into tremendous
difficulties very often.

I shall not discuss the critical behavior of quantum systems \cite{Racznewrev},
self-organized critical phenomena \cite{BTW} 
and neither show experimental realizations. The discussion of the applied
methods is also omitted due to the lack of space, although in section
(\ref{fieldthint}) I give a brief introduction of the field theoretical
approach. This section shows the formalism for defining nonequilibrium models. 
This is necessary to express the symmetry relations affecting critical 
behavior. Researchers from other branches of science are provided a kind of 
catalog of classes in which they can identify their models and find 
corresponding theories. 
To help navigating in the text and in the literature
I provided a list of the most common abbreviations at the end of the text.

Besides scaling exponents and scaling relations there are many other 
interesting features of universality classes like scaling distribution, 
extremal statistics, finite size effects, statistics of fluctuations in 
surface growth models etc., which I do not discuss in this review. 
Still I believe the shown material provides a useful frame for orientation 
in this huge field. There is no general theory of nonequilibrium
phase transitions, hence a widespread overview of known classes
can help theorists deducing the relevant factors determining 
universality classes.

There are two recent, similar reviews available. One of them is 
by \cite{DickMar}, which gives a pedagogical introduction to driven 
lattice gas systems and to fundamental particle systems with absorbing states.
The other one \cite{Hayeof} focuses more on basic absorbing state 
phase transitions, methods and experimental realizations.
However the field evolves rapidly and since the publication of this two
remarkable introductory works a series of new developments have come up. 
The present work aims to give a comprehensive overview of known nonequilibrium
dynamical classes, incorporating surface growth classes, classes of
spin models, percolation and multi-component system classes and damage 
spreading transitions. The relations and mappings of the corresponding
models are pointed out. The effects of boundary conditions, long-range
interactions and disorder are shown systematically for each class were it
is known. Since a debate on the conditions of the parity conserving class 
has not been settled yet I provide a discussion on it through some surface 
catalytic model examples.
Naturally this review can not be complete and I apologize for the omitted 
references.

\subsection{Critical exponents of equilibrium systems} \label{eqexpsec}

In this section I briefly summarize the definition of well known critical 
exponents of homogeneous equilibrium systems and show some scaling relations
\cite{F67,K67,ST71,Ma,Amit}.
The basic exponents are defined via the scaling laws:
\begin{eqnarray}
c_H &\propto&\alpha_H^{-1}\left(\left(|T-T_c|/T_c\right)^{-\alpha_H}-1\right) 
\ ,\\
m   &\propto& \left(T_c-T\right)^\beta \ , \\
\chi &\propto& |T-T_c|^{-\gamma} \ ,\\
m    &\propto&  H^{1/{\delta_H}} \ , \\
G_{c}^{(2)}(r) &\propto& r^{2-d-\eta_a} \ , \\
\xi &\propto& |T-T_c|^{-\nu_{\perp}}  \ .
\end{eqnarray}
Here $c_H$ denotes the specific heat, $m$ the order parameter,
$\chi$ the susceptibility and $\xi$ the correlation length.
The presence of another degree of freedom besides the temperature $T$, 
like a (small) external field (labeled by $H$), leads to
other interesting power laws when $H\,{\rightarrow}\,0$. The $d$
present in the expression of two-point correlation function
$G_{c}^{(2)}(r)$ is the space dimension of the system.

Some laws are valid both to the right and to the left of the critical 
point; the values of the relative proportionality constants, or 
{\it amplitudes}, are in general different for the two branches of the 
functions, whereas the exponent is the same. However there are
universal amplitude relations among them.
We can see that there are altogether six basic exponents
Nevertheless they are not independent of each other, but
related by some simple scaling relations:
\begin{eqnarray}
\label{scalrelat}
\alpha_H + 2\beta + \gamma=2,\,\,\,
\alpha_H + \beta(\delta_H + 1) = 2, \\ \nonumber
(2 - \eta_a)\nu_{\perp} = \gamma, \,\,\,
\nu_{\perp}d = 2- \alpha_H \ .
\end{eqnarray}
The last relation is a so-called hyper-scaling law, which depends
on the spatial dimension $d$ and is not valid above the upper critical
dimension $d_c$, for example by the Gaussian theory.
Therefore below $d_c$ there are only two independent exponents in equilibrium.
One of the most interesting aspects of second order
phase transitions is the so-called {\bf universality}, i.e.,
the fact that systems which can be very different from each other
share the same set of critical indices (exponents and some amplitude ratios).
One can therefore hope to assign all systems to {\bf classes}
each of them being identified by a set of critical indices.

\subsection{Static percolation cluster exponents} \label{clusexps}

Universal behavior may occur at percolation \cite{Stau,Grim}, which 
can be considered a purely geometrical phenomenon describing the occurrence 
of infinitely large connected clusters on lattices. On the other
hand such clusters emerge at critical phase transition of lattice models
indeed.
The definition of connected clusters is not unambiguous. It may mean the set 
of sites or bonds with variables in the same state, or connected
by bonds with probability $b=1-\exp(-2J/kT)$.

By changing the system control parameters ($p\to p_p$) (that usually is 
the temperature in equilibrium systems) the coherence length between sites 
may diverge as
\begin{equation}
\xi(p) \propto |p-p_p|^{-\nu_{\perp}} \ \ ,
\end{equation}
hence percolation at $p_p$ like standard critical phenomena exhibits
renormalizibility and universality of critical exponents. 
At $p_p$  the cluster size ($s$) distribution follows the 
scaling law:
\begin{equation}
n_s \propto s^{-\tau} f(|p-p_p|s^{\sigma}) \ .
\end{equation}
while moments of this distribution exhibit singular behavior the
exponents:
\begin{eqnarray}
\sum_s s n_s(p) &\propto& |p-p_p|^{\beta_p} \ ,     \\
\sum_s s^2 n_s(p) &\propto& |p-p_p|^{-\gamma_p} \ \ \ .
\end{eqnarray}
Further critical exponents and scaling relations among them are shown in 
\cite{Stau}. 
In case of completely random placement of (sites, bonds, etc) 
variables (with probability $p$) on lattices we find 
{\bf random isotropic (ordinary) percolation} (see Sect.\ref{Isopercsect}). 
Percolating clusters may arise at critical, thermal transitions or by 
nonequilibrium processes. If the critical point ($p_c$) of the order 
parameter does not coincide with $p_p$ than at the percolation transition 
the order parameter coherence length is finite and does not influence the 
percolation properties. We observe random percolation in that case. 
In contrast if $p_p=p_c$ percolation is influenced by the order
parameter behavior and we find different, {\bf correlated percolation} 
universality \cite{FK72,Con-Kl,Stau}
whose exponents may coincide with those of the order parameter.

According to the Fortuin-Kasteleyn construction of clusters \cite{FK72}
two nearest-neighbor spins of the same state belong to the same cluster 
with probability $b=1-\exp(-2J/kT)$. It was shown that using 
this prescription for $Z_n$ and $O(n)$ and symmetric models 
\cite{Con-Kl,Bial,Fort,Blanc} the {\bf thermal phase transition} point
coincides with the percolation limits of such clusters. 
On the other hand in case of ``pure-site clusters'' ($b=1$) different,
universal cluster exponents are reported in two dimensional 
models \cite{Fort,Fort2} (see Sects. \ref{Ipercs}, \ref{PpercsI}, \ref{PpercsO}).

\subsection{Dynamical critical exponents}\label{expdef}

Nonequilibrium systems were first introduced to study relaxation to 
equilibrium states \cite{Hohen} and phase ordering kinetics 
\cite{Bin,Mar}. Power-law time dependences were investigated
away from the critical point as well, example by the domain growth in 
a quench to $T=0$.
Later the combination of different heat-baths, different dynamics, external 
currents became popular investigation tools of fully nonequilibrium 
models. To describe the dynamical behavior of a critical system 
additional exponents were introduced. For example the relation of the 
divergences of the relaxation time $\tau$ and correlation length $\xi$
is described by the dynamical exponent $Z$ 
\begin{equation}
\tau \propto \xi^Z \ \ .
\end{equation}
Systems out of equilibrium may show anisotropic scaling of two (and n)
point functions
\begin{equation}
G(b{\bf r},b^{\zeta}t) = b^{-2x} G({\bf r},t) \  \,
\end{equation}
where ${\bf r}$ and $t$ denote spatial and temporal coordinates, $x$ is the
scaling dimension and $\zeta$ is the anisotropy exponent. 
As a consequence the temporal ($\nu_{||}$) and spatial ($\nu_{\perp}$) 
correlation length exponents may be different, described by $\zeta=Z$.
\begin{equation}
Z=\zeta=\frac{\nu_{||}} {\nu_{\perp}} \ \ .
\end{equation}
For some years it was believed that dynamical critical phenomena are 
characterized by a set of three critical exponents, comprising two independent 
static exponents (other static exponents being related to these by scaling 
laws) 
and the dynamical exponent $Z$. Recently, it was discovered that there is 
another dynamical exponent, the `non-equilibrium' or {\bf short-time exponent} 
$\lambda$, needed to describe two-time correlations in a spin system 
($\{s_i\}$) of size $L$ relaxing to the critical state from a disordered 
initial condition \cite{Janssen,Huse}.
\begin{equation}
A(t,0) = \frac{1}{L^d} < \sum_i s_i(0) s_i(t) > \propto t^{-\lambda/Z}
\end{equation}

More recently the {\bf persistence exponents} $\theta_l$ and $\theta_g$ 
were introduced by \cite{DBG94,Majumpers}. These are 
associated with the probability, $p(t)$, that the local or global order 
parameter has not changed sign in time $t$ following a quench to the critical 
point.
In many systems of physical interest these exponents decay algebraically as
\begin{equation}
p(t) \propto t^{-\theta} 
\end{equation}
(see however example Sect. \ref{AB}).
It turned out that in systems where the scaling relation
\begin{equation} \label{persscal}
\theta_g Z = \lambda - d + 1 - \eta_a/2 
\end{equation}
is satisfied the dynamics of the global order parameter is a Markov process. 
In contrast in systems with non-Markovian global order parameter 
$\theta_g$ is in general a new, non-trivial critical exponent \cite{Majumpers}.
For example it was shown that while in the $d=1$ Glauber Ising model the 
magnetization is Markovian and the scaling relation (\ref{persscal}) is 
fulfilled, at the critical point of the $d=1$ NEKIM Ising model this is not 
satisfied and the persistence behavior there is characterized by a
different, non-trivial $\theta_g$ exponent \cite{perscikk} 
(see discussion in section \ref{nekim}).
As we can see the {\it universality classes of static models are split by the
dynamical exponents}.

\subsection{Critical exponents of spreading processes}\label{spexps}

In the previous section I defined quantities describing dynamical properties
of the bulk of a material. In a dual way to this one may also consider
cluster properties arising by initiating a process from an ordered (correlated)
state with a small cluster of activity.
Here I define a basic set of critical exponents that occur in
spreading processes and show the scaling relations among them.
In such processes phase transition may exist to {\bf absorbing state(s)}
where the density of spreading entity (particle, agent, epidemic etc.) 
disappears. The order parameter is usually the density of active 
sites $\{s_i\}$
\begin{equation}
\rho(t)= \frac{1}{L^d} \langle \sum_i s_i(t) \rangle  \,,
\end{equation}
which in the supercritical phase vanishes as
\begin{equation}
\label{StatDensity}
\rho^{\infty} \propto |p-p_c|^\beta
\, ,
\end{equation}
as the control parameter p is varied.
The ``dual'' quantity is the ultimate survival probability $P_\infty$ 
of an infinite cluster of active sites that scales in the active phase as
\begin{equation}
\label{PScaling}
P_\infty \propto |p-p_c|^{\beta^\prime} \
\end{equation}
with some critical exponent $\beta'$ \cite{GrasTor}. 
In field theoretical description
of such processes $\beta$ is associated with the particle annihilation,
$\beta'$ with the particle creation operator and in case of time 
reversal symmetry (see Eq. (\ref{DPsymeq})) they are equal.
The critical long-time behavior of these quantities are described by
\begin{equation}
\label{DensityScaling}
\rho(t) \propto t^{-\alpha}  f( \Delta \, t^{1/\nu_{||}})
\ , \qquad
P(t) \propto t^{-\delta}  g( \Delta \, t^{1/\nu_{||}}) \ ,
\end{equation}
where $\alpha$ and $\delta$ are the critical exponents for decay and survival,
$\Delta=|p-p_c|$, $f$ and $g$ are {\em universal scaling functions}
\cite{GrasTor,MGT,Jansperc}. The obvious scaling relations among them are

\begin{equation}
\alpha = \beta/\nu_{||} \ , \qquad
\delta = \beta^\prime/\nu_{||} \ .
\end{equation}
For finite systems (of size $N=L^d$) these quantities scale as
\begin{eqnarray}
\label{FSDensityScaling}
\rho(t) & \propto & t^{-\beta/\nu_{||}} \,
f'( \Delta \, t^{1/\nu_{||}}, \, t^{d/Z}/N) \ ,
\\
\label{FSScalingSurvival}
P(t) & \propto & t^{-\beta^\prime/\nu_{||}} \,
g'(\Delta \, t^{1/\nu_{||}}, \, t^{d/Z}/N) \ .
\end{eqnarray}
For ``relatively short times'' or for initial conditions with a single 
active seed the the number of active sites $N(t)$ and its mean square of 
spreading distance ($x_i$) from the origin 
\begin{equation}
R^2(t) = \frac{1}{N(t)} \langle \sum_i x_i^2(t) \rangle \label{R2def}
\end{equation}
follow the ``initial slip'' scaling laws \cite{GrasTor}
\begin{equation}
N(t) \propto t^{\eta} \ \ ,
\end{equation}
\begin{equation}
R^2(t) \propto t^z \ \ ,
\end{equation}
and usually the $z=2/Z$ relation holds.

\subsubsection{Damage spreading exponents} \label{DSsect}

Phase transitions between chaotic and non-chaotic states may be described
by damage spreading (DS). While DS was first introduced in biology 
\cite{Kaufman} it has become an interesting topic in physics as well 
\cite{Creutz,Stanley,Derrida}.
The main question is if a damage introduced in a dynamical system survives 
or disappears. To investigate this the usual technique is to make replica(s) 
of the original system and let them evolve with the same dynamics and 
external noise.
This method has been found to be very useful to measure accurately dynamical
exponents of equilibrium systems \cite{GrasA}. It has turned out however, that
DS properties do depend on the applied dynamics. An example is the case of the
two-dimensional Ising model with heat-bath algorithm versus Glauber dynamics
\cite{Mariz,Jan,GrasJPA}.

To avoid dependency on the dynamics a definition of "physical" 
family of DS dynamics was suggested by \cite{HinWD} according to 
the active phase may be divided to a sub-phase in which DS occurs for 
every member of the family, another sub-phase where the damage heals 
for every member of the family and a third possible sub-phase, 
where DS is possible for some members and the damage disappears for other 
members. The family of possible DS 
dynamics is defined to be consistent with the physics of the 
single replicas (symmetries, interaction ranges etc.).

Usually the order parameter of the damage is the Hamming distance between
replicas
\begin{equation}
D(t) = \left < {1\over L} \sum_{i=1}^L \vert s(i) - s^,(i) \vert \right > 
\label{Dscal}
\end{equation}
where $s(i)$ and $s'(i)$ denote variables of the replicas.
At continuous DS transitions $D$ exhibits power-law singularities as
physical quantities at the critical point. For example one can follow
the fate of a single difference between two (or more) replicas and
measure the spreading exponents: 
\begin{equation}
D(t)\propto t^{\eta_d} \ ,
\end{equation}
Similarly the survival probability of damage variables behaves as:
\begin{equation}
P_D(t)\propto t^{-\delta_d} \,
\end{equation}
and similarly to eq.(\ref{R2def}) the average mean square spreading distance 
of damage variables from the center scales as:
\begin{equation}
R^2_D(t)\propto t^{z_d} \ .
\end{equation}
Grassberger conjectured, that DS transitions should belong to DP class
(see Sect.\ref{DPS}) unless they coincide with other transition points and 
provided the probability for a locally damaged state to become healed is 
not zero \cite{GrasJ}. This hypothesis has been confirmed by simulations 
of many different systems.

%
\subsection{Field theoretical approach to reaction-diffusion \\
systems}
\label{fieldthint}

In this review I define nonequilibrium systems formally
by their field theoretical action where it is possible.
Therefore in this subsection I give a brief introduction to (bosonic) 
field theoretical formalism. 
This will be through the simplest example of reaction-diffusion systems,
via the: $A+A\to\emptyset$ annihilating random walk (ARW) (see Sect.\ref{2A0}).
Similar stochastic differential equation formalism can also be set up for
growth processes in most cases.
For a more complete introduction see \cite{Cardyintro,Cardytalk,Uwetalk}.

A proper field theoretical treatment should start from the master equation
for the microscopic time evolution of probabilities $p(\alpha;t)$ of
states $\alpha$
\begin{equation}
\frac{d p(\alpha;t)}{dt} = \sum_{\beta} R_{\beta\to\alpha} p(\beta;t) -
\sum_{\beta} R_{\alpha\to\beta} p(\alpha;t) \ ,
\end{equation}
where $R_{\alpha\to\beta}$ denotes the transition matrix from state 
$\alpha$ to $\beta$.
In field theory this can be expressed in Fock space formalism with
annihilation ($a_i$) and creation ($c_i$) operators satisfying the commutation
relation
\begin{equation}
[ a_i,c_j ] = \delta_{ij} \ .
\end{equation}
The states are built up from the vacuum $|0>$ as the linear superposition
\begin{equation}
\Psi(t) = \sum_{\alpha} p(n_1,n_2,...;t) c_1^{n_1} c_2^{n_2} ... |0> \ ,
\end{equation}
with occupation number coefficients $p(n_1,n_2,...;t)$.
The evolution of states can be described by a Schr\"odinger-like equation
\begin{equation}
\frac{d\Psi(t)}{dt}= -H \Psi(t) 
\end{equation}
with a generally non-hermitian Hamiltonian, which in case of the ARW
process looks like
\begin{equation}
H = D \sum_{ij} (c_i - c_j)(a_i-a_j) -\lambda\sum_j (a_j^2 - c_j^2 a_j^2) \ ,
\end{equation}
here $D$ denotes the diffusion strength and $\lambda$ the annihilation rate.
By going to the continuum limit this turns into
\begin{equation}
H = \int d^d x \left[D(\nabla \psi)(\nabla \phi) -\lambda 
(\phi^2 - \psi^2 \phi^2) \right] \ ,
\end{equation}
and in the path integral formalism over fields $\phi(x,t)$, $\psi(x,t)$
with weight $e^{-S(\phi,\psi)}$ one can define an action, that in case of
ARW is
\begin{equation}
S = \int dt d^d x \left[ \psi\partial_t\phi + D\nabla\psi\nabla\phi
-\lambda(\phi^2-\psi^2\phi^2)\right] \ .
\label{ARWaction}
\end{equation}
The action is analyzed by renormalization group (RG) methods at criticality
\cite{Ma,Amit}, usually by perturbative epsilon expansion below the upper 
critical dimension $d_c$ --- that is the lower limit of the validity of the
mean-field (MF) behavior of the system.
The symmetries of the model can be expressed in terms of the $\phi(x,t)$
field and $\psi(x,t)$ response field variables and the corresponding
hyper-scaling relations can be derived \cite{MGT,Jansperc}.

By a Gaussian transformation one may set up an alternative formalism ---
integrating out the response field --- the Langevin equation, that in
case of ARW is
\begin{equation}
\partial_t\phi = D \nabla^2\phi - 2\lambda\phi^2 + \eta(x,t)
\end{equation}
with a Gaussian noise, exhibiting correlations
\begin{equation}
< \eta(x,t) \eta(x',t') > = - \lambda \phi^2 \delta(x-x')\delta(t-t')  \ .
\end{equation}
Here $\delta$ denotes the Dirac delta functional and 
$\lambda$ is the noise amplitude. 
From the Langevin equation -- if it exists -- one can deduce a naive upper
critical dimension ($d_c$) by power counting. However this estimate may be 
modified by fluctuations, which can be analyzed by the application of the 
RG method.

%
\section{Scaling at first order phase transitions} \label{1storder}

In nonequilibrium systems dynamical scaling of variables may occur even
when the order parameter jumps at the transition. We call such a transition 
first order, although the free energy is not defined.
First order phase transitions 
have rarely been seen in low dimensions. This is due to the fact that in lower 
dimension the fluctuations are more relevant and may destabilize the 
ordered phase.
Therefore fluctuation induced second ordered phase transitions are likely to
appear. Hinrichsen advanced the hypothesis \cite{Hayefirsto}
that first-order transitions do not exist in 1+1 dimensional systems 
without extra symmetries, conservation laws, special boundary conditions or 
long-range interactions (which can be generated by macroscopic currents or 
anomalous diffusion in nonequilibrium systems for instance).
Examples are the Glauber and the NEKIM Ising spin systems
(see sections \ref{Isingcl},\ref{nekim}) possessing $Z_2$ symmetry in one 
dimension \cite{gla63,cpccikk}, where the introduction of a 
``temperature like'' flip inside of a domain or an external field ($h$)
causes discontinuous jump in the magnetization order parameter ($m$). 
Interestingly enough the correlation length diverges at the transition 
point: 
$\xi\propto p_T^{-\nu_{\perp}}$ and static 
\begin{equation}
m\propto\xi^{-\beta_s/\nu_{\perp}} g(h\xi^{\Delta/\nu_{\perp}})
\end{equation}
as well as cluster critical exponents can be defined:
\begin{eqnarray}
P_s(t,h)\propto t^{-\delta_s} \\
R_s^2(t,h) \propto t^{z_s} \\
\vert m(t,h)-m(0)\vert \propto t^{\eta_s} \\
\lim_{t\to\infty} P_s(t,h) \propto h^{\beta_s^,}
\end{eqnarray}
Here $s$ refers to the spin variables.
The table \ref{1dIsingspinexps} summarizes the results obtained for these 
transitions.
\begin{table}
\begin{center}
\begin{tabular}{|l|l|l|l|l|l|l|l|}
\hline
  & $\beta_s$&$\nu_{\perp}$& ${\beta_s}'$&$\Delta$&$\eta_s$&$\delta_s$&$z_s$\\
\hline
Glauber &$0$&$1/2$& $.99(2)$ & $1/2$ & $.0006(4)$ & $.500(5)$&$1$ \\
\hline
NEKIM &$ .00(2)$&$.444$ &$.45(1)$&$.49(1)$ & $.288(4)$ & $.287(3)$&$1.14$ \\
\hline
\end{tabular} \label{1dIsingspinexps}
\caption{Critical 1d Ising spin exponents at the Glauber and NEKIM 
transition points \cite{cpccikk}}
\end{center} 
\end{table}
Other examples for first order transition are known in driven diffusive
systems \cite{JS86}, in the 1d asymmetric exclusion process 
\cite{D98}, in bosonic annihilation-fission models
(Sect.\ref{binsp}), in asymmetric triplet and quadruplet models
\cite{tripcikk} (Sect.\ref{pkarwsect}) and DCF models for 
$D_A>D_B$ and $d>1$ \cite{OWLH} (Sect.\ref{DCF}).
By simulations it is quite difficult to decide whether a transition is
really discontinuous. The order parameter of weak first order 
transitions -- where the jump is small -- may look very similar to
continuous transitions. The hysteresis of the order parameter that 
is considered to be the indication of a first order transition is a 
demanding task to measure. There are some examples with debates over 
the order of the transition 
(see for example \cite{Dick1st,Hayefirsto,TO1,Sz1,Lipo1,Lipo2,Haye1}).
In some cases  the mean-field solution results in first order transition
\cite{boccikk,meorcikk}. 
In two dimensions there are certain stochastic cellular automata
for which systematic cluster mean-field techniques combined with simulations
made it possible to prove first oder transitions \cite{OdSzo} firmly
(see table \ref{GMF2}).
\begin{table}
\begin{center}
\begin{tabular}{|c|c|c|c|c|c|}
\hline
$n$& $y=1$ &\multicolumn{2}{c|}{$y=2$} &\multicolumn{2}{c|}{$y=3$} \\
   & $p_c$ & $p_c$ & $\rho(p_c)$ & $p_c$ & $\rho(p_c)$ \\
\hline
$1$& $0.111$ & ${\bf 0.354}$ & ${\bf 0.216}$ & ${\bf 0.534}$ & ${\bf 0.372}$\\
\hline
$2$& $0.113$ & ${\bf 0.326}$ & ${\bf 0.240}$ & ${\bf 0.455}$ & ${\bf 0.400}$\\
\hline
$4$& $0.131$ & ${\bf 0.388}$ & ${\bf 0.244}$ & ${\bf 0.647}$ & ${\bf 0.410}$\\
\hline
simulation& $0.163$ & ${\bf 0.404}$ & ${\bf 0.245}$ & ${\bf 0.661}$ & ${\bf
0.418}$\\ \hline
\end{tabular}
\caption{Convergence of the critical point estimates in various ($y=1,2,3$) 
two dimensional SCA calculated by n-cluster (GMF) approximation
(see section \ref{sectsca}).
First order transitions are denoted by boldface numbers.
The gap sizes ($\rho(p_c)$) of the order parameter shown for $y=2,3$
increase with $n$, approximating the simulation value \cite{OdSzo}.}
\label{GMF2}
\end{center}
\end{table}

\section{Out of equilibrium classes} \label{eqintro}

In this section I begin introducing basic nonequilibrium classes starting 
with the simplest dynamical extensions of equilibrium models. These dynamical
systems exhibit hermitian Hamiltonian and starting from a nonequilibrium state 
they evolve into a Gibbs state. Such nonequilibrium models include, 
for example, phase ordering systems, spin glasses, glasses etc. 
In these cases one is usually interested in the `nonequilibrium dynamics' 
at the equilibrium critical point. 

It is an important and new universal phenomenon that scaling behavior can be
observed far away from criticality as well. In quenches to zero temperature
of {\bf model A} systems (which do not conserve the order parameter)
the characteristic length in the late time regime grows with an universal 
power-law $\xi\propto t^{1/2}$, while in case of {\bf model B} systems 
(which conserve the order parameter) $\xi\propto t^{1/3}$. 
In {\bf model C} systems a conserved secondary density is coupled to the 
non-conserved order parameter. Such models may exhibit {\bf model A} behavior
or $\xi\propto t^{1/(2+\alpha_{H}/\nu_{\perp})}$ depending on the model 
parameters. 
The effects of such conservation laws in critical systems without hermitian 
Hamiltonian have also been investigated and will be discussed in 
later sections.

Since percolation is a central topic in reaction-diffusion systems 
discussed in Sects.~\ref{genchap} and ~\ref{multichap}, for the sake of 
completeness I show recent percolation results obtained for systems of 
Section \ref{eqintro} too. 
Another novel feature of dynamic phase transitions is the emergence of a
chaotic state, therefore I shall discuss damage spreading transitions and
behavior in these systems.

Then I continue towards such nonequilibrium
models which have no hermitian Hamiltonian and equilibrium Gibbs state.
In this section I show cases when this is achieved by combining different 
competing dynamics (for example by connecting two reservoirs with different 
temperatures to the system) or by generating current from outside.
Field theoretical investigations have revealed that {\bf model A} systems
are {\bf robust} against the
introduction of various competing dynamics, which are {\it local} and
{\it do not conserve the order parameter} \cite{GJY}.
Furthermore it was shown, that this robustness of the critical behavior
persists if the competing dynamics breaks the discrete symmetry of the
system \cite{BS94} or it comes from reversible mode coupling to a
non-critical conserved field \cite{TR97}.
On the other hand if a competing dynamics is coupled to {\bf model B} systems
by an external drive \cite{S-Z} or by local,
anisotropic order-parameter conserving process \cite{S-Z91,S93,B-R94,RaczXY}
long-range interactions are generated in the steady state with angular
dependence. The universality class will be the same as that of the kinetic
version of the equilibrium Ising model with {\bf dipolar long-range
interactions}.

As the number of neighboring interaction sites decreases by lowering the
spatial dimensionality of a system with short-ranged interactions the relevance
of fluctuations increases. In equilibrium models finite-range interactions
cannot maintain long-range order in $d<2$. This observation is known as the 
Landau-Peierls argument \cite{LP}. According to the Mermin-Wagner theorem 
\cite{MW}, for systems with continuous symmetry long-range order do not 
exist even in $d=2$. 
Hence in equilibrium models phase transition universality classes exist
for $d\ge 2$ only.  
One of the main open questions to be answered is whether there exist a 
class of nonequilibrium systems with restricted dynamical rules for which 
the Landau-Peierls or Mermin-Wagner theorem can be applied. 

   \subsection{Ising classes} \label{Isingcl}

The equilibrium Ising model was introduced by \cite{Ising} as the
simplest model for an uniaxial magnet but it is used in different settings 
for example binary by fluids or alloys as well. It is defined in terms of spin 
variables $s_i=\pm 1$ attached to sites $i$ of some lattice with the 
Hamiltonian
\begin{equation} \label{HIsing}
H = -J \sum_{i,i^,} s_i s_{i^,} - B \sum_i s_i \ \,
\end{equation}
where $J$ is the coupling constant and $B$ is the external field.
In one and two dimensions it is solved exactly \cite{Onsager},
hence it plays a fundamental test-ground for understanding phase 
transitions.
The Hamiltonian of this model exhibits a global, so called $Z_2$ 
(up-down) symmetry of the state variables.
While in one dimension a first order phase transition occurs at $T=0$ only 
(see section \ref{1storder}) in two dimensions there is a continuous phase 
transition where the system exhibits conformal symmetry \cite{Ising2D} 
as well. The critical dimension is $d_c=4$.
The following table (\ref{Isinge}) summarizes some of the known 
critical exponents of the Ising model.
\begin{table} 
\begin{center} 
\begin{tabular}{|c|c|c|c|}
\hline
exponent  & $d=2$   & $d=3$   & $d=4$(MF)  \\ \hline
$\alpha_H$  & 0(log)  &0.1097(6)& 0   \\ \hline
$\beta$   & 1/8     &0.3265(7)& 1/2 \\ \hline
$\gamma$  & 7/4     &1.3272(3)& 1   \\ \hline
$\nu_{\perp}$     & 1       &0.6301(2)& 1/2 \\ 
\hline
\end{tabular}
\caption{Static critical exponents of the Ising model}
\label{Isinge}
\end{center}
\end{table}
The quantum version of the Ising model, which in the simplest cases 
might take the form (in 1d)
\begin{equation}
H = -J \sum_i (t\sigma_i^x + \sigma_i^z\sigma_{i+1}^z + h\sigma_i^z) \ ,
\end{equation}
-- where $\sigma^{x,z}$ are Pauli matrices and $t$ and $h$ are couplings --
for $T>0$ has been shown to exhibit the same critical behavior as the
classical one (in the same dimension). For $T=0$ however quantum effects 
become important and the quantum Ising chain can be associated with the two 
dimensional classical Ising model such that the transverse field $t$ plays 
the role of the temperature.
In general a mapping can be constructed between classical $d+1$ dimensional 
statistical systems and $d$ dimensional quantum systems without changing the
universal properties, which has been widely utilized \cite{Suz71,FS}.
The effects of disorder and boundary conditions are not discussed here 
(for recent reviews see \cite{AlMu01,IPT93}).

\subsubsection{Correlated percolation clusters at $T_C$}\label{Ipercs}

If we generate clusters in such a way that we join nearest neighbor spins
of the same sign we can observe percolation at $T_c$ in 2d.
While the order parameter percolation exponents $\beta_p$ and $\gamma_p$ of
this percolation (defined in section \ref{Isopercsect}) were found to be 
different from the exponents of the magnetization ($\beta$, $\gamma$) 
the correlation length exponent is the same: $\nu=\nu_p$. 
For 2d models with $Z_2$ symmetry the universal percolation exponents are 
\cite{Fort,Fort2}:
\begin{equation}
\beta_p=0.049(4), \qquad  \gamma_p=1.908(16) \ .
\end{equation}
These exponents are clearly different from those of the ordinary percolation 
classes (Table \ref{Operct}) or from Ising class magnetic exponents.

On the other hand by Fortuin-Kasteleyn cluster construction \cite{FK72} 
the percolation exponents of the Ising model at $T_c$ coincide with those 
of the magnetization of the model.

\subsubsection{Dynamical Ising classes} \label{dynissect}

Kinetic Ising models such as the spin-flip Glauber Ising model \cite{gla63}
and the spin-exchange Kawasaki Ising model \cite{kawa66} were originally 
intended to study relaxational processes near equilibrium states.
In order to assure the arrival to an equilibrium state the detailed balance 
condition for transition rates ($w_{i\to j}$) and probability distributions
($P(s)$) are required to satisfy
\begin{equation} \label{DBal}
w_{i\to j} P(s(i)) = w_{j\to i} P(s(j)) \ \ .
\end{equation}
Knowing that $P_{eq}(s)\propto \exp(-H(s)/(k_B T))$ this entails the
\begin{equation}
\frac {w_{i\to j}} {w_{j\to i}} = \exp(-\Delta H(s)/(k_B T))
\end{equation}
condition which can be satisfied in many different ways. 
Assuming {\bf spin-flips} (which do not conserve the magnetization
({\bf model A})) Glauber formulated the most general form in a magnetic 
field ($h$)
\begin{equation}
{w_i}^h=w_i(1-\tanh{h}s_i)\approx w_i(1-hs_i)
\end{equation}
\begin{equation}
w_i = {\frac{\Gamma}{2}}(1+\tilde {\delta} s_{i-1}s_{i+1})\left(1 - 
{\gamma\over2}s_i(s_{i-1} + s_{i+1})\right)
\label{eq:wi}
\end{equation}
where $\gamma=\tanh{{2J}/{kT}}$, $\Gamma$ and $\tilde\delta$ are further
parameters.
The $d=1$ Ising model with Glauber kinetics is exactly solvable.
In this case the critical temperature is at $T=0$ and the transition is of 
first order. We recall that $p_T=e^{-\frac{4J}{kT}}$ plays the role of 
$\frac{T-T_c}{T_c}$ in 1d and in the vicinity of $T=0$ critical
exponents can be defined as powers of $p_T$, thus e.g. that of the 
coherence length, $\nu_{\perp}$, via  $\xi\propto {p_T}^{-\nu_{\perp}}$ 
(see Section \ref{1storder}).
In the presence of a  magnetic field $B$, the magnetization is known exactly. 
At $T=0$ 
\begin{equation}
m(T=0,B)=sgn(B). 
\label{eq:sgh}
\end{equation}
Moreover, for $\xi \gg 1$ and
$B/{kT} \ll 1$ the the exact solution reduces to
\begin{equation}
m\sim 2h\xi\,; \qquad h=B/k_{B}T .
\label{eq:m1}
\end{equation}
In scaling form one writes:
\begin{equation}
m\sim \xi^{-\frac{\beta_s}{\nu_{\perp}}}g(h\xi^{\frac{\Delta}{\nu_{\perp}}})
\label{eq:m2}
\end{equation}
where $\Delta$ is the static magnetic critical exponent.
Comparison of eqs. (\ref{eq:m1}) and (\ref{eq:m2}) results in $\beta_s=0$ and 
$\Delta=\nu_{\perp}$ . These values are well known for the 1d Ising model. 
It is clear that the transition is discontinuous at $B=0$, also when changing 
$B$ from positive to negative values, see eq.(\ref{eq:sgh}). 
The order of limits are meant as: $B\to 0$ and then $T\to 0$.
The $\tilde\delta=0$, $\Gamma=1$ case is usually referred as the 
Glauber-Ising model. The dynamical exponents are \cite{gla63,Majumpers}: 
\begin{equation}
Z_{1d \ Glauber} = 2 \ \ , \ \ \theta_{g,1d \ Glauber} = 1/4 
\end{equation} 

Applying {\bf spin-exchange} Kawasaki dynamics, which conserves the 
magnetization ({\bf model B})
\begin{equation}
w_i=\frac{1}{2\tau}\left[1-\frac{\gamma_2}{2}(s_{i-1}s_i+s_{i+1}s{i+2})\right]
\end{equation}
where $\gamma_2=\tanh(2J/k_B T)$, the dynamical exponent is different. 
According to linear response theory \cite{linr} in one dimension, at the 
critical point ($T_c=0$) it is:
\begin{equation}
Z_{1d \ Kaw} = 5
\end{equation}
Note however, that in case of fast quenches to $T=0$ coarsening with scaling
exponent $1/3$ is reported \cite{CKS91}. 
Hence another dynamic Ising universality class appears with the same static 
but different dynamical exponents.

Interestingly while the two dimensional equilibrium Ising model is 
solved the exact values of the dynamical exponents are not known. 
Table \ref{Isinged} summarizes the known dynamical exponents of the Ising
model in $d=1,2,3,4$. The $d=4$ results are mean-field values.
\begin{table} 
\begin{center} 
\begin{tabular}{|c|c|c|c|c|c|c|c|}
\hline
 & \multicolumn{2}{c|}{$d=1$}&\multicolumn{2}{c|}{$d=2$}& $d=3$&
\multicolumn{2}{c|}{$d=4$} \\
    & A  & B &  A   & B    & A     & A  & B  \\
\hline
$Z$ & 2 & 5  & 2.165(10) & 2.325(10) & 2.032(4) & 2& 4
\\ \hline
$\lambda$&1 &    &0.737(1) & 0.667(8) &1.362(19) & 4
& \\ \hline
$\theta_g$ &  1/4 & &0.225(10)&           & 0.41(2) & 1/2 & \\ 
\hline
\end{tabular}
\caption{Critical dynamical exponents in the Ising model. Columns
denoted by (A) and (B) refer, model A and model B dynamics.
Data are from \cite{linr,ZK,ZK2,GrasA,J,St96,BZ}}.
\label{Isinged}
\end{center}
\end{table}
In Section \ref{nekim} I shall discuss an another fully nonequilibrium critical
point of the $d=1$ Ising model with competing dynamics (the NEKIM), where the 
dynamical exponents break the scaling relation (\ref{persscal}) and therefore
the magnetization is a non-Markovian process. For $d>3$ there is no 
non-Markovian effect (hence $\theta$ is not independent) but for $d=2,3$
the situation is still not completely clear \cite{BZ}.

In one dimension the domain walls (kinks) between up and down regions can 
be considered as particles. The spin-flip dynamics can be mapped onto 
particle movement
\begin{equation}
\uparrow\downarrow\downarrow\rightleftharpoons\uparrow\uparrow\downarrow \ \
\sim \ \ \bullet\circ \rightleftharpoons \circ\bullet
\end{equation}
or to the creation or annihilation of neighboring particles
\begin{equation}
\uparrow\uparrow\uparrow\rightleftharpoons\uparrow\downarrow\uparrow \ \
\sim \ \ \circ\circ \rightleftharpoons \bullet\bullet
\end{equation}
Therefore the $T=0$ Glauber dynamics is equivalent to the diffusion limited
annihilation (ARW) mentioned already in Section. \ref{1storder}.
By mapping the spin-exchange dynamics in the same way more complicated 
particle dynamics emerges, example:
\begin{equation}
\uparrow\uparrow\downarrow\downarrow \rightleftharpoons 
\uparrow\downarrow\uparrow\downarrow \ \
\sim \ \ \circ\bullet\circ \rightleftharpoons \bullet\bullet\bullet
\end{equation}
one particle may give birth of two others or three particle may coagulate
to one. Therefore these models are equivalent to branching and annihilating
random walks to be discussed in Section \ref{BARWe}.

\subsubsection{Competing dynamics added to spin-flip} \label{spinflipsect}

Competing dynamics in general break the detailed balance symmetry (\ref{DBal}) 
and make the kinetic Ising model to relax to a nonequilibrium steady state 
(if it exists). Generally these models become unsolvable, for an overview 
see \cite{Raczof}. It was argued by \cite{GJY}
that stochastic {\bf spin-flip} models with two states per site and updating 
rules of a short-range nature with $Z_2$ symmetry should belong to the 
(kinetic) Ising model universality class. Their argument rests 
on the stability of the dynamic Ising fixed point in $d=4-\epsilon$ 
dimensions with respect to perturbations preserving both the spin 
inversion and the lattice symmetries. This hypothesis has received 
extensive confirmation by Monte Carlo simulations \cite{mario}--\cite{nos},
\cite{tamayo,zia2, santos} as well as from analytic calculations 
\cite{ceu1,ceu2,tome}. The models investigated include Ising models 
with a competition of two (or three \cite{tamayo}) Glauber-like rates 
at different temperatures \cite{ceu1,tome,gonzalez,zia1}, or a 
combination of spin-flip and spin-exchange dynamics \cite{garrido}, 
majority vote models \cite{mario,santos} and other types of transition 
rules with the restrictions mentioned above \cite{nos}.
 
Note that in all of the above cases the ordered state is fluctuating and
non-absorbing. By relaxation processes this allows fluctuations in the
bulk of a domain. It is also possible (and in 1d it is the only choice)
to generate nonequilibrium two-state spin models (with
short ranged interactions) where the ordered states are frozen 
(absorbing), hence by the relaxation to the steady state fluctuations 
occur at the boundaries only. In this case non-Ising universality
appears which is called the voter model (VM) universality class 
(see Sect.\ref{glau}).

\begin{figure}
\begin{center}
\leavevmode
\epsfxsize=70truemm
\epsfbox{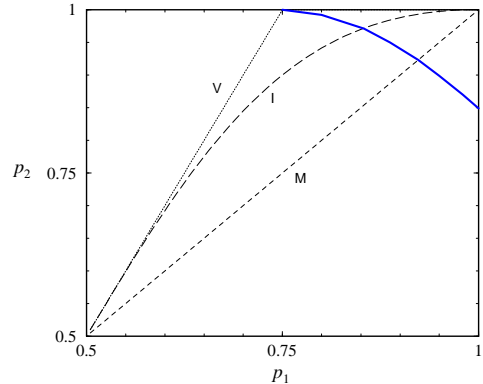}
\end{center}
\caption{Phase diagram of 2d, $Z_2$ symmetric nonequilibrium spin models
from \cite{Drou}. Broken lines correspond to the noisy voter model (V),
the Ising model (I) and the majority vote model (M). The low temperature
phase is located in the upper right corner, above the transition line 
(full line).
\label{fig_phase}} 
\end{figure}

In two dimensions {\bf general $Z_2$ symmetric update rules} were 
investigated by \cite{nos,Drou,Dor,AAM96}. The essence of this 
-- following \cite{Drou} -- is described below.
Let us consider a two dimensional lattice of spins $s_{i}=\pm 1$, 
evolving with the following dynamical rule. 
At each evolution step, the spin to be updated flips with
the heat bath rule: the probability that the spin $s_i$ takes the 
value $+1$ is $P(s_i=1)=p(h_i)$, where the local field
$h_{i}$ is the sum over neighboring sites $\sum_{j} s_{j}$ and
\begin{equation}
p(h)=\frac{1}{2}\left(1+\tanh [\beta(h) h]\right)
.\end{equation}
The functions $p(h)$ and $\beta(h)$ are defined over integral values of $h$.
For a square lattice, $h$ takes the values 4, 2, 0, $-2$, $-4$.
We require that $p(-h)=1-p(h)$, in order to keep the up down symmetry,
hence $\beta(-h)=\beta(h)$ and this fixes $p(0)=1/2$.
The dynamics therefore depends on two parameters 
\begin{equation}
p_{1}=p(2), \quad p_{2}=p(4)
,\end{equation}
or equivalently on two effective temperatures 
\begin{equation}
\label{defT}
T_1=\frac{1}{\beta(2)}, \quad T_2=\frac{1}{\beta(4)}
.\end{equation}
Defining the coordinate system
\begin{equation}
t_1=\tanh \frac{2}{T_1}, \quad
t_2=\tanh \frac{2}{T_2}
\end{equation}
with $0\le t_1,t_2\le1$ this yields
\begin{equation}
p_{1}=\frac{1}{2}\left( 1+t_1\right),
\quad
p_{2}=\frac{1}{2}\left( 1+\frac{2 t_2}{1+t_2^2}\right)
,\end{equation}
with $1/2\le p_1,p_2\le1$.
One can call $T_1$ and $T_2$ as two temperatures, respectively
associated with {\it interfacial noise}, and to {\it bulk noise}.
Each point in the parameter plane $(p_1,p_2)$, or alternatively in the
temperature plane $(t_1,t_2)$, corresponds to a particular model.
The class of models thus defined comprises as special cases the Ising model, 
the voter and anti-voter models \cite{Ligget}, as well as the majority vote 
\cite{Ligget,mario} model (see Fig. \ref{fig_phase}).
The $p_2=1$ line corresponds to models with no bulk noise ($T_2=0$),
hence the dynamics is only driven by interfacial noise, defined above.
The $p_1=1$ line corresponds to models with no interfacial noise
($T_1=0$), hence the dynamics is only driven by bulk noise.
In both cases effects due to the curvature of the interfaces is 
always present.
For these last models, the local spin aligns in the
direction of the majority of its neighbors with probability one, 
if the local field is $h=2$, 
i.e. if there is no consensus amongst the neighbors. 
If there is consensus amongst them, i.e. if $h=4$, the local
spin aligns with its neighbors with a probability $p_2<1$. 

Simulations \cite{nos} revealed that the transition line between the 
low and high temperature regions is Ising type except for the endpoint
($p_1=1$,$p_2=0.75$), that is first-order and corresponds to the
voter model class. The local persistence exponent was also found to be constant
along the line $\theta_l \sim 0.22$ \cite{Drou} in agreement with that of 
the A-model (see Table \ref{Isinged}) except for the VM point.
The dynamics of this class of models may be described formally in terms of 
reaction diffusion processes for a set of coalescing, annihilating, and 
branching random walkers \cite{Drou}. 
There are simulation results for other models exhibiting absorbing
ordered state indicating VM critical behavior\cite{Lipp,Hin97}.

It is important that nontrivial, nonequilibrium phase transition may occur 
even in one dimension if spin-exchange is added to spin-flip dynamics, 
the details will be discussed in Section \ref{nekim}.

\subsubsection{Competing dynamics added to spin-exchange} \label{DDSsect}

As mentioned in Section \ref{eqintro} {\bf model-B} systems are
more sensitive to competing dynamics. {\it Local and anisotropic} order 
parameter conserving processes generate critical behavior that coincides 
with that of the kinetic dipolar-interaction Ising model. In two dimensions
both simulations and field theory \cite{Prae,Prae2} predicts the critical 
exponents
\begin{table} 
\begin{center} 
\begin{tabular}{|c|c|c|c|}
\hline
$\beta$   & $\gamma$ & $\eta_a$   & $\nu_{\perp}$  \\ \hline
0.33(2)   & 1.16(6)  & 0.13(4)  & 0.62(3)   \\ 
\hline
\end{tabular}
\caption{Critical exponents of the $d=2$ randomly driven 
lattice gas}
\label{dipoleIe}
\end{center}
\end{table}
The critical dimension is $d_c=3$. It is shown that the Langevin equation 
(and therefore the critical behavior) of the anisotropic diffusive system 
coincides with that of the randomly driven lattice gas system as well.
Other systems in this universality class are the two-temperature model 
\cite{GLMS}, the ALGA model \cite{ALGA} and the infinitely fast driven
lattice gas model \cite{IDLG}. In the randomly driven lattice gas model 
particle current does not occur but an anisotropy can be found, 
therefore it was argued \cite{IDLG} that the particle current 
is not a relevant feature for this class.
This argument gives the possibility to understand why some set of 
simulations of driven lattice systems \cite{Val} leads to different 
critical behavior than that of the canonical coarse-grained 
representative of this class, in which an 
explicit particle current $j{\bf\hat x}$ is added to the continuous, 
model-B Ising model Hamiltonian: 
\begin{equation}
\frac {\partial \phi({\bf r},t)} {\partial t} = -\nabla \left[\eta \frac
{\delta H}{\delta\phi} + j {\bf\hat x} \right] + \nabla \zeta
\end{equation}
\cite{DickMar} (here $\eta$ is a parameter, $\zeta$ is a Gaussian
noise). In this model one obtains mean-field exponents for $2\ge d \ge 5$ 
(with weak logarithmic corrections at $d=2$) with $\beta=1/2$ exactly.
To resolve contradictions between simulation results of \cite{Val}
and \cite{Leu} ref. \cite{ZSS} raised the possibility of the existence 
of another, extraordinary, ``stringy'' ordered phase in Ising type 
driven lattice gases that may be stable in {\it square} systems.

\subsubsection{Long-range interactions and correlations} \label{IsinglongS}

Universal behavior is due to the fact at criticality long-range correlations
are generated that make details of short ranged interactions irrelevant.
However one can also investigate the scaling behavior in systems with 
long-range interactions or with dynamically generated long-range correlations.
If the Glauber Ising model (with non-conserving dynamics) changed to a 
nonequilibrium one in such a way that one couples a {\bf nonlocal dynamics} 
\cite{DRT} to it long-range isotropic interactions are generated and 
mean-field critical behavior emerges. For example if the nonlocal 
dynamics is a random Levy flight with spin exchange probability distribution
\begin{equation}
P(r) \propto \frac {1} {r^{d+\sigma}}
\end{equation}
effective long range interactions of the form $V_{eff}\propto r^{-d-\sigma}$
are generated and the critical exponents change continuously
as the function $\sigma$ and $d$ \cite{BR}. Similar conclusions for other 
nonequilibrium classes will be discussed later (Sect.\ref{sectlevydp}).

The effect of power-law correlated initial conditions 
$\langle \phi (0) \phi (r) \rangle \sim r^{-(d-\sigma)}$ 
in case of a {\bf quench to the ordered phase in 
systems with non-conserved order parameter} 
was investigated by \cite{bray}. 
An important example is the (2+1)-dimensional Glauber-Ising model 
quenched to zero temperature.
It was observed that long-range correlations are relevant
if $\sigma$ exceeds a critical value $\sigma_c$. Furthermore,
it was shown that the relevant regime is characterized by a continuously
changing exponent in the autocorrelation function
$A(t) = \left[ \phi (r,t) \phi (r,0) \right] \sim t^{-(d-\sigma)/4}$,
whereas the usual short-range scaling exponents 
could be recovered below the threshold. 
These are in agreement with the simulations of the two-dimensional Ising model
quenched from $T=T_c$ to $T=0$.

\subsubsection{Damage spreading behavior}

The high temperature phase of the Ising model is chaotic. By lowering
$T$ a non-chaotic phase may emerge at $T_d$ with different types of
transitions depending on the dynamics.
The dynamics dependent DS critical behavior in different Ising models
is in agreement with a conjecture by \cite{GrasJ}.
Dynamical simulations with heat-bath algorithm in 2 and 3 dimensions
\cite{GrasA,Gropen,WS96} resulted in $T_d=T_c$ with a DS dynamical 
exponent coinciding with that of the $Z$-s of the replicas. 
In this case the DS transition picks up the Ising class universality
of its replicas. 
With Glauber dynamics in 2d $T_d < T_c$ and DP class DS exponents 
were found \cite{GrasJPA}.
With Kawasaki dynamics in 2d on the other hand the damage always spreads
\cite{Vojta}. With Swendsen-Wang dynamics in 2d $T_d > T_c$ and
DP class DS behavior was observed \cite{HDS}.

In one dimensional, nonequilibrium Ising models it is possible to design 
different dynamics showing either PC or DP class DS transition as the 
function of some control parameter. This depends on the DS transition 
coincides or not with the critical point. 
In the PC class DS case damage variables follow BARW2 
dynamics (see Sects.\ref{BARWe} and \ref{nekim}) and $Z_2$ symmetric 
absorbing states occur \cite{HD97,odme98}.

   \subsection{Potts classes}

\begin{figure}
\epsfxsize=70mm
\epsffile{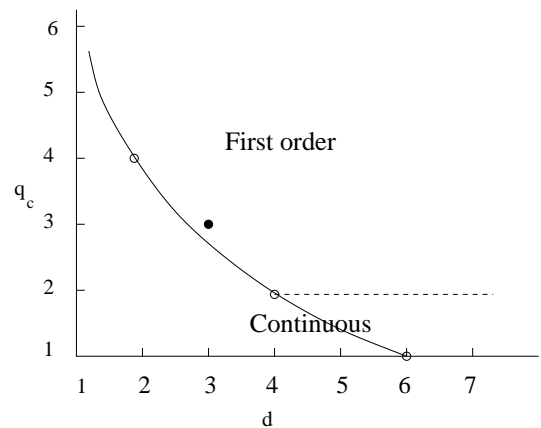}
\caption{Schematic plot from \cite{Wu} for $q_c(d)$ (solid line). 
Open symbols correspond to continuous phase transition, filled symbol 
to a known first order transition. Below the dashed line the transition
is continuous too.
\label{Pottspt}
}
\end{figure}

The generalization of the two-state equilibrium Ising model was introduced
by \cite{Potts}, for an overview see \cite{Wu}. In the $q$-state Potts 
model the state variables can take $q$ different values $s_i\in (0,1,2,...q)$ 
and the Hamiltonian is a sum of Kronecker delta function of states over 
nearest neighbors
\begin{equation}
H= -J \sum_{<i,i^,>} \delta(s_i-s_{i^,})
\end{equation}
This hamiltonian exhibits a global symmetry described by the permutation 
group of $q$ elements ($S_q$). The Ising model is recovered in the $q=2$ case 
(discussed in Section \ref{Isingcl}).
The $q$-state Potts model exhibits a disordered high-temperature
phase and an ordered low-temperature phase. The transition is 
first-order, mean-filed-like for $q$-s above the $q_c(d)$ curve (shown
in Figure \ref{Pottspt}) (and for $q>2$ in high dimensions).
The $q=1$ limit can be shown \cite{FK72} to be equivalent to the isotropic 
percolation (see Section \ref{Isopercsect}) that is known to exhibit 
a continuous phase transition with $d_c = 6$.
The problem of finding the effective resistance
between two node points of a network of linear resistors was solved by
Kirchhoff in 1847. Ref. \cite{FK72} showed that Kirchhoff's
solution can be expressed as a $q=0$ limit of the Potts partition function. 
Further mappings were discovered between the spin glass \cite{EA75} and the 
$q=1/2$ Potts model and the two dimensional $q=3,4$ cases to vertex models 
(see \cite{Baxter}).
From this figure we can see that for $q>2$ Potts models continuous 
transitions occur in two dimensions only ($q=3,4$). 
Fortunately these models are exactly solvable (see \cite{Baxter}) 
and exhibit conformal symmetry as well as topological, Yang-Baxter invariance. 
The static exponents in two dimensions are known exactly (Table \ref{Pottse})
\begin{table} 
\begin{center} 
\begin{tabular}{|c|c|c|c|c|c|}
\hline
exponent  & $q=0$     & $q=1$     & $q=2$   & $q=3$   & $q=4$ \\ \hline
$\alpha_H$  & $-\infty$ & $-2/3$    & 0(log)  & 1/3     & 2/3   \\ \hline
$\beta$   & 1/6       & 5/36      & 1/8     & 1/9     & 1/12 \\ \hline
$\gamma$  & $\infty$  & 43/18     & 7/4     & 13/9    & 7/6   \\ \hline
$\nu_{\perp}$     & $\infty$  & 4/3       & 1       & 5/6     & 2/3    \\ 
\hline
\end{tabular}
\caption{Static exponents of the $q$ states Potts model in two dimension}
\label{Pottse}
\end{center}
\end{table}

\subsubsection{Correlated percolation at $T_c$}\label{PpercsI}

In 2d models with $Z_3$ symmetry the critical point coincides with
the percolation of site connected clusters and the following percolation 
exponents are reported \cite{Fort}:
\begin{equation}
\beta_p=0.075(14), \qquad  \gamma_p=1.53(21) \ \ .
\end{equation}
In case of Fortuin-Kasteleyn cluster construction \cite{FK72} the percolation
exponents of the $q$ state Potts model at $T_c$ coincide with those of the 
magnetization of the model.

\subsubsection{Dynamical Potts classes}

The {\bf model-A} dynamical exponents of the Potts classes
are determined in two dimensions for 
$q=3,4$ by short-time Monte Carlo simulations \cite{ZPottse} 
(Table \ref{Pottsed}). The exponents were found to be the same for
heat-bath and Metropolis algorithms.
\begin{table} 
\begin{center} 
\begin{tabular}{|c|c|c|}
\hline
exponent  & $q=3$     & $q=4$ \\ \hline
$Z$       & 2.198(2)  & 2.290(3) \\ \hline
$\lambda$ & 0.836(2)  &          \\ \hline
$\theta_g$  & 0.350(8)&          \\ 
\hline
\end{tabular}
\caption{Model-A dynamical exponents of the $q$ states Potts model in 
two dimensions}
\label{Pottsed}
\end{center}
\end{table}
For the zero temperature {\bf local persistence} exponent in one dimension 
exact formulas have been determined. For sequential dynamics \cite{1dpers}
\begin{equation}
\theta_{l,s} = -\frac{1}{8} + \frac{2}{\pi^2} 
\left[ \cos^{-1}(\frac{2-q}{\sqrt 2q}) \right]^2
\end{equation}
while for parallel dynamics $\theta_{l,p}=2\theta_{l,s}$ \cite{Gautam}. 
In a deterministic coarsening it is again different (see \cite{BDG,Aj}.
As we can see {\it the dynamical universality class characterized by the 
$\xi\propto t^{1/2}$ characteristic length growth is split by different
dynamics, as reflected by the persistence exponent}. 

Similarly to the Ising model case in nonequilibrium systems the Potts 
model symmetry turned out to be a relevant factor for determining
the universal behavior of transitions to {\bf fluctuating ordered states} 
\cite{Brun,Cris,SzaboC}.
On the other hand in case of nonequilibrium transitions to 
{\bf absorbing states} a simulation study \cite{Lipp} suggests 
first order transition for all $q>2$ state Potts models in $d>1$ dimensions.
The $q=2$, $d=2$ case corresponds to the Voter model class 
(Sect. \ref{glau}) and in 1d either PC class transition ($q=2$) or
N-BARW2 class transition ($q=3$) (Sect.\ref{NBARWS}) occurs. 

The {\bf DS transition} of a $q=3$, $d=2$ Potts model with heat-bath 
dynamics was found to belong to the DP class (Sect.\ref{DPS}) because
$T_d > T_c$ \cite{STM97}.

      \subsubsection{Long-range interactions} \label{PottslongS}

The effect of long-range interaction has also been investigated in case
of the one dimensional $q=3$ Potts model \cite{1dPottsl} with the
Hamiltonian
\begin{equation}
H = - \sum_{i<j} \frac{J} {|i-j|^{1+\sigma}} \delta(s_i,s_j) \ .
\end{equation}
For $\sigma < \sigma_c \sim 0.65$ a crossover from second order to
first order (mean-field) transition was located by simulations. Similarly
to the Ising model here we can expect to see this crossover if we generate
the long-range interactions by the addition of a Levy type Kawasaki dynamics.

 \subsection{XY model classes} \label{XYsec}

The classical XY model is defined by the Hamiltonian
\begin{equation}
H = - J \sum_{i,i^,} \cos(\Theta_i-\Theta_{i^,}) \ \ .
\end{equation}
with continuous $\Theta_i\in\left[0,2\pi\right]$ state variables.
This model has a global $U(1)$ symmetry. 
Alternatively the XY model can be defined as a special $N=2$ case of 
$O(N)$ symmetric models such that the spin vectors are two dimensional
${\bf S}_i$ with absolute value ${\bf S}_i^2=1$
\begin{equation}
H = -J \sum_{<i,i^,>} {\bf S}_i {\bf S}_i^, \ \ \ .
\end{equation}
In this continuous model in {\bf two dimensions} no local order parameter 
can take zero value according to the Mermin-Wagner theorem \cite{MW}.
The appearance of free vortexes (which are non-local) cause an unusual 
transition mechanism that implies that most of the thermodynamic 
quantities do not show power-law singularities. The singular behavior 
of the correlation length ($\xi$) and the susceptibility ($\chi$) 
is described by the forms for $T>T_c$
\begin{equation}
\xi\propto\exp\left( C (T-T_c)^{-1/2} \right) \ \ ,
 \ \ \ \chi\propto\xi^{2-\eta_a}  \ \ ,
\end{equation}
where $C$ is a non-universal positive constant. 
Conventional critical exponents cannot be used, but one can define 
scaling dimensions. At $T_c$ the two-point correlation
function has the following long-distance behavior
\begin{equation}
G(r) \propto r^{-1/4} (\ln r)^{1/8}
\end{equation}
implying $\eta_a=1/4$, and in the entire low-temperature phase
\begin{equation}
G(r) \propto r^{-\eta_a(T)}
\end{equation}
such that the exponent $\eta_a$ is a continuous function of the temperature,
i.e. the model has a line of critical points starting from $T_c$ to $T=0$.
This is the so called Kosterlitz-Thouless critical behavior \cite{KT} and
corresponds to the conformal field theory with $c=1$ \cite{ID}.
This kind of transition can experimentally observed in many effectively
two-dimensional systems with $O(2)$ symmetry, such as thin films of 
superfluid helium and describes roughening transitions of SOS models at
crystal interfaces.
\begin{table} 
\begin{center} 
\begin{tabular}{|c|c|c|c|c|}
\hline
$\alpha_H$  & $\beta$   & $\gamma$ & $\nu_{\perp}$   & $\eta_a$ \\ \hline
-0.011(4) & 0.347(1)  & 1.317(2) & 0.670(1)& 0.035(2)\\ \hline
\end{tabular}
\caption{Static exponents of the XY model in three dimensions}
\label{XYe}
\end{center}
\end{table}
In {\bf three dimensions} the critical exponents of the O(N) ($N=0,1,2,3,4$)
symmetric field theory have been determined by perturbative expansions up to 
seventh loop order \cite{GZ}. The Table \ref{XYe} summarizes these results for 
the XY case (for a more detailed overview see \cite{PV}).

Exponents with {\bf model A} dynamics in 2d have been determined by 
short-time simulations and logarithmic corrections to scaling
were found \cite{YZYT}. Table \ref{XYed} summarizes the known dynamical 
exponents
\begin{table} 
\begin{center} 
\begin{tabular}{|c|c|c|}
\hline
$Z$  & $\lambda$   & $\eta$ \\ \hline
2.04(1)& 0.730(1)  & 0.250(2) \\ \hline
\end{tabular}
\caption{Model-A dynamical exponents of XY model in two dimensions}
\label{XYed}
\end{center}
\end{table}

\subsubsection{Long-range correlations}

Similarly to the Ising model \cite{RaczXY} studied the 
validity of the Mermin-Wagner theorem by transforming the two dimensional 
XY model to a non-equilibrium one using {\bf two-temperature, model-A} 
dynamics. They found that the Mermin-Wagner theorem does not apply for this 
case since effective long-range interactions are generated by the local 
nonequilibrium dynamics. The universality class of the
phase transition of the model coincides with that of the two 
temperature driven Ising model (see Table \ref{dipoleIe}).

\subsubsection{Self-propelled particles } \label{sppsect} 

An other XY-like nonequilibrium model that exhibits ordered state for
$d\le 2$ dimensions is motivated by the description of the `flocking" 
behavior among living things, such as birds, slime molds and bacteria.
In the simplest version of the self-propelled particle model 
\cite{VCBCS95} each particle's velocity is set to a fixed magnitude, $v_0$.
The interaction with the neighboring particles changes only the direction 
of motion: the particles tend to align their orientation to the local average 
velocity. In one dimension it is defined on the lattice as
\begin{eqnarray}
x_i(t+1) = x_i(t) + v_0 u_i(t), \nonumber\\
u_i(t+1) = G\Bigl(\langle u(t) \rangle_i\Bigr) + \xi_i,
\label{EOMD}
\end{eqnarray}
where the particles are characterized by their coordinate
$x_i$ and dimensionless velocity $u_i$ and the function $G$ that
incorporates both the propulsion and friction forces which set the 
velocity in average to a prescribed value $v_0$:
$G(u)>u$ for $u<1$ and $G(u)<u$ for $u>1$. 
The distribution function $P(x=\xi)$ of the noise $\xi_i$ is uniform
in the interval $[-\eta/2,\eta/2]$.
Keeping $v_0$ constant, the adjustable control parameters of the model 
are the average density of the particles,
$\rho$, and the noise amplitude $\eta$. 
The order parameter is the average velocity $\phi\equiv\langle u \rangle$
which vanishes as
\begin{equation}
\phi(\eta,\rho)\sim \cases{
         \Bigl({\eta_c(\rho) - \eta\over \eta_c(\rho)}\Bigr)^{\beta}
                & for $\eta<\eta_c(\rho)$ \cr
        0  & for $\eta>\eta_c(\rho)$ \cr
    },
\label{scale}
\end{equation}
at a critical $\eta_c(\rho)$ value.

This model is similar to the $XY$ model of classical
magnetic spins because the velocity of the particles, like the
local spin of the $XY$ model, has fixed length and continuous rotational
$U(1)$ symmetry. In the $v_0=0$ and low noise limit the
model reduces {\it exactly} to a Monte-Carlo dynamics of the $XY$ model.

A field theory that included in a self-consistent way the non-equilibrium 
effects was proposed by \cite{TT}. 
They have shown that their model is different from the XY model for $d<4$.
The essential difference between the self-propelled particle model and 
the equilibrium XY model 
is that at different times, the "neighbors" of one particular "bird" 
will be different depending on the velocity field itself. Therefore, two 
originally distant "birds" can interact with each other at some later time.
They found a critical dimension $d_c=4$, below which linearized hydrodynamics 
breaks down, but owing to a Galilean invariance they could obtain exact
scaling exponents in $d=2$. For the dynamical exponent they got $Z=6/5$.
Numerical simulations \cite{VCBCS95,CSV97} indeed found a long range ordered 
state with a continuous transition characterized by $\beta=0.42(3)$ in two
dimensions.

In one dimension the field theory and simulations \cite{CBV99} provided 
evidence for a continuous phase transition with $\beta=0.60(5)$, 
which is different from the mean-field value $1/2$ \cite{ST71}.

     \subsection{O(N) symmetric model classes}

As already mentioned in the previous section the $O(N)$ symmetric models
are defined on spin vectors ${\bf S}_i$ of unit length ${\bf S}_i^2=1$ with the
Hamiltonian
\begin{equation}
H = -J \sum_{<i,i^,>} {\bf S}_i {\bf S}_i^, \ \ \ .
\end{equation}
The most well known of them is the classical Heisenberg model that corresponds
to $N=3$ being the simplest model of isotropic ferromagnets. The $N=4$ case 
corresponds to the Higgs sector of the Standard Model at finite temperature.
The $N=0$ case is related to polymers and the $N=1$ and $N=2$ cases are the
Ising and XY models respectively. 
The critical dimension is $d_c=4$ and by the Mermin-Wagner 
theorem we cannot find finite temperature phase transition in the short range 
equilibrium models for $N>2$ below $d=3$. 
The static critical exponents have been determined by $\epsilon=4-d$
expansions up to five loop order \cite{Gor}, by exact RG methods 
(see \cite{EXRG} and the references therein), by simulations \cite{ONMC} 
and by series expansions in 3D (see \cite{GZ} and the references there).
In Table \ref{ONe} I show the latest estimates from \cite{GZ} in three 
dimensions for $N=0,3,4$.
\begin{table} 
\begin{center} 
\begin{tabular}{|c||c|c|c|c|c|}
\hline
N &  $\alpha_H$ & $\beta$    & $\gamma$ & $\nu_{\perp}$  & $\eta_a$ \\ \hline
0 &  0.235(3) & 0.3024(1)  & 1.597(2) & 0.588(1)& 0.028(2)\\ \hline
3 &  -0.12(1) & 0.366(2)   & 1.395(5) & 0.707(3)& 0.035(2) \\ \hline 
4 &  -0.22(2) & 0.383(4)   & 1.45(1)  & 0.741(6)& 0.035(4) \\ \hline  
\end{tabular}
\caption{Static exponents of the O(N) model in three dimensions}
\label{ONe}
\end{center}
\end{table}
The $N\to\infty$ limit is the exactly solvable spherical model \cite{Sph,Sph2}.
For a detailed discussion of the static critical behavior of the $O(N)$ models
see \cite{PV}.

The dynamical exponents for {\bf model A} are known exactly for the 
($N\to\infty$) spherical model \cite{Janssen,Majumpers} case. 
For other cases $\epsilon=4-d$ expansions up to two loop order 
exist \cite{Majumpers,OCB}.
\begin{table} 
\begin{center} 
\begin{tabular}{|c|c|c|c|}
\hline
N          & $Z$      & $\lambda$   & $\theta_g$ \\ \hline
$\infty$   & 2        & 5/2         & 1/4       \\ \hline
3          & 2.032(4) & 2.789(6)    & 0.38      \\ \hline
\end{tabular}
\caption{Model A dynamical exponents of the O(N) model in three dimensions}
\label{ONed}
\end{center}
\end{table}
For a discussion about the combination of different dynamics see the general
introduction Section \ref{eqintro} and \cite{TSR}.

\subsubsection{Correlated percolation at $T_c$}\label{PpercsO}

In three dimensions for $O(2)$, $O(3)$ and $O(4)$ symmetric models the 
Fortuin-Kasteleyn cluster construction \cite{FK72} results in
percolation points and percolation exponents which coincide with the 
corresponding $T_c$-s and magnetization exponent values \cite{Blanc}.

\section{Genuine, basic nonequilibrium classes} \label{genchap}

In this section I introduce such ``genuine nonequilibrium'' universality
classes that do not occur in dynamical generalizations of equilibrium systems.
Naturally in these models there is no hermitian Hamiltonian and they
are defined by transition rates not satisfying the detailed balance
condition (\ref{DBal}).  
They can be described by a master equation and the
deduced stochastic action or Langevin equation if it exists.
The most well known cases are reaction-diffusion systems with order-disorder
transitions in which the ordered state may exhibit only small fluctuations, 
hence they trap a system falling in it (absorbing state).
They may occur in models of population \cite{Alb}, 
epidemics \cite{Ligget,Mollison}, catalysis \cite{ZGB}
or enzyme biology \cite{Berry} for example.
There are also other nonequilibrium phase transitions for
example in lattice gases with currents \cite{EFGM95,KSKS98,EKKM98,E00}
or in traffic models \cite{Chowd}, but in these systems the critical 
universality classes have not been explored yet.

Phase transitions in such models may occur in low dimensions in contrast 
with equilibrium ones \cite{DickMar}. As it was already shown in
Section \ref{dynissect} reaction-diffusion particle systems may be 
mapped onto spin-flip systems stochastic cellular automata \cite{Chop} or 
interface growth models (see Sect. \ref{growth}). The mapping however
may lead to nonlocal systems that does not bear real physical relevance.
The universality classes of the simple models presented in this section
constitute the fundamental building blocks of more complex systems.

For a long time phase transitions with completely frozen absorbing states
were investigated. A few universality classes of this kind were known 
\cite{GrasWu},\cite{Hayeof}, the most prominent and the first 
one that was discovered is that of the directed percolation (DP) \cite{DP}.
An early hypothesis \cite{DPuni,DPuni2,DPuni3} 
was confirmed by all examples up to now.
This {\bf DP hypothesis} claims that 
{\bf in one component systems exhibiting continuous phase 
transitions to a single absorbing state (without extra symmetry and 
inhomogeneity or disorder) short ranged interactions can generate DP
class transition only}. Despite the robustness of this class experimental
observation is still lacking \cite{GrasWu,HayeDPexp} probably owing to 
the sensitivity to disorder that cannot be avoided in real materials.

A major problem of these models is that they are usually far from the 
critical dimension and critical fluctuations prohibit mean-field (MF)
like behavior. Further complication is that bosonic field theoretical 
methods cannot describe particle exclusion that may obviously happen 
in $d=1$. The success of the application of bosonic field theory in many 
cases is the consequence of the asymptotically low density of particles 
near the critical point.
However in multi-component systems, where the exchange between different 
types is non-trivial, bosonic field theoretical descriptions may fail.
In case of the binary production (BP) models (Sect. \ref{binsp}) bosonic 
RG predicts diverging density in the active phase contrary to the lattice 
model version of hard-core particles 
\cite{Odo00,Carlon,Hayepcpd,Odo01,HayeDP-ARW,binary}. 
Fermionic field theories on the other hand have the disadvantage that 
they are non-local, hence results exist for very simple reaction-diffusion 
systems only \cite{SPark,Wij,BOW99}.
Other techniques like independent interval approximation \cite{IIA},
empty interval method \cite{EIm}, series expansion \cite{essam-etal:1996} 
or density matrix renormalization (DMRG) are currently under development.

The universal scaling-law behavior in these models is described by
the critical exponents in the neighborhood of a steady state, hence the
generalization of dynamical exponents introduced for "Out of equilibrium 
classes" (Sect. \ref{eqintro}) (like $Z$, $\theta$, $\lambda$ ...etc.) 
is used. Besides that there are genuinely non-equilibrium dynamical 
exponents as well to characterize spreading behavior, 
defined in Sect.\ref{spexps}. For each class I discuss the
damage spreading transitions, the effects of different boundary conditions,
disorder and long-correlations generated by anomalous diffusion or by
special initial states.

\subsection{Directed percolation classes} \label{DPS}

\begin{figure}
\epsfxsize=70mm
\epsffile{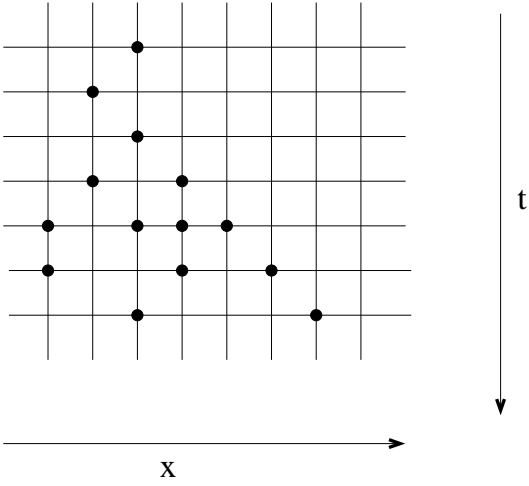}
\caption{Directed site percolation in $d=1+1$ dimensions
\label{DPfig}}
\end{figure}

The directed percolation (DP) introduced by \cite{BH57} is an anisotropic 
percolation with a preferred direction $t$. This means that this problem
should be $d\ge 2$ dimensional. If there is an object (bond, site etc.) 
at $(x_i,y_j,...,t_k)$ it must have a nearest neighboring object at 
$t_{k-1}$ unless $t_k=0$ (see Figure \ref{DPfig}).
If we consider the preferred direction as the time we recognize a
spreading process of an agent $A$ that can not have spontaneous source: 
$\emptyset\not\to A$. This results in the possibility of a completely frozen,
so called {\it absorbing state} from which the system cannot escape if it
has fallen into it. As a consequence these kinds of models may have phase
transitions in $d=1$ spatial dimension already. By increasing the branching
probability $p$ of the agent we can have a phase transition 
between the absorbing state and an active steady state with finite density
of $A$-s. If the transition is continuous it is very likely that it belongs 
to a robust universality (DP) class. For a long time all examples 
of such absorbing phase transitions were found to belong 
to the DP class and a conjecture was advanced by \cite{DPuni,DPuni2,DPuni3}.
This claims that in one component systems exhibiting continuous phase 
transitions to a single absorbing state (without extra symmetry, inhomogeneity 
or disorder) short ranged interactions can generate DP class transition only.
This hypothesis has been confirmed by all examples up to now, moreover DP class 
exponents were discovered in some systems with multiple absorbing states. 
For example in systems with infinitely many frozen absorbing states 
\cite{PCP,PCP2,dimerr,TTP}
the static exponents were found to coincide with those of DP. 
Furthermore by models without any special symmetry of the absorbing states 
DP behavior was reported \cite{Parkh,meod96,odme98} too. 
So although the necessary conditions for the DP 
behavior seem to be confirmed the determination of sufficient conditions is 
an open problem. There are many introductory works available now to DP 
\cite{DP,DickMar,GrasWu,Hayeof} therefore I shall not go very deeply
into the discussions of details of various representations.

In the reaction-diffusion language the DP is built up from the 
following processes
\begin{equation}
A\stackrel{\gamma}{\to}\emptyset \ \ \ 
A\emptyset\stackrel{D}{\leftrightarrow}\emptyset A 
\ \ \ A\stackrel{\sigma}{\to} 2A \ \ \
2A \stackrel{\lambda}{\to} A \label{DPproc}
\end{equation}
The mean-field equation for the coarse-grained particle density 
$\rho(t)$ is
\begin{equation}
\frac{d \rho}{d t} = (\sigma-\gamma)\rho - (\lambda+\sigma)\rho^2 \ \ .
\label{DPSMFeq}
\end{equation}
This has the stationary stable solution
\begin{equation}
\rho(\infty) = \left\{ \begin{array}{r@{\quad{\rm for \ :}\quad}l}
            \frac{\sigma-\gamma}{\lambda+\sigma} & \sigma > \gamma \\
                                               0 & \sigma\le\gamma
                          \end{array} \right.
\end{equation}
exhibiting a continuous transition at $\sigma=\gamma$. A small variation of
$\sigma$ or $\gamma$ near the critical point implies a linear change of $\rho$,
therefore the order parameter exponent in the mean-field approximation is
$\beta=1$. Near the critical point the $O(\rho)$ term is the dominant one,
hence the density approaches the stationary value exponentially. 
For $\sigma=\gamma$
the remaining $O(\rho^2)$ term causes power-law decay: $\rho\propto t^{-1}$
indicating $\alpha=1$. 
To get information about other scaling exponents we have to take
into account the diffusion term $D \nabla^2$ describing local density 
fluctuations. Using rescaling invariance two more independent exponents can 
be determined: $\nu_{\perp}=1/2$ and $Z=2$ if $d\geq 4$. Therefore the upper 
critical dimension of directed percolation is $d_c=4$. 

\begin{table}
\footnotesize
\begin{center}
\begin{tabular}{|c|c|c|c|c|}
\hline\\[-3mm] critical & $d=1$ & $d=2$ & $d=3$ & $d=4-\epsilon$ \\
exponent & &  &  & \\ \hline
$\beta=\beta^\prime$ & $0.276486(8)$ & 0.584(4) & 0.81(1) &
       $1-\epsilon/6-0.01128\,\epsilon^2$ \\
$\nu_{\perp}$ & $1.096854(4)$ & 0.734(4) & 0.581(5) &
       $1/2+\epsilon/16+0.02110\,\epsilon^2$ \\
$\nu_{||}$    & $1.733847(6)$ & 1.295(6) & 1.105(5) &
       $1+\epsilon/12+0.02238 \,\epsilon^2$ \\
$Z=2/z$ & $1.580745(10)$ & 1.76(3) & 1.90(1) &
       $2-\epsilon/12-0.02921 \,\epsilon^2$ \\ 
$\delta=\alpha$ & $0.159464(6)$ & $0.451$ & $0.73$ &
       $1-\epsilon/4-0.01283 \,\epsilon^2$ \\
$\eta$ & $0.313686(8)$ & $0.230$ & $0.12$ &
       $\epsilon/12+0.03751 \,\epsilon^2$ \\
$\gamma_p$ & $2.277730(5)$ & $1.60$ & $1.25$ &
       $1+\epsilon/6+0.06683 \,\epsilon^2$ \\
\hline
\end{tabular}
\end{center}
\caption{\label{DPe} Estimates for the critical exponents of directed 
percolation. Data are from: 
\cite{Jensen99a} (1d),\cite{VoigtZiff97}(2d),\cite{Jensen92}(3d),
\cite{BronzanDash74,DPuni}($4-\epsilon$).}
\end{table}

Below the critical dimension the RG analysis of the Langevin equation
\begin{eqnarray} \label{DPLangeq}
\frac{\partial\rho(x,t)}{\partial t} &=& D \nabla^2 \rho(x,t) 
+ (\sigma-\gamma)\rho(x,t) - \nonumber \\ 
&-& (\lambda+\sigma)\rho^2(x,t) + \sqrt{\rho(x,t)} \eta(x,t)
\end{eqnarray}
is necessary \cite{DPuni}. 
Here $\eta(x,t)$ is the Gaussian noise field, defined by the correlations
\begin{eqnarray}
<\eta(x,t)> & = & 0 \\
<\eta(x,t) \eta(x^,,t^,)> &=& \Gamma \delta^d(x-x^,)\delta(t-t^,)
\end{eqnarray}
The noise term is proportional $\sqrt{\rho(x,t)}$ ensuring that in the 
absorbing state ($\rho(x,t)=0$) it vanishes. The square-root behavior stems
from the definition of $\rho(x,t)$ as a coarse-grained density of active
sites averaged over some mesoscopic box size.
Note that DP universality occurs in many other processes like 
in odd offspring, branching and annihilating random walks (BARWo) 
(see Section \ref{BARWo}) or in models described by field theory with 
higher order terms like $\rho^3(x,t)$ or $\nabla^4\rho(x,t)$, 
which are irrelevant under the RG transformation. 
This stochastic process can through standard techniques 
\cite{Jan76} be transformed into a Lagrangian formulation with the action
\begin{equation}
S = \int d^dx dt \left[\frac{D}{2}\psi^2\phi
+ \psi(\partial_t\phi - \nabla^2\phi - r\phi + u\phi^2) \right]
\label{DPaction}
\end{equation}
where $\phi$ is the density field and $\psi$ is the response field 
(appearing in response functions) and the action is invariant under
the following time-reversal symmetry
\begin{equation}
\phi(x,t) \to -\psi(x,-t)  \ \ , \ \  \psi(x,t) \to -\phi(x,-t) \ \ .
\label{DPsymeq}
\end{equation}
This symmetry yields \cite{GrasTor,MGT} the scaling relations
\begin{eqnarray}
\beta & = & \beta^, \\
4\delta + 2\eta & = & d z
\label{dphyper}
\end{eqnarray}
This field theory was found to be equivalent \cite{CardySug80} to the Reggeon 
field theory \cite{RFT1,RFT2}, which is a model of scattering elementary 
particles at high energies and low-momentum transfers. 

Perturbative $\epsilon=4-d$ renormalization group analysis 
\cite{BronzanDash74,DPuni} up to two-loop order resulted in estimates for 
the critical exponents shown in the Table \ref{DPe}.
The best results obtained by approximative techniques for DP like 
improved mean-field \cite{BenNaimKrapivsky94}, coherent anomaly method 
\cite{OdDPCAM}, Monte Carlo simulations \cite{GrasTor,Gras2D,Gras3D,Dick-Jen}, 
series expansions \cite{DPser,DPser2,DPser3,DPser4,DPser5,DPser6,DPser7,Jensen99a}, DMRG \cite{DPDMRG,DPDMRG2}, and numerical
integration of eq.\ref{DPLangeq} \cite{D94} are also shown in Table \ref{DPe}.
  
The {\bf local persistence} probability may be defined as the probability
($p_l(t)$) that a particular site never becomes active up to time $t$.
Numerical simulations \cite{HKod} for this in the 1+1 d Domany-Kinzel
SCA (see Sect.\ref{sectsca}) found power-law with exponent
\begin{equation}
\theta_l = 1.50(1)
\end{equation}
The {\bf global persistence} probability defined here as the probability 
($p_g(t)$) that the deviation of the global density from its mean value 
does not change its sign up to time $t$.
The simulations of \cite{HKod} in 1+1 dimensions claim 
$\theta_g \ge \theta_l$. This agrees with field theoretical 
$\epsilon=4-d$ expansions \cite{OW98} that predict $\theta_g=2$ for 
$d\ge 4$ and for $d<4$:
\begin{equation}
\theta_g = 2-\frac{5\epsilon}{24} + O(\epsilon^2) \ .
\end{equation}

{\bf The crossover from isotropic to directed percolation} was investigated by
perturbative RG \cite{FTS94} up to one loop order. They found that while
for $d>6$ the isotropic, for $d>5$ the directed Gaussian fixed point
is stable. For $d<5$ the asymptotic behavior is governed by the
DP fixed point.
On the other hand in $d=2$ exact calculations and simulations \cite{FN97}
found that the isotropic percolation is stable with respect to DP if 
we control the crossover by a spontaneous particle birth parameter. 
It is still an open question what happens for $2 < d < 5$.
Crossovers to mean-field behavior generated by long-range interactions
\ref{long} and to compact directed percolation \ref{glau} will be discussed
later.

\begin{figure}
\epsfxsize=70mm  
\epsffile{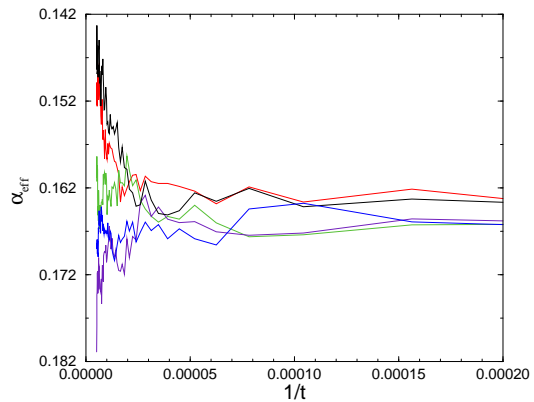}
\caption{Local slopes of the density decay in a bosonic BARW1 model
(Sect. \ref{BARWo}). Different curves correspond to 
$\lambda=0.12883$, $0.12882$, $0.12881$ $0.1288$, $0.12879$ 
(from bottom to top) \cite{OdMe02}.}
\label{bdp}
\end{figure}
It was conjectured \cite{Wijcon} that in 1d ``fermionic'' (single site 
occupancy) and bosonic (multiple site occupancy) models may exhibit 
different critical behavior. 
An attempt for a fermionic field theoretical treatment of the DP in 1+1 d 
was shown in \cite{Wij,BOW99}. This run into severe convergence problems 
and has not resulted in precise quantitative estimates for the 
critical exponents. 
Although the bosonic field theory is expected to describe the fermionic case
owing to the asymptotically low density at criticality it has never been 
proven rigorously. 
Since only bosonic field theory exists that gives rather inaccurate critical 
exponent estimates, in \cite{OdMe02} simulations of a a BARW1 process 
(\ref{BARWodef}) with unrestricted site occupancy were performed to 
investigate the density ($\rho(t)$) decay of a DP process from random 
initial state. Figure \ref{bdp} shows the local slopes of density decay
defined as
\begin{equation}
\alpha_{eff} = - \frac{d \log \rho(t)}{d \log t}
\label{aeff}
\end{equation} 
around the critical point for several annihilation rates ($\lambda$).
The critical point is estimated at $\lambda_c=0.12882(1)$ 
(corresponding to straight line) with the extrapolated decay exponent 
$\alpha=0.165(5)$, which agrees well with fermionic model simulation and 
series expansion results 0.1595(1) \cite{DPser}.

Note that in site restricted models already the 
\begin{equation}
A\stackrel{\gamma}{\to}\emptyset \ \ \ 
A\emptyset\stackrel{D}{\leftrightarrow}\emptyset A 
\ \ \ A\stackrel{\sigma}{\to} 2A \label{FDPproc}
\end{equation}
processes generate a DP class phase transition, while in the bosonic version
the $2A \stackrel{\lambda}{\to} A$ process is also necessary to ensure an
active steady state (without it the density blows up for $\sigma > \gamma$).

Models exhibiting DP transitions have been reviewed in great detail in
\cite{DickMar} and in \cite{Hayeof}. In the next subsections I recall 
only three important examples.

 \subsubsection{The Contact process} \label{CPsect}

The contact process is one of the earliest and simplest lattice
model for DP with {\bf asynchronous} update introduced by Harris 
\cite{ContactProcess,GrasTor} to model epidemic spreading without immunization.
Its dynamics is defined by nearest-neighbor processes that occur
spontaneously due to specific rates (rather than probabilities).
In numerical simulations models of this type are usually realized
by random sequential updates. In one dimension this means that a 
pair of sites  $\{s_i,s_{i+1}\}$ is chosen at random and an update is
attempted according to specific transition rates
$w(s_{i,t+dt}\,,\,s_{i+1,t+dt}\,|\,
s_{i,t}\,,\,s_{i+1,t})$. 
Each attempt to update a pair of sites increases
the time $t$ by $dt=1/N$, where $N$ is the total number of sites. One
time step (sweep) therefore consists of $N$ such attempts. The
contact process is defined by the rates
\begin{eqnarray}
\label{gl1}
w(A,I\,|\,A,A)=w(I,A\,|\,A,A)&=&\mu\,,\\
\label{gl2}
w(I,I\,|\,A,I)=w(I,I\,|\,I,A)&=&\lambda\,,\\
\label{gl3}
w(A,A\,|\,A,I)=w(A,A\,|\,I,A)&=&1\,,
\end{eqnarray}
where $\lambda>0$ and $\mu>0$ are two parameters (all other rates are zero).
Equation (\ref{gl1}) describes the creation of inactive ($I$) spots
within active ($A$) islands. Equations (\ref{gl2}) and (\ref{gl3}) 
describe the shrinkage and growth of active islands. In order to fix 
the time scale, we chose the rate in Eq. (\ref{gl3}) to be equal to one. 
The active phase is restricted to the region $\lambda < 1$ where
active islands are likely to grow. In one dimension series expansions 
and numerical simulations determined the critical point and critical 
exponents precisely 
\cite{Dick-Jen,DPser,DPser2,DPser3,DPser4,DPser5,DPser6,DPser7,Dick-Sil}. 
In two dimensions the order parameter moments and the cumulant
ratios were determined by \cite{Dick-Sil}.

 \subsubsection{DP-class stochastic cellular automata} \label{sectsca}

\begin{figure}
\epsfxsize=70mm
\epsffile{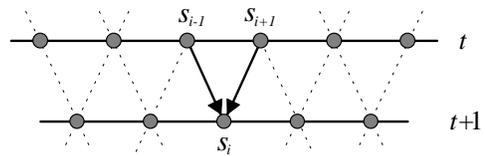}
\caption{Update in the Domany-Kinzel model}
\label{DomanyKinzelFigure}
\end{figure}

Cellular automata as the simplest systems exhibiting synchronous dynamics
have extensively been studied \cite{Wolfram}. By making the update rules
probabilistic phase transitions as the function some control parameter may
emerge.
There are many stochastic cellular automata (SCA) that exhibit DP transition
\cite{BocRog} perhaps the first and simplest one is the (1+1)-dimensional
Domany-Kinzel (DK) model \cite{DoKi}.
In this model the state at a given time $t$ is specified by binary variables
$\{s_i\}$, which can have the values $A$ (active) and $I$ (inactive).
At odd times odd-indexed, whereas at even times the rest of the sites 
are updated according to specific conditional probabilities. 
This defines a cellular automaton
with {\bf parallel} updates (discrete time evolution) acting on two
independent triangular sub-lattices (Fig.\ref{DomanyKinzelFigure}).
The conditional probabilities in the Domany-Kinzel model
$P(s_{i,t+1} \,|\, s_{i-1,t}\,,\,s_{i+1,t})$ are given by
\begin{eqnarray}
\label{eq1}
P(I\,|\,I,I)&=&1\,,\\
\label{eq2}
P(A\,|\,A,A)&=&p_2\,,\\
\label{eq3}
P(A\,|\,I,A)=P(A|A,I)&=&p_1\,,
\end{eqnarray}
and $P(I|s_{i-1},s_{i+1})+P(A|s_{i-1},s_{i+1})=1$, where $0\leq p_1\leq 1$ 
and $0\leq p_2\leq 1$ are two parameters. Equation (\ref{eq1}) ensures 
that the configuration $\ldots,I,I,I,\ldots$ is the absorbing state.
The process in Eq. (\ref{eq2}) describes the creation of inactive       
spots within active islands with probability $1-p_2$.
The random walk of boundaries between active and inactive domains
is realized by the processes in Eq.~(\ref{eq3}).
A DP transitions can be observed only if $p_1>\frac12$,
when active islands are biased to grow \cite{Wolfram}.
\begin{figure}
\epsfxsize=70mm
\epsffile{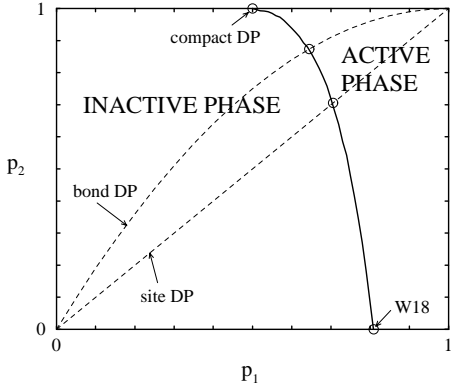}
\caption{Phase diagram of the 1d Domany-Kinzel SCA \cite{Hin97}.}
\label{DKpd}
\end{figure}
The phase diagram of the 1d DK model is shown in Fig. \ref{DKpd}.
It comprises an active and an inactive phase, separated by a phase
transition line (solid line) belonging to DP class. The dashed lines
corresponds to {\it directed bond percolation} ($p_2=p(2-p_1)$) and 
{\it directed site percolation}
($p_1=p_2$) models. At the special symmetry endpoint
($p_1=\frac12$, $p_2=1$) compact domain growth occurs (CDP) and the 
transition becomes first order (see Section \ref{glau}).
The transition on the $p_2=0$ axis corresponds to the transition of
the stochastic version of Wolfram's rule-18 cellular automaton
\cite{Wolfram}.
This range-1 SCA generates A at time $t$ only when the right or left 
neighbor was A at $t-1$:
\begin{verbatim}
            t-1:        AII        IIA     
             t:          A          A        
 \end{verbatim}
with probability $p_1$ \cite{BocRog}.
The critical point was determined by precise simulations 
($p_1^*=0.80948(1)$) \cite{OASP}. In the $t\to\infty$ limit the steady 
state is built up from II and IA blocks \cite{Elo}. 
This finding permits us to map this model onto an even simpler one, the
rule-6/16 SCA with new variables: IA $\to$ A and II $\to$ I:
\begin{verbatim}
   t-1:         I I     I A     A I      A A
    t:           I       A       A        I
\end{verbatim}
By solving GMF approximations and applying Pad\'e approximations 
\cite{SzaOd94} or CAM method \cite{OdDPCAM} very precise order 
parameter exponent estimates were found: $\beta=0.2796(2)$.
The {\bf DS phase structure} of the 1d DK model was explored
by \cite{HinWD} and DP class transitions were found.

Another SCA example I mention is the family of {\bf range-4 SCA} with an 
acceptance rule
\[ s(t+1,j) = \left\{ \begin{array}{ll}
X & \mbox{if} \ \ y \leq \sum_{j-4}^{j+4}s(t,j) \leq 6 \\
0 & \mbox{otherwise \ \ ,} 
\end{array} \right. \]
where $X\in\{0,1\}$ is a two valued random variable such that
$Prob(X=1)=p$ \cite{OdSzo}. The $y=3$ case was introduced and 
investigated by \cite{Bid} in $d=1,2,3$. The very first simulations in 
one dimension \cite{Bid} suggested a counter-example to the DP conjecture. 
More precise spreading simulations of this model \cite{Jen91B},
GMF + CAM calculations and simulations of the $y<6$ family in one 
and two dimensions have proven that this does not happen for
any case \cite{OdSzo}. The transitions are either belong to DP class or 
first order.

 \subsubsection{Branching and annihilating random walks with \\
 odd number of offspring} \label{BARWo}

Branching and annihilating random walks (BARW) introduced by \cite{Taka} 
can be regarded as generalizations of the DP process. They are
defined by the following reaction-diffusion processes:
\begin{equation}
A\stackrel{\sigma}{\to}(m+1)A \ \ \ \ k A\stackrel{\lambda}{\to}\emptyset
\ \ \ \ A\emptyset\stackrel{D}{\leftrightarrow}\emptyset A \ .
\label{BARWodef}
\end{equation}
The $2A\to A$ and $2A\to\emptyset$ reactions dominating in the
inactive phase are shown to be equivalent \cite{Peliti} (see Sect.\ref{2A0}).
Therefore the $k=2$ and $m=1$ (BARW1) model differs from the DP process
(\ref{DPproc}) that spontaneous annihilation of particles is not
allowed.
The action of the BARW process was set up by \cite{Cardy-Tauber,Cardy-Tauber2}
\begin{eqnarray}
S = \int d^d x dt & [ \psi(\partial_t-D\nabla^2)\phi 
- \lambda(1-\psi^k)\phi^k + \nonumber \\ 
& +\sigma(1-\psi^m) \psi \phi ] 
\end{eqnarray}
The bosonic RG analysis of BARW systems \cite{Cardy-Tauber} proved that 
for $k=2$ all the lower branching reactions with $m-2,m-4,...$ are generated 
via fluctuations involving combinations of branching and annihilation 
processes. As a consequence for {\bf odd $m$} the $A\to\emptyset$ reaction 
appears (via $A\to 2A\to\emptyset$).
Therefore after the first coarse graining step in the BARW1 
(and in general in the odd $m$ BARW (BARWo) cases) the action becomes the
the same as that of the DP process. The fluctuations are relevant for
$d\le 2$ and the universal behavior is DP type, with $d_c=2$. 
This prediction was confirmed by simulations (see for example \cite{Jen92}).

For even $m$ (BARWe), when the parity of the number of particles is conserved
the spontaneous decay $A\to\emptyset$ is not generated, 
hence there is an absorbing state with a lonely wandering particle. 
This systems exhibits a non-DP class critical transition, which will be 
discussed in Sect.\ref{BARWe}.

  \subsubsection{DP with spatial boundary conditions} \label{sectsbon}

For a review of critical behavior at surfaces of {\em equilibrium} models
see \cite{IPT93}). Ref. \cite{Cardy83b} suggested that surface critical 
phenomena may be described by introducing an additional 
{\em surface exponent} for 
the order parameter field which is generally independent of the other 
bulk exponents. In nonequilibrium statistical physics one can introduce
spatial, temporal (see Sect.\ref{long}) or mixed (see Sect.\ref{sectmbond})
boundary conditions.

In DP an absorbing wall may be introduced by cutting all bonds 
({\bf IBC}) crossing a given ($d$-1)-dimensional hyperplane in space. In case 
of reflecting boundary condition ({\bf RBC}) where the wall acts like a mirror 
so that the sites within the wall are always a mirror image of those 
next to the wall. A third type of boundary condition is the active 
boundary condition ({\bf ABC}) where the sites within the wall are forced 
to be active.
The density at the wall is found to scale as
\begin{equation}
\rho_s^{stat} \sim (p-p_c)^{\beta_1} \,
\end{equation}
with a surface critical exponent $\beta_1 > \beta$.
Owing to the time reversal symmetry of DP (\ref{DPsymeq}) only one extra 
exponent is needed to describe surface effects, hence the cluster survival
exponent is the same
\begin{equation}
\beta^{\prime}_1= \beta_1 \ \ .
\end{equation}
The mean lifetime of finite clusters at the wall is defined as
\begin{equation}
        \left< t \right> \sim |\Delta|^{-\tau_1}
\label{taudef1}
\end{equation}
where $\Delta_s=(p-p_c)$, and is related to $\beta_1$ by the scaling 
relation
\begin{equation}
\tau_1 = \nu_{||} - \beta_1
\end{equation}
The average size of finite clusters grown from seeds on the wall is
\begin{equation}
        \label{DPsize_wall}
        \langle s \rangle \sim |\Delta|^{-\gamma_1} ,
\end{equation}
Series expansions \cite{essam-etal:1996} and numerical simulations 
\cite{lauritsen-etal} in 1+1 dimensions indicate that the presence of 
the wall alters several exponents. However, the scaling properties 
of the correlation lengths (as given by $\nu_\parallel$ and $\nu_\perp$) 
are {\it not\/} altered. 

\begin{figure}[tbhp]
\begin{center}
\leavevmode
\vbox{
\epsfxsize=50mm
\epsffile{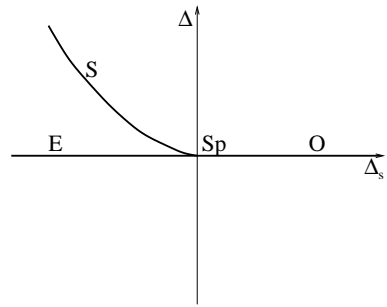}
}
\end{center}
\caption{Schematic mean field phase diagram for boundary DP \cite{FHL01}. 
The transitions are labeled by O=ordinary, E=extraordinary, S=surface, and
Sp=special.}
\label{dpps}
\end{figure}

The field theory for DP in a semi-infinite geometry
was first analyzed by \cite{janssen-etal}. They showed 
that the appropriate action for DP with a wall at $x_{\perp}=0$ is 
given by $S=S_{\rm bulk}+S_{\rm surface},$ where
\begin{equation}
        S_{\rm surface}= \int d^{d-1}x\int dt ~ \Delta_s\,
        \psi_s\, \phi_s, 
        \label{actionsurface}
\end{equation}
with the definitions 
        $\phi_s=\phi ({\bf x_{\parallel}},x_{\perp}=0,t)$ and
        $\psi_s =\psi({\bf x_{\parallel}},x_{\perp}=0,t)$.
The surface term $S_{\rm surface}$ corresponds to the most relevant 
interaction consistent with the symmetries of the problem and which 
also respects the absorbing state condition.
The appropriate surface exponents were computed to first order
in $\epsilon = 4-d$ using renormalization group techniques:
\begin{equation}
        \label{beta_1}
        \beta_{1}= {3\over 2}-{7\epsilon\over 48}+O(\epsilon^2). \\
\end{equation}
They also showed that the corresponding hyperscaling relation is
\begin{equation}
  \nu_{\parallel}+d\nu_{\perp}=\beta_1+\beta+\gamma_{1} 
        \label{DPsurf_hyperscaling}
\end{equation}
relating $\beta_1$ to
\begin{equation}
        \gamma_{1} = {1\over 2}+{7\epsilon\over 48}+O(\epsilon^2).
\end{equation}
The schematic phase diagram for boundary DP is shown in Fig.~\ref{dpps}, 
where $\Delta$ and $\Delta_s$ represent, respectively, the deviations 
of the bulk and surface from criticality. 
For $\Delta>0$ and for $\Delta_s$ sufficiently negative the boundary 
orders even while the bulk is disordered, there is a {\it surface transition}. 
For $\Delta_s<0$ and $\Delta\to 0$, the bulk orders in the presence of
an already ordered boundary, there is an {\it extraordinary transition} 
of the boundary. Finally at $\Delta=\Delta_s=0$, where all the critical lines
meet, and where both the bulk and isolated surface are critical,
we find a multi-critical point, i.e.\ the {\it special transition}.

For $\Delta_s>0$ and $\Delta\to 0$ there is an {\it ordinary transition}, 
since the bulk orders in a situation in which the boundary, if isolated, 
would be disordered. At the ordinary transition, one 
finds just {\bf one extra independent exponent} related to
the boundary: this can be the surface density exponent
$\beta_{1,{\rm dens}}$.
In 1d the IBC and RBC cases belong to the same universality
class \cite{LFH98} that was identified as the ordinary transition.
There are numerical data for the exponents of the extraordinary and
special transitions (however see \cite{janssen-etal} for an RG analysis).

The best exponent estimates currently available were summarized in 
\cite{FHL01}. Some of them are shown in Table \ref{table-dp}. 
In $d=1$ the best results are from series expansions 
\cite{essam-etal:1996,newjensen}; in all other cases are from 
Monte-Carlo data.
\cite{lauritsen-etal,dp-wall-edge,LFH98,HFL00}
\begin{table}[bhtbp]
\centerline{
\begin{tabular}{|l|l|l|c|}
\hline
              & \makebox[20mm]{$d=1$} & \makebox[16mm]{$d=2$}
              & \makebox[16mm]{Mean Field} \\
\hline
$\beta_1$ & ~0.733 71(2)   & ~1.07(5)   &  3/2  \\
\hline
$\delta_1=\alpha_1$    & ~0.423 17(2)   & ~0.82(4)   &  3/2  \\
\hline
$\tau_1$               & ~1.000 14(2)   & ~0.26(2)   &    0  \\
\hline
$\gamma_1$             & ~1.820 51(1)   & ~1.05(2)   &  1/2 \\
\hline
\end{tabular}}
\caption{Critical exponents for DP in $d=1$ and $d=2$ for the ordinary 
        transition at the boundary.}
\label{table-dp}
\end{table}
The exponent $\tau_1$ was conjectured to equal unity,
\cite{essam-etal:1996} although this has now been challenged by the 
estimate $\tau_1 = 1.00014(2)$ \cite{newjensen}. 

It has been known for some time that the presence of an {\bf edge} 
introduces new exponents, independent of those associated with the bulk 
or with a surface (see \cite{Cardy83b}). For an investigation showing 
numerical estimates in 2d and mean-field values see \cite{dp-wall-edge}.
Table \ref{table-beta} summarizes results for the ordinary edge exponents.
A closely related application is the study of spreading processes in 
narrow channels~\cite{Albano97}.
\begin{table}[htb]
\begin{center}
\begin{tabular}{|l|c|c|c|c|} \hline
Angle ($\alpha$)  &  \makebox[16mm]{$\pi/2$}   & \makebox[16mm]{$3\pi/4$}
                  &  \makebox[16mm]{$\pi$}     & \makebox[16mm]{$5\pi/4$}
                  \\ \hline
${\beta_{2}^{\rm O}}_{~}^{~}~(d=2)$  & $1.6(1)$    & $1.23(7)$
                     & $1.07(5)$ & $0.98(5)$ \\ 
\hline
${\beta_{2}^{\rm O}}_{~}^{~}~({\rm MF})$
                      &    2    &   5/3  &   3/2   &   7/5  \\
\hline
\end{tabular}
\caption{Numerical estimates for the ordinary $\beta_2^{\rm O}$ exponents 
        for edge DP together with the mean field values.
        Note that $\beta_2^{\rm O}(\pi)=\beta_1^{\rm O}$}
\end{center}
\label{table-beta}
\end{table}

  \subsubsection{DP with mixed (parabolic) boundary condition \\
  scaling} \label{sectmbond}

\begin{figure}
\epsfxsize=80mm
\epsffile{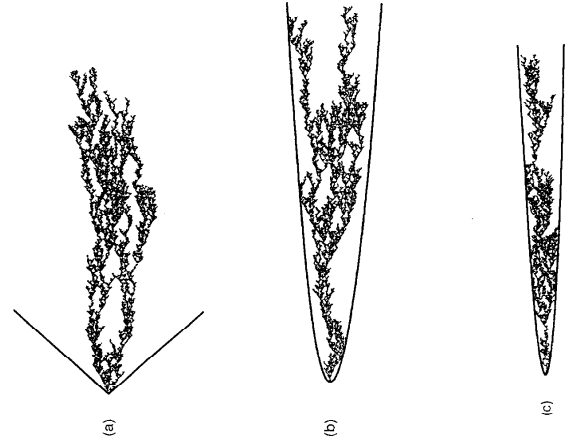}
\caption{
\label{parabDP}
The space-time evolution of the critical, 1+1 dimensional directed site 
percolation process confined by parabola \cite{Turban}.
(a) $\sigma<1/Z$, (b) $\sigma = 1/Z$, (c) $\sigma>1/Z$.
}
\end{figure}

Boundary conditions, which act both in space and time direction can also
been investigated in dynamical systems. These turn out to be related to
hard-core repulsion effects of 1d systems (Sect.\ref{2}).
\cite{Turban,Turban2} investigated the 1+1 d DP process 
confined in a parabola-shaped geometry. Assuming an absorbing boundary of 
the form $x=\pm C t^\sigma$ they proposed a general scaling theory. 
It is based on the observation that the coefficient of the parabola ($C$)
scales as $C \rightarrow \Lambda^{Z \sigma -1} C$ under
rescaling
\begin{equation}
\label{Rescaling}
x \rightarrow \Lambda x
\, , \quad
t \rightarrow \Lambda^Z t
\, , \quad
\Delta \rightarrow \Lambda^{-1/\nu_{\perp}} \Delta
\, , \quad
\rho \rightarrow \Lambda^{-\beta/\nu_{\perp}} \rho
\, ,
\end{equation}
where $\Delta=|p-p_c|$ and $Z$ is the dynamical exponent of DP. 
By referring to conformal mapping of the parabola to straight lines
and deriving it in the mean-field approximation they claimed that this 
boundary is a relevant perturbation for $\sigma>1/Z$, irrelevant for 
$\sigma<1/Z$ and marginal for $\sigma = 1/Z$ (see Fig. \ref{parabDP}). 
The marginal case results in $C$ dependent non-universal power-law decay,
while for the relevant case stretched exponential functions have been
obtained. The above authors have given support to these claims
by numerical simulations.

  \subsubsection{L\'evy flight anomalous diffusion in DP} \label{sectlevydp}

\begin{figure}
\epsfxsize=65mm
\epsffile{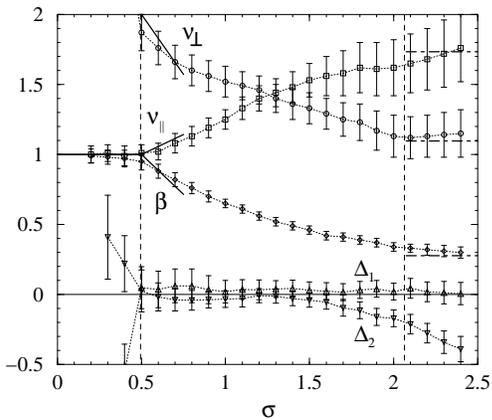}
\caption{
\label{levyfig}
Estimates for the exponent $\beta$ and the derived exponents $\nu_\perp$ 
and $\nu_{||}$ in comparison with the field-theoretic results (solid lines)
and the DP exponents (dot-dashed lines) \cite{HH99}.
The quantities $\Delta_1$ and $\Delta_2$ represent deviations from the
scaling relations (\ref{HyperScalingRelation}) and (\ref{NewScalingRelation}), 
respectively.
}
\end{figure}

L\'evy flight anomalous diffusion generating long-range correlations
was already mentioned in Sections \ref{IsinglongS} and \ref{PottslongS} 
in case of equilibrium models.
In non-equilibrium systems following the suggestion of \cite{Mollison} 
ref. \cite{Grassberger} introduced a variation of the
epidemic processes with infection probability distribution 
$P\left(R\right)$, which decays with the distance $R$ as a power-law like
\begin{equation}
P(R)\propto \frac{1}{R^{d+\sigma }} \ .
\label{LongRange}
\end{equation}
This can model long-range epidemics mediated by flies, wind, ... etc. 
He claimed that the critical exponents should depend continuously on
$\sigma$. This result was confirmed in 1d by estimating $\beta$ based on CAM 
calculations in \cite{MF94}. The study of anomalous diffusion was extended 
for GEP processes (see Sect. \ref{dynperc}) and for annihilating random walks
(see Sect. \ref{LARWs}) too. The effective action of the L\'evy flight
DP model is
\begin{eqnarray}
\label{EffectiveActionDP}
S[\psi,\phi]
&=& 
\int d^dx ~ dt ~
\Bigl[\psi(\partial_t - \tau - D_N \nabla^2 - D_A
\nabla^\sigma)\phi 
\nonumber
\\ && \hspace{20mm}
+  \frac{g}{2}(\psi\phi^2-\psi^2\phi) \Bigr] \ ,
\end{eqnarray}
where $D_N$ denotes the normal, $D_A$ the anomalous diffusion constant
and $g$ is the interaction coupling constant.
Field theoretical RG method up to first order in $\epsilon=2\sigma -d$ 
expansion \cite{JOWH99} gives:
\begin{eqnarray}
\label{CriticalExponents}
\beta &=& 1-\frac{2\epsilon}{7\sigma} + O\left( \epsilon^{2}\right) \ ,
\nonumber \\
\nu_{\perp} &=& \frac{1}{\sigma} + \frac{2\epsilon}{7\sigma^2}
 + O\left( \epsilon^{2}\right) \ ,
\nonumber \\
\nu_{||} &=& 1 + \frac{\epsilon}{7\sigma} + O\left( \epsilon^{2}\right)
\ ,
\\
Z={\nu_{\parallel}\over\nu_{\perp}}
&=&\sigma-\epsilon/7  + O\left( \epsilon^{2}\right)\nonumber \ .
\end{eqnarray}
Moreover, it was shown that the hyperscaling relation
\begin{equation}
\label{HyperScalingRelation}
\eta + 2 \delta = d/Z 
\, \qquad
(\delta = \beta/\nu_{||}) \ ,
\end{equation}
for the so-called critical initial slip exponent $\eta$ and the relation
\begin{equation}
\label{NewScalingRelation}
\nu_{||} - \nu_\perp(\sigma-d)-2\beta = 0 \ .
\end{equation}
hold exactly for arbitrary values of $\sigma$. 
Numerical simulations on 1+1 d bond percolation confirmed these results
except in the neighborhood of $\sigma=2$ \cite{HH99} 
(see Fig.\ref{levyfig}).
%
%
    \subsubsection{Long-range correlated initial conditions in DP} 
    \label{long}

It is well known that initial conditions influence the temporal
evolution of nonequilibrium systems. The ``memory'' of systems
for the initial state usually depends on the dynamical rules.
For example, stochastic processes with a finite temporal correlation 
length relax to their stationary state in an exponentially short time.
An interesting situation emerges when a system undergoes a 
nonequilibrium phase transition in which the temporal correlation
length diverges. This setup motivates the question whether
it is possible construct initial states that affect the 
{\it entire} temporal evolution of such systems.

Monte-Carlo simulations of critical models with absorbing states 
usually employ two different types of initial conditions.
On the one hand {\it uncorrelated random initial conditions} (Poisson
distributions) are used to study the relaxation of an initial state 
with a finite particle density towards the absorbing state. 
In this case the particle density $\rho(t)$
{\it decreases} on the infinite lattice asymptotically as
\begin{equation}
\label{Decrease}
\rho(t) \sim t^{-\beta/\nu_{||}} \,.
\end{equation}
On the other hand, in spreading simulations ~\cite{GrasTor},
each run starts with a {\it single particle} as a localized
active seed from where a cluster originates (this is a long-range
correlated state).
Although many of these clusters survive for only a short time, 
the number of particles $n(t)$ averaged over many
independent runs {\em increases} as
\begin{equation}
\label{Increase}
\langle n(t) \rangle \sim t^{+ \eta} \,,
\end{equation}
These two cases seem to represent extremal situations where the
average particle number either decreases or increases.

A {\em crossover} between these two extremal cases takes place 
in a critical spreading process that starts from a random initial 
condition of very low density. Here the particles are initially 
separated by empty intervals of a certain typical size, wherefore
the average particle number first increases according to 
Eq.~(\ref{Increase}). Later, when the growing clusters begin to interact 
with each other, the system crosses over to the algebraic decay of 
Eq.~(\ref{Decrease}) -- a phenomenon which is referred to as the 
``critical initial slip'' of nonequilibrium systems~\cite{Janssen}.

In \cite{HaOd98,MeOdof} it was investigated whether it is possible
to interpolate {\em continuously} between the two extremal cases
in cases of 1+1 dimensional DP and PC processes.
It was shown that one can in fact generate certain initial states in a way
that the particle density on the infinite lattice varies as
\begin{equation}
\label{Decay}
\rho(t) \sim t^\kappa
\end{equation}
with a continuously adjustable exponent $\kappa$ in the range
\begin{equation}
\label{KappaRange}
-\beta/\nu_{||} \leq \kappa \leq +\eta \,.
\end{equation}
To this end artificial initial configurations with algebraic long-range 
correlations of the form
\begin{equation}
\label{TwoPointCorrelations}
C(r) = \langle s_i \, s_{i+r} \rangle
\sim r^{-(d-\sigma)} \,,
\end{equation}
were constructed, where $\langle\rangle$ denotes the average over 
many independent realizations, $d$ the spatial dimension, and $s_i=0,1$ 
inactive and active sites. 
The exponent $\sigma$ is a free parameter and can be varied continuously
between $0$ and $1$. This initial condition can be taken into
account by adding the term
\begin{equation}
S_{ic} = \mu \int d^dx \psi(x,0)\phi_0(x)
\end{equation}
to the action, where $\phi_0(x)$ represents the initial particle 
distribution. The long-range correlations limit $\sigma \rightarrow d$ 
corresponds to a constant particle density and thus we expect 
Eq.~(\ref{Decrease}) to hold ($\phi_0(x)=$ const. is irrelevant under
rescaling). On the other hand, the short-range limit $\sigma \rightarrow 0$ 
represents an initial state where active sites are separated by 
infinitely large intervals ($\phi_0(x)=\delta^d(x)$) so that the particle 
density should increase according to Eq.~(\ref{Increase}). 
In between we expect 
$\rho(t)$ to vary algebraically according to Eq.~(\ref{Decay}) with an 
exponent $\kappa$ depending continuously on $\sigma$. 

In case of the 1+1 d {\bf Domany-Kinzel SCA} (see Sect.~\ref{sectsca})
field-theoretical renormalization group calculation and simulations 
proved \cite{HaOd98} the exact functional dependence
\begin{equation}
\label{ExactSolution}
\kappa(\sigma) = \left\{
\begin{array}{lll}
\eta & \mbox{for} & \sigma<\sigma_c \\[3mm]
\frac{1}{z} (d-\sigma-\beta/\nu_\perp) & \mbox{for} & \sigma>\sigma_c
\end{array}
\right.
\end{equation}
with the critical threshold $\sigma_c=\beta/\nu_\perp$.

\subsubsection{Quench disordered DP systems}

Perhaps the lack of experimental observation of the robust DP class lies
in the fact that even weak disorder changes the critical behavior of such 
models. 
Therefore this section provides a view onto interesting generalizations 
of DP processes that may be observable in physical systems.
First \cite{Noest,Noest2} showed using Harris criterion \cite{Harris}
that {\bf spatially quenched disorder} (frozen in space) changes the critical 
behavior of DP systems for $d<4$. 
Ref.~\cite{Janssen97b} studied the problem by field theory taking into 
account the disorder in the action by adding the term
\begin{equation}
S\rightarrow S + \gamma \int d^dx \left[ \int dt\,\psi\phi \right]^2 \, .
\end{equation}
This additional term causes {\bf marginal perturbation} and the stable 
fixed point is shifted to an unphysical region, leading to runaway 
solutions of the flow equations in the physical region of interest. 
This means that spatially quenched disorder changes the critical behavior 
of DP. This conclusion is supported by the simulation results of 
\cite{MoreiraDickman96} who reported logarithmic spreading 
behavior in two-dimensional contact process at criticality.
In the sub-critical region they found Griffiths phase in which the time 
dependence is governed by non-universal power-laws, while in the active
phase the relaxation of $P(t)$ is algebraic.

In 1+1 dimension \cite{Noest,Noest2} predicted generic scale invariance.
Ref. \cite{WACH98} reported glassy phase with non-universal 
exponents in a 1+1 d DP process with quenched disorder.
Ref. \cite{CGM98} showed that DP with spatially
quenched randomness in the large time limit can be mapped onto a 
non-Markovian spreading process with memory, in agreement with 
previous results. They showed that the time reversal symmetry of
the DP process (\ref{DPsymeq}) is not broken therefore
\begin{equation}
\delta=\delta^,
\end{equation}
and derived a hyper-scaling law for the inactive phase
\begin{equation}
\eta = d z/2 
\end{equation}
and for the absorbing phase
\begin{equation}
\eta + \delta = d z/2 \ . 
\end{equation}
They confirmed these relations by simulations and found 
that the dynamical exponents change continuously as the function of 
the disorder probability.
An RG study by \cite{HIV02} showed that
in case of strong enough disorder the critical behavior is controlled 
by an infinite randomness fixed point (IRFP), the static exponents of 
which in 1d are
\begin{equation}
\beta =(3- \sqrt 5)/2 \ \ \ , \ \ \nu_{\perp}=2
\end{equation}
and $\xi^{1/2}\propto \ln \tau$. For disorder strengths outside the 
attractive region of the IRFP disorder dependent critical exponents
are detected.

The {\bf temporally quenched disorder} can be taken into the action by adding
the term: 
\begin{equation}
S\rightarrow S + \gamma \int dt \left[ \int d^dx\,\psi\phi \right]^2 \, .
\end{equation}
This is a {\bf relevant} perturbation for the DP processes. 
Ref. \cite{Jensen96} investigated the 1+1 d directed bond percolation
(see Sect. \ref{sectsca}) with temporal disorder via series expansions and 
Monte Carlo simulations. The temporal disorder was introduced by allowing 
time slices to become fully deterministic ($p_1=p_2=1$), with probability
$\alpha$. He found $\alpha$ dependent, continuously changing critical point 
and critical exponent values between those of the the 1+1 d DP class and
those of the deterministic percolation. This latter class is defined by the
exponents:
\begin{equation}
\beta=0, \ \ \ \delta=0, \ \ \ \eta=1, \ \ \ Z=1, \ \ \ \nu_{||}=2, \ \ \ 
\nu_{\perp}=2 \ .
\end{equation}
For small disorder parameter values violation of the Harris criterion is 
reported. 

If {\bf quenched disorder takes place in both space and time} the 
corresponding term to action is
\begin{equation}
S\rightarrow S + \gamma \int dt d^dx\,\left[\psi\phi\right]^2 \, .
\end{equation}
and becomes an {\bf irrelevant} perturbation to the Reggeon field theory.
This has the same properties as the intrinsic noise in the system and can
be considered as being annealed.

\subsection{Dynamical percolation (DyP) classes} \label{dynperc}

If we allow memory in the unary DP spreading process (Sect.\ref{DPS}) such 
that the infected sites may have a different re-infection probability ($p$) 
than the virgin ones ($q$) we obtain different percolation behavior \cite{GCR}.
The model in which the re-infection probability is zero is called the
General Epidemic Model (GEP) \cite{Mollison}. In this case the epidemic stops in
finite systems but an infinite epidemic is possible in the form of a solitary
wave of activity. When starting from a single seed this leads to annular
growth patterns. The transition between survival and extinction is a
critical phenomenon called dynamical percolation \cite{G82}. Clusters
generated at criticality are the ordinary percolation clusters of 
the lattice in question. Field theoretical treatment was given by 
\cite{C83,CG85,J85,MDG,AH03}. The action of the model is
\begin{equation}
S=\int d^d x dt \frac{D}{2}\psi^2\phi - \psi\left(\partial_t\phi
-\nabla^2\phi - r\phi + w\phi\int_0^t ds \phi(s) \right)
\end{equation}
This is invariant under the non-local symmetry transformation
\begin{equation}
\phi(x,t)\leftrightarrow -\partial_t\psi(x,-t) \ ,
\end{equation}
which results in the hyperscaling relation \cite{MGT}:
\begin{equation}
\eta + 2\delta + 1 = \frac{d z}{2}
\end{equation}
As in case of the DP, the relations $\beta=\beta'$ and $\delta=\alpha$
again hold. The upper critical dimension is $d_c=6$.
The dynamical critical exponents as well as spreading and avalanche 
exponents are summarized in \cite{munoz}. The dynamical exponents are
$Z=1.1295$ for $d=2$, $Z=1.336$ for $d=3$ and $Z=2$ for $d=6$.
Dynamical percolation was observed in forest fire models \cite{Dros,Alb}
and in some Lotka-Volterra type lattice prey-predator models \cite{ADLO}
as well.

\subsubsection{Isotropic percolation universality classes} \label{Isopercsect}

\begin{figure}
\epsfxsize=55mm
\epsffile{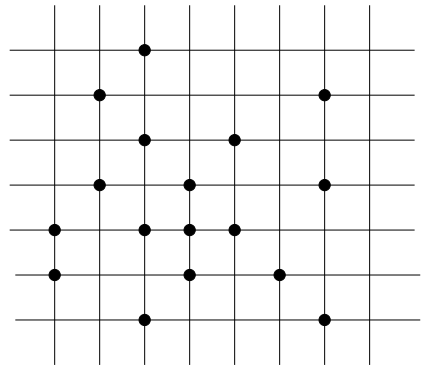}
\caption{Isotropic site percolation in $d=2$ dimensions
\label{Pfig}}
\end{figure}

The ordinary percolation \cite{Stau,Grim} is a geometrical 
phenomenon that describes the occurrence of infinitely large connected 
clusters by {\bf completely random} displacement of some variables 
(sites, bonds, etc) (with probability $p$) on lattices 
(see Fig.\ref{Pfig}).

The dynamical percolation process is known to generate such
percolating clusters (see Sect.\ref{dynperc}). At the transition point 
moments of the $s$ cluster size distribution $n_s(p)$ show singular behavior.
The ordinary percolation corresponds to the $q=1$ limit of the 
Potts model. That means its generating functions can be expressed 
in terms of the free energy of the $q\to 1$ Potts model.
In the low-temperature dilute Ising model the occupation probability ($p$)
driven magnetization transition is an ordinary percolation 
transition as well. 
As a consequence the critical exponents of the magnetization can be 
related to the cluster-size exponents. 
For example the susceptibility obeys simple homogeneity form with
$p-p_c$ replacing $T-T_c$
\begin{equation}
\chi\propto |p-p_c|^{-\gamma}
\end{equation}

Table \ref{Operct} summarizes the known critical exponents of the 
ordinary percolation. The exponents are from the overview \cite{Bunde}.
Field theoretical treatment \cite{Benz} provided an upper critical 
dimension $d_c=6$.
The $d=1$ case is special : $p_c=1$ and the order parameter jumps 
($\beta=0$). Furthermore here some exponents exhibit non-universal 
behavior by increasing the interaction length unless the we redefine 
the scaling variable (see \cite{Stau}).
\begin{table}
\begin{center} 
\begin{tabular}{|c|c|c|c|c|c|}
\hline
d & $\beta=\beta'$  &$\gamma_p$  & $\nu_{\perp}$ & $\sigma$ &$\tau$ \\ 
\hline
1 & 0       & 1        & 1     & 1        & 2     \\
\hline
2 & 5/36    & 43/18    & 4/3   & 36/91    & 187/91  \\
\hline
3 & 0.418(1)& 1.793(4) &0.8765(17)& 0.452(1)&2.189(1)\\
\hline
4 & 0.64    & 1.44     & 0.68  & 0.48     & 2.31    \\
\hline
5 & 0.84    & 1.18     & 0.57  & 0.49     & 2.41  \\
\hline
6 & 1       &  1       & 1/2   & 1/2      & 3/2   \\
\hline
\end{tabular}
\caption{Critical exponents of the ordinary percolation}
\label{Operct}
\end{center}
\end{table}

  \subsubsection{DyP with spatial boundary conditions} \label{sectsdbon}

There are very few numerical results exists for surface critical exponents
of the dynamical percolation.
The GEP process in 3d was investigated numerically by Grassberger
\cite{grassberger-3d}. The surface and edge exponents 
(for angle $\pi/2$) were determined in case of IBC. Different 
measurements (density and cluster simulations) resulted in a single
surface $\beta_1=0.848(6)$ and a single edge $\beta_e=1.36(1)$ exponent.

  \subsubsection{L\'evy flight anomalous diffusion in DyP}

To model long-range epidemic spreading in a system with immunization
the effect of L\'evy flight diffusion (\ref{LongRange}) was investigated by
\cite{JOWH99}.
The renormalization group analysis of the GEP with anomalous diffusion
resulted in the following $\epsilon=3\sigma-d$ expansion
results for the critical initial slip exponent:
\begin{equation}
\eta  = \frac{3{{\epsilon}}}{16\sigma }{+O}\left({\epsilon}
^{2}\right) ,  \label{GEPtheta}
\end{equation}
for the order parameter (density of removed (immune) individuals) exponent:
\begin{equation}
\beta = 1-\frac{{{\epsilon}}}{4\sigma }{+O}\left({\epsilon}
^{2}\right) ,  \label{GEPbeta}
\end{equation}
for the spatial correlation exponent
\begin{equation}
\nu_{\perp} = \frac{1}{\sigma }+\frac{{{\epsilon}}}{
4\sigma ^{2}}{+O}\left( \bar{\varepsilon}^{2}\right) ,
\end{equation}
and for the temporal correlation length exponent
\begin{equation}
\nu_{||} = 1+\frac{{{\epsilon}}}{16\sigma }{
+O}\left({\epsilon}^{2}\right) .
\end{equation}

\subsection{Voter model (VM) classes} \label{glau}

Now we turn to models which can describe the spreading of opinion of
voters arranged on regular lattices. They exhibit first order transition
still dynamical scaling can be observed in them.
The voter model \cite{Ligget,Durr} is defined by the following spin-flip 
dynamics. A site is selected randomly which takes the ``opinion'' (or spin) 
of one of its nearest neighbors (with probability $p$). 
This rule ensures that the model has two homogeneous absorbing states 
(all spin up or down) and is invariant under the $Z_2$ symmetry.
General feature of these models is that dynamics takes place only at the 
boundaries. The action, which describes this behavior was proposed in 
\cite{Dick-Tre,Peliti} 
\begin{equation}
S = \int d^d x dt \left[\frac{D}{2} \phi (1-\phi)\psi^2 
- \psi(\partial_t\phi-\lambda\nabla^2\phi)\right] 
\end{equation}
is invariant under the symmetry transformation:
\begin{equation}
\phi \leftrightarrow 1-\phi, \ \ \ \ \psi \leftrightarrow -\psi \ \ \ .
\end{equation} 
This results in the ``hyperscaling'' relation\cite{MGT}
\begin{equation}
\delta + \eta = d z / 2
\end{equation}
that is valid for all first order transitions ($\beta=0$) with $d\le 2$, 
hence $d_c=2$ is the upper critical dimension. It is also valid for all
{\bf compact growth processes} (where ``compact'' means that the density in 
surviving colonies remains finite as $t\to\infty$).

In {\bf one dimension} at the upper terminal point of the DK SCA 
(Fig.\ref{DKpd} $p_1=\frac12$, $p_2=1$) an extra $Z_2$ symmetry exists
between 1-s and 0-s, hence the scaling behavior is not DP class like 
but corresponds to the fixed point of the inactive phase of 
PC class models (Sect. \ref{PCS}). 
As a consequence {\it compact domains} of 0-s and 1-s grow such as the 
domain walls follow annihilating random walks (ARW) (see Sect.\ref{2A0}) 
and belong to the 1d VM class. 
In 1d the compact directed percolation (CDP) is also equivalent to the 
$T=0$ Glauber Ising model (see Sections \ref{1storder},\ref{Isingcl}). 
By applying non-zero temperature (corresponding to spin flips in domains) 
or symmetry breaking (like changing $p_2$ or adding an external magnetic 
field) a first order transition takes place ($\beta=0$).

In {\bf two (and higher) dimensions} the $p=1$ situation corresponds to 
the $p_1=3/4$, $p_2=1$ point in the phase diagram of $Z_2$ symmetric 
models (see Fig.\ref{fig_phase}).
This model has a ``duality'' symmetry with coalescing random walks:
going backward in time, the successive ancestors of a given spin follow 
the trail of a simple random walk; comparing the values of several spins
shows that their associated random walks necessarily merge upon encounter
\cite{Ligget}. This correspondence permits us to solve many aspects of the
kinetics. In particular, the calculation of the density of interfaces 
$\rho_m(t)$ (i.e. the fraction of $+-$ nearest neighbour (n.n.) pairs) 
starting from random initial conditions of magnetization $m$, 
is ultimately given by the probability that a random walk 
initially at unit distance
from the origin, has not yet reached it at time $t$.
Therefore, owing to the recurrence properties of random walks, the VM shows 
coarsening for $d \le 2$ (i.e. $\rho_{m}(t)\to 0$ when $t \to \infty$).
For the the `marginal' $d = 2$  case one finds the slow logarithmic 
decay \cite{Sch88,Kra92,FK96}:
\begin{equation}
\rho_m(t) = (1-m^2)\left[\frac{2 \pi D}{\ln t} +
{\cal O}\left(\frac{1}{\ln^2{t}}\right)\right] \;,
\label{eq_rho}
\end{equation}
with $D$ is being the diffusion constant of the underlying random walk
($D=1/4$ for the standard case of n.n., square lattice walks,
when each spin is updated on average once per unit of time).

Simulating general, $Z_2$ symmetric spin-flip rules in 2d \cite{Dor} 
conjectured that all critical $Z_2$-symmetric rules without 
bulk noise form a co-dimension-1 `voter-like' manifold separating order 
from disorder, characterized by the logarithmic decay of both $\rho$ and $m$.
The critical exponents for this class are summarized in Table \ref{VMexps}.
\begin{table} 
\begin{center} 
\begin{tabular}{|c|c|c|c|c|c|c|c|c|}
\hline
d & $\beta$ & $\beta'$& $\gamma$ & $\nu_{||}$ & $\nu_{\perp}$ & $Z$ & 
$\delta$ &$\eta$ \\ 
\hline
1 &  0.0    & 1       & 2       & 2           &  1            & 2   & 
1/2      & 0      \\
\hline
2 &  0.0    & 1       & 1       & 1           & 1/2           & 2   & 
1        & 0      \\
\hline
\end{tabular}
\caption{Critical exponents of VM classes}
\label{VMexps}
\end{center}
\end{table}
Furthermore ref. \cite{Dor} found that this $Z_2$ symmetry is not a necessary
condition, the VM behavior can also be observed in systems without bulk
fluctuations, where the total magnetization is conserved. Field theoretical
understanding of these results are still lacking.

\subsubsection{The $2A\to\emptyset$ (ARW) and the $2A\to A$ models} \label{2A0}

As it was mentioned in Sect. \ref{glau} in one dimension the annihilating 
random walk and the voter model are equivalent.
In higher dimensions this is not the case (see Sect. \ref{kA0}).
The simplest reaction-diffusion model -- in which identical particles 
follow random walk and annihilate on contact of a pair -- is adequately 
described by mean-field-type equations in $d_c>2$ dimensions 
\begin{equation}
\rho(t) \propto t^{-1} \ ,
\end{equation}
but in lower dimensions fluctuations become relevant. Omitting boundary and 
initial condition terms, the field theoretical action is
\begin{equation}
S = \int d^d x dt \left[\psi(\partial_t\phi- D \nabla^2\phi)
- \lambda (1-\psi^2)\phi^2 \right] 
\end{equation}
where, $D$ denotes the diffusion coefficient and $\lambda$ is 
the annihilation rate.

For $d=d_c=2$ the leading order decay of the ARW was derived exactly 
by Lee using field theoretical RG method \cite{Lee}:
\begin{equation}
\rho(t)= \frac{1}{8\pi D} \ln(t)/t + O(1/t) \ \ .
\end{equation}
For $d=1$ \cite{Racz85,Lush87} predicted that the particle density decays as
\begin{equation} \label{ARWlaw}
\rho(t) = A_2 (D t)^{-1/2} \ \ .
\end{equation}
This scaling law was confirmed by $\epsilon$ expansion and the universal
amplitude $A_2$ was found to be
\begin{equation}
\frac{1}{4\pi\epsilon} + \frac{2 \ln 8\pi - 5}{16\pi} + O(\epsilon) \ \ .
\end{equation} 

The universal scaling behavior of the ARW was shown to be equivalent to 
that of the $A+A\to A$ coagulation random walk process 
by \cite{Peliti}. 
The renormalization group approach provided universal decay amplitudes 
(different from those of the ARW) to all orders in epsilon expansion. 
It was also shown \cite{DoKi} that the motion of kinks in the compact 
version of directed percolation (CDP) \cite{Essam} and the Glauber-Ising 
model \cite{gla63} at the $T=0$ transition point are also described exactly
by (\ref{ARWlaw}).
These reactions have also intimate relationship to the EW interface growth
model (see Sect. \ref{EWsect}).

  \subsubsection{Compact DP (CDP) with spatial boundary \\
  conditions} \label{sectscbo}

By introducing a wall in CDP, the survival probability is altered and one 
obtains surface critical exponents just as for DP. 
With {\bf IBC}, the cluster is free to approach and leave the wall, but not 
cross. For $d=1$, this gives rise to $\beta_1^\prime = 2$. 
On the other hand, for {\bf ABC}, the cluster is 
stuck to the wall and therefore described by a single random walker for $d=1$.
By reflection in the wall, this may be viewed as {\em symmetric} compact DP 
which has the same $\beta^\prime$ as normal compact DP,
giving $\beta_1^\prime = 1$ \cite{EssamTanlaKishani94,EssamGuttmann95}.

  \subsubsection{CDP with parabolic boundary conditions} \label{sectscbon}

\begin{figure}
\begin{center}
\epsfxsize=65mm
\epsffile{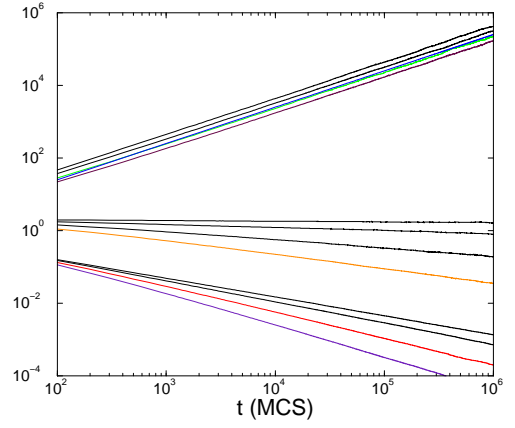}
\caption{Parabola boundary confinement cluster simulations for CDP 
\cite{gdkcikk}.
Middle curves: number of active sites ($C=2,1.5,1.2,1$ top to bottom);
Lower curves: survival probability ($C=2,1.5,1.2,1$ top to bottom);
Upper curves: $R^2(t)$ ($C=2,1.5,1.2,1$ top to bottom).}
\label{parab}
\end{center}
\end{figure}

Space-time boundaries are also of interest in CDP.
Cluster simulations in 1+1 d and MF approximations \cite{gdkcikk,DA01} for 
CDP confined by repulsive parabolic boundary condition of the form 
$x=\pm C t^\sigma$ resulted in $C$ dependent $\delta$ and $\eta$ exponents 
(see Fig.\ref{parab}) similarly to the DP case (see Sect. \ref{sectmbond}) 
in case of marginal condition: $\sigma = 1/2$.
In the mean-field approximations \cite{gdkcikk} similar results were obtained 
as for the DP \cite{Turban2}. Analytical results can be obtained only 
in limiting cases. For narrow systems (small $C$) one obtains the following 
asymptotic behavior for the connectedness function to the origin:
\begin{equation}
P(t,x)\sim t^{-\pi^2/8C^2}\cos\left({\pi x\over 2C\sqrt{t}}\right) \ .
\label{diff10}
\end{equation}
Recently an analytical solution was derived for a related problem \cite{DA01}.
For a one-dimensional lattice random walk with an absorbing boundary at the 
origin and a movable partial reflector (with probability $r$) $\delta$ varies 
continuously between $1/2$ and $1$ as $r$ varies between $0$ and $1$.

  \subsubsection{L\'evy flight anomalous diffusion in ARW-s} \label{LARWs}

Long-range interactions generated by non-local diffusion in annihilating 
random walks results in the recovery of mean-field behavior. This has
been studied by different approaches.
Particles performing simple random walks subject to the reactions 
$A+B\rightarrow \emptyset$ (Sect.\ref{2A0}) and $A+A\rightarrow \emptyset$ 
(Sect.\ref{AB}) in the presence of a quenched velocity field were investigated
in \cite{ZumofenKlafter8}. The quenched velocity field enhances the diffusion 
in such a way that the effective action of the velocity field is reproduced 
if L\'{e}vy flights are substituted for the simple random walk motion. 
In the above mentioned reactions the particle density decay is
algebraic with an exponent related to the step length distribution of 
the L\'{e}vy flights defined in
Eq.~(\ref{LongRange}). These results have been confirmed by
several renormalization group calculations
\cite{Oerding8,DeemPark8,DeemPark8b}.

\begin{figure}
\epsfxsize=70mm
\epsffile{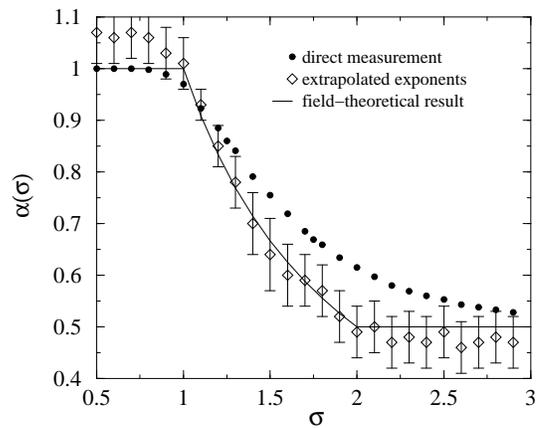}
\caption{
\label{FigAnnh}
The anomalous annihilation process: the graph from \cite{HH99} shows direct
estimates and extrapolations for the decay exponent $\alpha$,
as a function of $\sigma$. The solid line represents
the exact result (neglecting $\log$ corrections at $\sigma=1$).}
\end{figure}

The $A+A\rightarrow\emptyset$ process with anomalous diffusion 
was investigated by field theory \cite{HH99}. 
The action of this model is
\begin{eqnarray}
\label{AnnihilationAction}
S[\bar{\psi},\psi]
&=& 
\int d^dx \, dt \,\Bigl\{
\bar{\psi} (
\partial_t - D_N \nabla^2 - D_A \nabla^\sigma 
)\psi
\nonumber
\\ && \hspace{12mm}
+  2 \lambda \bar{\psi}\psi^2 + \lambda \bar{\psi}^2\psi^2 -
n_0\bar{\psi}\delta(t)   
\Bigr\} \ ,
\end{eqnarray}
where $n_0$ is the initial (homogeneous) density at $t=0$.
The density decays for $\sigma<2$ as :
\begin{equation}
\label{AnnhilationResult}
n(t) \sim \left\{ \begin{array}{ll}
    t^{-d/\sigma}       & {\rm for \ } d<\sigma \ , \\
    t^{-1} \ln t             & {\rm for \ } d=d_c=\sigma \ , \\
    t^{-1}              & {\rm for \ } d>\sigma \ .
    \end{array} \right.
\end{equation}
The simulation results of the corresponding 1+1 d lattice model
\cite{HH99} can be seen in Fig. \ref{FigAnnh}.
It was also shown in \cite{HH99} that L\'evy flight {\bf annihilation and
coagulation processes} ($A+A\to A$) are in the same universality class.

\subsection{Parity conserving (PC) classes} \label{PCS}

In an attempt to generalize DP and CDP like systems we arrive to new models, 
in which a conservation law is relevant.
A new universality class appears among 1+1 dimensional, single component, 
reaction-diffusion models.
Although it is usually named parity conserving class (PC) examples 
have proved that the parity conservation itself is not a sufficient 
condition for the PC class behavior. 
For example in \cite{ITT95} a 1d stochastic cellular 
automaton with a global parity conservation was shown to exhibit DP class 
transition. Binary spreading process (see Sect.\ref{binsp}) in one 
\cite{binary} and two dimensions \cite{OSC02} were also found to be
insensitive to the presence of the parity conservation. 
Multi-component BARW2 models in 1d (see Sect.\ref{NBARWS}) generate
different, robust classes again \cite{Cardy-Tauber,Od01b,Hoy}. 
By this time it is known that the BARW2 dynamics in
{\bf single-component, single-absorbing} state systems 
(without inhomogeneities,
long-range interactions and other symmetries) 
provides sufficient condition for the PC class \cite{Cardy-Tauber}.
In {\bf single-component, multi-absorbing} state systems the $Z_2$ symmetry 
ensures necessary but not sufficient condition, 
the BARW2 dynamics (Sect.\ref{BARWe}) of domain walls is also a necessary 
condition. 
Some studies have shown \cite{Parkh,meod96,HKPP98} that an external field 
that destroys the $Z_2$ symmetry of absorbing states (but preserves the
the BARW2 dynamics) yields a DP instead of a PC class transition in the system.
Some other names for this class are also used, like directed Ising (DI) class, 
or BARW class.

   \subsubsection{Branching and annihilating random walks with \\
   even number of offspring} \label{BARWe}

The generic models of the PC class, for which field theoretical treatment
exists are branching and annihilating random walks -- introduced in 
Sect. \ref{BARWo} -- with $k=2$ and even number ($m$) of offspring (BARWe)
\cite{ZA95,L96,Ja97,ALR94,Jen93,Jensen}.
These conserve the particle number mod 2, hence there are two distinct 
sectors in these models, an odd and an even parity one. In the
even sector particles finally die out ($\delta\not=0$, $\eta=0$), 
while in the odd one at least one particle always remain alive 
($\delta=0$, $\eta\not=0$). BARWe dynamics may also appear in multi-component 
models exhibiting $Z_2$ symmetric absorbing states in terms of the kinks 
between ordered domains. Such systems are the NEKIM (Sect. \ref{nekim}) and 
GDK models (Sect. \ref{GDKmod}) for example. It was conjectured \cite{MeOdof}
that in all models with $Z_2$ symmetric absorbing states
an underlying BARWe process is a necessary condition for 
phase transitions with PC criticality. 
Sometimes it is not so easy to find the underlying BARWe process and can 
bee seen on the coarse grained level only (see example the GDK model, 
in which the kinks are spatially extended objects). This might have lead 
some studies to the conjecture that the $Z_2$ symmetry is a sufficient 
condition for the PC class \cite{HP99}. 
However the example of CDP (see Sect.\ref{glau}) shows that this cannot
be true. The field theory of BARWe models was investigated
by \cite{Cardy-Tauber,Cardy-Tauber2}. 
For this case the action
\begin{eqnarray}
S=\int d^d x dt &[&\psi(\partial_t-D\nabla^2)\phi - 
\lambda(1-\psi^2)\phi^2 + \nonumber \\ 
&+&\sigma(1-\psi^2) \psi \phi \ ]
\end{eqnarray}
is invariant under the simultaneous transformation of fields
\begin{equation}
\psi \leftrightarrow -\psi, \ \ \ \phi \leftrightarrow -\phi
\end{equation}
Owing to the non-recurrence of random walks in $d\ge2$ the system is in 
the active phase for $\sigma > 0$ and mean-field transition occurs with 
$\beta=1$. However the survival probability of a particle cluster is 
finite for any $\sigma > 0$ implying $\beta'=0$. Hence contrary to the 
DP class $\beta \ne \beta'$ for $d\ge 2$. At $d=2$ random walks are barely 
recurrent and logarithmic corrections can be found. In this case the 
generalized hyperscaling law \cite{TTP} is valid among the exponents
\begin{equation}
2 \left(1 + \frac{\beta}{\beta'}\right)\delta' + 2\eta' = d z .
\label{ghypers}
\end{equation}
In $d=1$ however $\beta=\beta'$ holds owing to an exact duality mapping
\cite{Mussa} and the hyperscaling is the same that of the DP 
(eq. \ref{dphyper}).

The RG analysis of BARWe for $d<2$ run into difficulties. These stem from, 
the presence of another critical dimension $d_c'=4/3$ (above which the 
branching reaction is relevant at $\sigma=0$, and irrelevant for $d<d_c'$) 
hence the $d=1$ dimension cannot be accessed by controlled 
expansions from $d_c=2$. The truncated one-loop expansions \cite{Cardy-Tauber}
for $d=1$ resulted in
\begin{equation}
\beta=4/7 \ \ ,\ \nu_{\perp}=3/7 \ \ , \ \nu_{||}=6/7 \ \ , \ Z=2
\end{equation}
which are quite far from the numerical values determined by Jensen's 
simulations \cite{Jensen}
(Table \ref{1dBAWeexp}).
\begin{table}
\begin{center} 
\begin{tabular}{|c|c|c|c|c|c|c|c|}
\hline
  $d$  & $\beta$ & $\beta'$ & $\gamma$ & $\delta$ &$Z$ &$\nu_{||}$ & $\eta$ \\ 
\hline 
 1     & 0.92(3) & 0.92(3)  & 0.00(5) & 0.285(2) & 1.75 &3.25(10) & 0.000(1) \\
\hline
 2     &  1     &   0       &  1      &   0      & 2    & 1     &    -1/2    \\
\hline
\end{tabular}
\caption{Critical exponents of BARWe.}
\label{1dBAWeexp}
\end{center}
\end{table}
Here the cluster exponents $\delta$ and $\eta$ corresponding to 
the sector with even number of initial particles are shown. 
In case of odd number of initial particles they exchange values.
\begin{figure}
\epsfxsize=70mm  
\epsffile{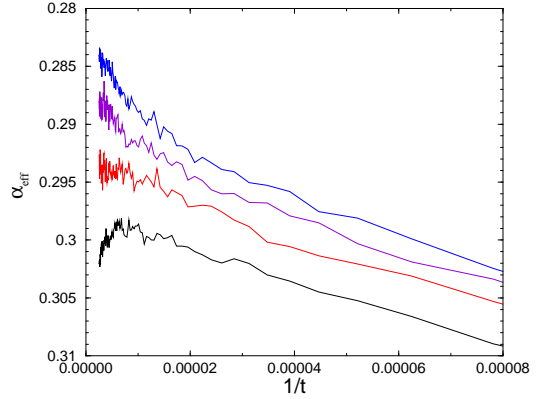}
\caption{Local slopes (\ref{aeff}) of the density decay in a bosonic 
BARW2 model. Different curves correspond to $\sigma=0.466$, $0.468$ $0.469$, 
$0.47$ (from bottom to top) \cite{OdMe02}.}
\label{bpc}
\end{figure}

It was conjectured \cite{Wijcon} that in 1d fermionic (single occupancy) and 
bosonic  (multiple occupancy) models may have different critical behavior. 
Since only bosonic field theory exists, which gives rather inaccurate critical 
exponent estimates in \cite{OdMe02} bosonic simulations were performed to 
investigate the density decay of BARW2 from random initial state.
Figure \ref{bpc} shows the local slopes of density decay ($\alpha_{eff}$
(\ref{aeff})) around the critical point for several branching rates ($\sigma$).
The critical point is estimated at $\sigma_c=0.04685(5)$, with the 
corresponding decay exponent $\alpha=0.290(3)$. This value agrees with that 
of the PC class.

If there is no explicit diffusion of particles besides the $AA\to\emptyset$,
$A\to 3A$ processes (called DBAP by \cite{Sud}) an implicit
diffusion can still be generated. By spatially asymmetric branching:
$A\emptyset\emptyset \to AAA$ and $\emptyset\emptyset A \to AAA$ a 
diffusion may go on by two lattice steps:
$A\emptyset\emptyset \to AAA \to \emptyset\emptyset A$. As a consequence
the decay process slows down, a single particle can not join a domain, 
hence domain sizes exhibit a parity conservation. This results in additional
new sectors (besides the existing two BARW2 sectors) that depend on the initial
conditions. For example in case of random initial distribution 
$\delta \sim 0.13(1)$ was measured by simulations \cite{HiOd99}.
Similar sector decomposition has been observed in diffusion of 
$k$-mer models (see for example \cite{Barma,Barma2}).

   \subsubsection{The NEKIM model} \label{nekim}

An other important representative of PC class appear among nonequilibrium 
Ising models, in which the steady state is generated 
by kinetic processes in connection with heat baths at different temperatures 
\cite{dem85,dem86,gonzalez,wan88,Droz89}.
The research of them have shown that phase transitions are possible 
even in 1d under nonequilibrium conditions 
(for a review see \cite{Raczof}). 
In short ranged interaction models any non-zero temperature spin-flip dynamics
cause disordered steady state. Menyh\'ard proposed a class of general 
nonequilibrium kinetic Ising models (NEKIM) with combined spin flip dynamics 
at $T=0$ and Kawasaki spin exchange dynamics at $T=\infty$ in which, for a 
range of parameters of the model, a PC-type transition takes place 
\cite{Men94}. 

A general form \cite{gla63} of the Glauber spin-flip transition rate 
in one-dimension for spin $s_i = \pm 1$ sitting at site $i$ is:
\begin{equation}
w_i = {\Gamma\over{2}}(1+\tilde\delta s_{i-1}s_{i+1})\left(1 - 
{\tilde\gamma\over2}s_i(s_{i-1} + s_{i+1})\right) \ .
\label{Gla}
\end{equation}
Here $\tilde\gamma=\tanh{({2J}/{kT}})$, $J$ denotes the coupling constant 
in the ferromagnetic Ising Hamiltonian, $\Gamma$ and $\tilde\delta$ are 
further parameters, which can in general, also depend on temperature. 
The Glauber model is a special case corresponding to $\tilde\delta=0$, 
$\Gamma=1$. There are three independent rates:
\begin{eqnarray}
w_{\uparrow\uparrow\uparrow} &=&
{\frac{\Gamma}2}(1+\tilde\delta)(1-\tilde\gamma), \ \ \ \
w_{\downarrow\uparrow\downarrow} = 
{\frac{\Gamma}2}(1+\tilde\delta)(1+\tilde\gamma) \nonumber \\
w_{\uparrow\uparrow\downarrow} &=& {\frac{\Gamma}2}(1-\tilde\delta).
\label{rates}
\end {eqnarray}
In the NEKIM model $T=0$ is taken, thus $\tilde\gamma =1$, 
$w_{\uparrow\uparrow\uparrow}=0$
and $\Gamma$, $\tilde\delta$ are the control parameters to be varied. 

The Kawasaki spin-exchange rate of neighboring spins is:
\begin{equation}
w_{ii+1}(s_i,s_{i+1})={p_{ex}\over2}(1-s_{i}s_{i+1})[1-{\tilde\gamma\over2}
(s_{i-1}s_i+
s_{i+1}s_{i+2})].
\label{Kaw}
\end{equation}
At $T=\infty$ ($\tilde\gamma=0$) the above exchange is simply an unconditional
nearest neighbor exchange:
\begin{equation}
w_{ii+1}={1\over2}p_{ex}[1-s_is_{i+1}]
\label{ex}
\end{equation}
where $p_{ex}$ is the probability of spin exchange.
The transition probabilities in eqs.(\ref{Gla}) and (\ref{ex}) 
are responsible for the basic elementary processes of kinks ($K$). 
Kinks separating two ferromagnetically
ordered domains can carry out random walks with probability 
\begin{equation}
p_{rw}\propto 2w_{\uparrow\uparrow\downarrow}={\Gamma}(1-\tilde\delta) \ ,
\end{equation}
while two kinks getting into neighboring positions will annihilate with 
probability 
\begin{equation}
p_{an}\propto w_{\downarrow\uparrow\downarrow}={\Gamma}(1+\tilde\delta)
\end{equation}
($w_{\uparrow\uparrow\uparrow}$ is responsible for creation of kink pairs 
inside of ordered domains at $T\not=0$).
In case of the spin exchanges, which act only at domain boundaries,
the process of main importance here is that a kink can produce two
offspring at the next time step with probability 
\begin{equation}
p_{K\rightarrow 3K}\propto{p_{ex}}.
\end{equation}
The abovementioned three processes compete, and it depends on 
the values of the parameters $\Gamma$, $\tilde\delta$ and $p_{ex}$ what the 
result of this competition will be. It is important to realize that the 
process $K \rightarrow 3K$ can develop into propagating offspring production 
only if $p_{rw} > p_{an}$, i.e. the new kinks are able to travel on the 
average some lattice points away from their place of birth and can thus avoid 
immediate annihilation. It is seen from the above definitions that 
$\tilde\delta < 0$ is necessary for this to happen. In the opposite case 
the only effect of the $K\rightarrow 3K$ process on the usual Ising 
kinetics is to soften domain walls. In the NEKIM model investigations the 
normalization condition $p_{rw}+p_{an}+p_{k\rightarrow3k}=1$ was set. 

\begin{figure}
\epsfxsize=75mm  
\epsffile{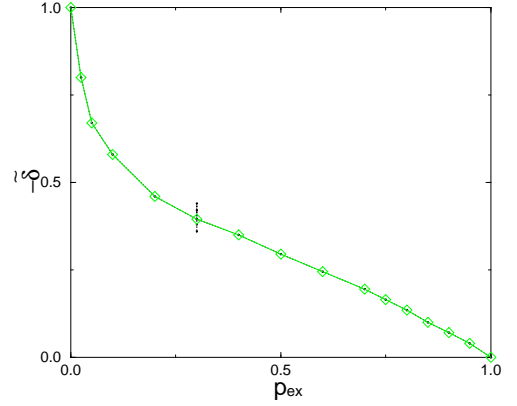}
\caption{Phase diagram of the two-parameter model. The transversal dotted 
line indicates the critical point that was investigated in more detail
\cite{MeOdof}.}
\label{nekimphasedia}
\end{figure}

\begin{figure}
\epsfxsize=70mm
\epsffile{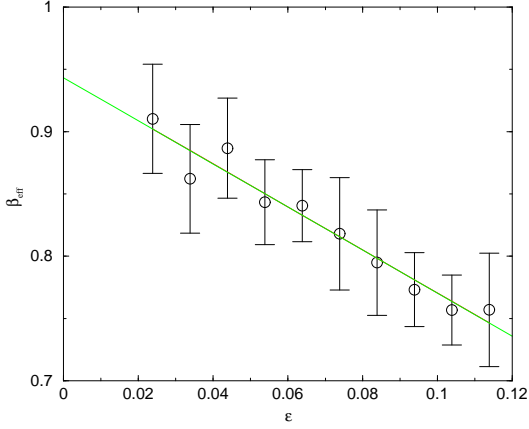}
\caption{$\beta_{eff}=\frac{d\log \rho_{\infty}}{d\log\epsilon}$ 
of kinks (circles) near the critical point 
($\epsilon=|\tilde\delta-\tilde\delta'|$) and linear extrapolation 
(dashed line) to the asymptotic value ($\beta=0.95(2)$).
Simulations were performed on a 1d NEKIM ring of size $L=24000$ \cite{MeOdof}.}
\label{betaf}
\end{figure}

The phase diagram determined by simulations and GMF calculations
\cite{meorcikk,MeOdof} is shown in Fig. \ref{nekimphasedia}.
The line of phase transitions separates two kinds of steady states reachable 
by the system for large times: in the Ising phase, supposing that an even
number of kinks are present in the initial states, the system orders in one 
of the possible ferromagnetic states of all spins up or all spins down, while 
the active phase is disordered from the point of view of the underlying spins.
The cause of disorder is the steadily growing number of kinks with time. 
While the low-level, $N=1,2$ GMF solutions for the SCA version of NEKIM
exhibit first order transitions, for $N>2$ this becomes continuous. 
GMF approximations (up to $N=6$) with CAM extrapolation found 
$\beta\simeq 1$ \cite{meorcikk}.
Recent high precision Monte Carlo simulations \cite{MeOdof} resulted in
critical exponents $\beta=0.95(2)$ (see Fig. \ref{betaf}) and 
$\delta=0.280(5)$ at the dotted line of the phase diagram.\\
Ref. \cite{Mussa} have shown that an exact duality mapping exist
in the phase diagram of the NEKIM:
\begin{eqnarray}
p'_{an} = p_{an} , \nonumber \\ 
p'_{rw} =  p_{an} + 2 p_{ex} ,\nonumber \\
2 p'_{ex} = p_{rw} - p_{an} \ .
\label{duality}
\end{eqnarray}
The regions mapped onto each other have the same physical 
properties. In particular, the line $p_{ex} = 0$ maps onto the
line $p_{rw}=p_{an}$ and the fast-diffusion limit to the limit 
$p_{ex} \to \infty$. There is a self-dual line at
\begin{equation}
\tilde\delta=\frac{-2p_{ex}}{1-p_{ex}} \ \ \ .
\end{equation}
By various static and dynamical simulations spin and kink density
critical exponents have been determined in \cite{meod96} and as the
consequence of the generalized hyperscaling law for the structure factor
\begin{equation}
S(0,t)=L[<M^2>-<M>^2]\propto t^x ,
\end{equation}
and kink density 
\begin{equation}
n(t)=\frac{1}{L}<\sum_{i}\frac{1}{2}(1-s_{i}s_{i+1})> \propto t^{-y}
\end{equation}
the exponent relation
\begin{equation}
2 y = x
\end{equation}
is established.
Spins-clusters at the PC point grow by compact domains as in
the Glauber point albeit with different exponents \cite{cpccikk}. 
The spin-cluster critical exponents in magnetic field 
are summarized in Table (\ref{1dIsingspinexps}).
\begin{table}
\begin{center}
\begin{tabular}{|l|l|l|l|l|l|l|}
\hline
& $\beta_s$ & $\gamma_s$ &$\nu_{\perp,s}$ & Z & $\theta_{g}$ & $\lambda_s$ \\
\hline
Glauber-Ising& 0 & $1/2$ & $1/2$ & $2$   & $1/4$ & $1$ \\
\hline
PC &$ .00(1)$ &.444(2) &$.444(2)$ & 1.75(1) & $.67(1)$ & 1.50(2)\\
\hline
\end{tabular}
\caption{Simulation data for static and dynamic critical spin exponents
for NEKIM.}
\label{nekimpersexps}
\end{center}
\end{table}
The global persistence ($\theta_g$) and time autocorrelation exponents 
($\lambda$) were determined both at the Glauber an at the PC critical points 
\cite{perscikk} and are shown in Table (\ref{nekimpersexps}). 
While at the Glauber point the scaling relation (\ref{persscal}) is satisfied
by these exponents it is not the case at PC criticality, therefore
the magnetization is non-Markovian process here.
 
By applying an external magnetic field $h$ that breaks the $Z_2$ symmetry
the transition type of the model changes to DP type (see Table \ref{tablex}).
\cite{meod96} 
\begin{table}
\begin{center}
\begin{tabular}{|l|r|r|r|r|r|l|}
\hline
h       & 0.0 & 0.01 & 0.05 & 0.08 &  0.1 & DP \\
\hline
$\beta$ & 1.0 & 0.281& 0.270& 0.258& 0.285& 0.2767(4) \\
\hline
$\gamma$&     & 0.674& 0.428& 0.622& 0.551& 0.5438(13) \\
\hline
\end{tabular}
\caption{CAM estimates for the kink density and its fluctuation exponents.}
\label{tablex}
\end{center}
\end{table}

In \cite{ujcikk} the symmetry of the spin updates was broken in such a way 
that two different types of domain walls emerged.
The following changes to the Glauber spin-slip rates (\ref{rates})
with $\Gamma=1$, $\tilde\delta=0$ were introduced:
\begin{eqnarray}
w_{\uparrow\downarrow\uparrow} &=& 0, \label{-++} \\
w_{\uparrow\uparrow\downarrow} &=& w_{\downarrow\uparrow\uparrow} = 
p_{+} < 1/2 \label{++-} \ ,
\end{eqnarray}
while the spin-exchange part remains the same. In the terminology of domain
walls as particles the following reaction-diffusion picture arises.
Owing to the symmetry breaking there are two kinds of domain walls
$\downarrow\uparrow\equiv A$ and $\uparrow\downarrow\equiv B$, 
which can only occur alternately ( ...B..A..B..A..B...A...) owing to the 
spin background. Upon meeting $AB\to\emptyset$ happens, while in the 
opposite sequence, $BA$, the two domain-walls are repulsive due to (\ref{-++}).
The spin exchange leads to $A\leftrightarrow ABA$ and $B\leftrightarrow BAB$
type of kink reactions, which together with the diffusion of $A$-s and $B$-s
leads to a kind of two-component, coupled branching and annihilating random
walk (see Sect.\ref{NBARWS}). There are two control parameters in this 
model: $p_{ex}$ that regulates the kink production-annihilation and 
$p_+$ that is responsible for the local symmetry breaking (\ref{++-}). 
Simulations show that for $p_{ex}\to 0$, $p_+<0.5$ an absorbing phase emerges
with N-BARW2 class exponents (owing to the pairwise order of kinks hard-core
effects cannot play a role), while the transition on the $p_{ex}>0$
line belongs to the 1+1 d DP class. Since the $AB\to\emptyset$ reaction
breaks the parity conservation of species (but preserves the global
parity conservation) the necessary conditions for N-BARW2 class can be
eased. On the other hand the occurrence of the DP transition introduces
a zero branching rate condition for N-BARW2 universal behavior.
This study and the results for the generalized contact process
(Sect.\ref{BGDP}) emphasize that the
conditions for the N-BARW2 class should be further investigated.

A generalization is the probabilistic cellular automaton version of 
NEKIM, which consists in keeping the spin-flip rates given in 
eqs.(\ref{rates}) and prescribing {\it synchronous updating}. 
In this case the $K\to 3K$ 
branching is generated without the need of additional, explicit spin-exchange 
process and for certain values of parameter-pairs $(\Gamma,\tilde\delta)$ with
$\tilde\delta<0$ PC-type transition takes place. 
The phase boundary of NEKIM-CA in the $(\Gamma,-\tilde\delta)$ plane is
similar to that in Fig. \ref{nekimphasedia} except for the highest value of
$\Gamma=1$, $\tilde\delta_c=0$ cannot be reached, the limiting value is
$\tilde\delta_c=-.065$.

An other possible variant of NEKIM was introduced in \cite{MeOdof} in 
which 
the Kawasaki rate eq.(\ref{Kaw}) is considered at some finite temperature,
instead of $T=\infty$, but  keeping $T=0$ in the Glauber-part of the rule. 
By lowering the temperature the spin exchange process acts
against the kink production and a PC class transition occurs. 
In this case the active phase part of the phase diagram shrinks. 
For more details see (\cite{Geza}).

The {\bf DS transition} of this model coincides with the critical
point and the scaling behavior of spin and kink damages is the same as
that of the corresponding NEKIM variables \cite{odme98}.

   \subsubsection{Parity conserving stochastic cellular automata} 
   \label{GDKmod}

The first models in which non-DP class transition to absorbing state 
was firmly established were 1d stochastic cellular automata defined by 
\cite{Gras84}. In these models the $00$ and $11$ pairs follow 
BARW2 dynamics and their density is the order parameter that vanishes at
some critical point. The PC class critical exponents estimated by simulations 
for these models by \cite{GrasAB}. While in the {\bf ``A'' model}
the critical point coincides with the DS transition point and
both of them are PC type, in the {\bf ``B'' model} the DS transition 
occurs in the active phase - where the symmetry of replicas is broken -
and therefore the DS exponents belong to the DP class \cite{odme98}.

Another stochastic CA, which may exhibit PC class transition and
which is studied later from different directions is also introduced here. 
It points out that the underlying BARW2 dynamics
of domain walls in $Z_2$ symmetric systems can sometimes be seen on
coarse grained level only.
A generalization of the Domany-Kinzel stochastic cellular automaton 
\cite{DoKi} (see Section \ref{sectsca}) was introduced by \cite{Hin97}. 
This model has $n+1$ states per site: one active state $Ac$ 
and $n$ different inactive states $I_1,I_2,\ldots,I_n$.
The conditional updating probabilities are given by
$(k,l=1,\ldots,n; \,\, k\neq l)$
\begin{eqnarray}
\label{eqa}
P(I_k\,|\,I_k,I_k)&=&1\,,\\
\label{eqb}
P(Ac|Ac,Ac)=1-n\,P(I_k|Ac,Ac)&=&q\,, \\
\label{eqc}
P(Ac|I_k,Ac)=P(Ac|Ac,I_k)&=&p_k\,,  \\ \nonumber
P(I_k|I_k,Ac)=P(I_k|Ac,I_k)&=&1-p_k\,,\\
\label{eqd}
P(Ac|I_k,I_l)&=&1\,, 
\end{eqnarray}
and the symmetric case $p_1,\ldots,p_n=p$ was explored.
Equations (\ref{eqa})--(\ref{eqc}) are straightforward
generalizations of Eqs. (\ref{eq1})--(\ref{eq3}).
The only different process is the creation of active sites between
two inactive domains of different colors in Eq.~(\ref{eqd}). 
For simplicity the probability of this process was chosen to be equal 
to one.
\\
\indent
For $n=1$ the model defined above reduces to the original Domany-Kinzel
model. For $n=2$ it has two $Z_2$ symmetrical absorbing states. 
The phase diagram of this model is very similar to
that of the DK model (Fig. \ref{DKpd}) except the transition line is PC type. 
If we call the regions separating inactive domains $I_1$ and $I_2$ as
domain walls (denoted by $K$), they follow BARW2 process 
\begin{equation}
K \to 3K \ \ \ \ 2 K \to\emptyset \ \ \ \ 
K\emptyset\leftrightarrow\emptyset K \label{BARW2proc}
\end{equation}
but for $1>q>0$ the size of active regions, hence the domain walls stays 
finite. Therefore a the observation
of the BARW2 process is not so obvious and can be done on coarse grained 
level only (except at the endpoint at $q=0$, where active sites really look 
like kinks of the NEKIM model).
Series expansions for the transition point and for the order parameter
critical exponent resulted in $\beta=1.00(5)$ \cite{JenGDK} that is
slightly higher than the most precise simulation results \cite{MeOdof}
but agrees with the GMF+CAM estimates \cite{meorcikk}.
Similarly to the NEKIM model the application of external symmetry breaking 
field changes the PC class transition into a DP class one \cite{Hin97}.
The other symmetrical endpoint ($q=1$, $p=\frac12$) in the phase diagram
again shows different scaling behavior 
(here three types of compact domains grow in competition, 
the boundaries perform annihilating random walks with exclusion (see. 
Section \ref{2}).

Models with $n>3$ symmetric absorbing states in 1d do not show phase 
transitions (they are always active). In terms of domain walls as 
particles they are related to $N>1$ component N-BARW2 processes, which
exhibit phase transition for zero branching rate only 
(see Sect.\ref{NBARWS}) \cite{Cardy-Tauber,Hoy}.

GDK type models -- exhibiting $n$ symmetric absorbing states -- 
can be generalized to higher dimensions. In two dimensions Hinrichsen's
spreading simulations for the $n=2$ case yielded mean-field like behavior with
$\delta=1$, $\eta=0$ and $z=1$ leading to the conjecture that $1< d_c < 2$. 
A similar model exhibiting Potts-like $Z_n$ symmetric 
absorbing states in $d=2$ yielded similar spreading exponents
but a first order transition ($\beta=\alpha=0$) for $n=2$ \cite{Lipp}.
In three dimensions the same model seems to exhibit mean-field like
transition with $\beta=1$. The verification of these findings would require
further research.

   \subsubsection{PC class surface catalytic models}

In this subsection I discuss some one-dimensional, surface catalytic-type 
reaction-diffusion models exhibiting PC class transition. Strictly speaking 
they are multi-component models, but I show that the symmetries among species 
enables us to interpret the domain-wall dynamics as a simple BARW2 process.

\begin{figure}
\epsfxsize=70mm
\epsffile{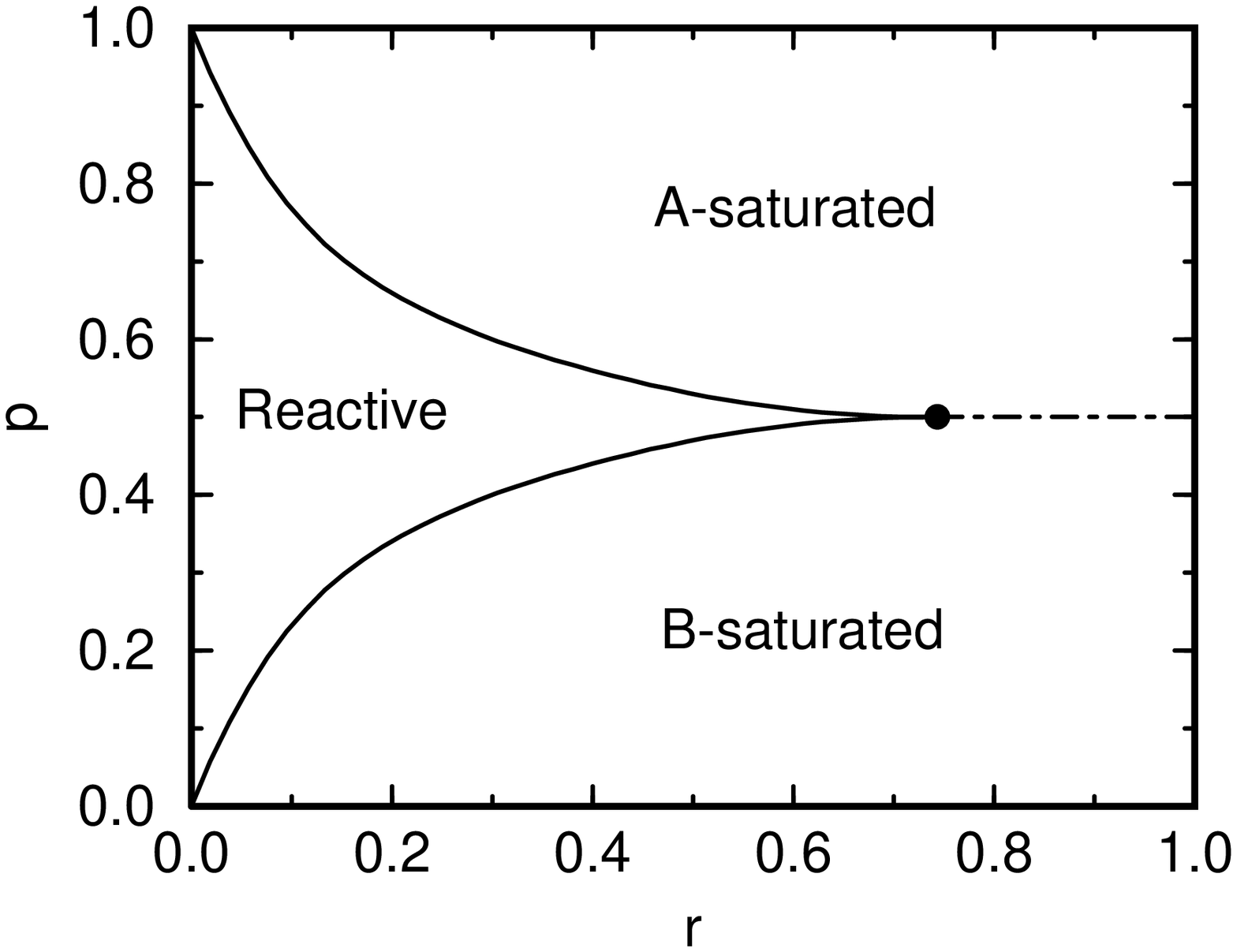}
\caption{Phase diagram of the MM from \cite{BBBE}.}
\label{mmpd}
\end{figure}

\begin{figure}
\epsfxsize=70mm
\epsffile{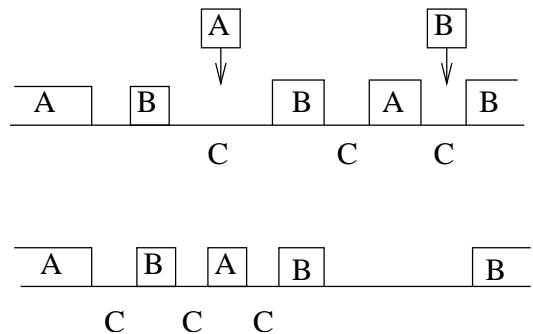}
\caption{Domain-wall dynamics in the interacting monomer-monomer model. 
Left part: branching; right part: annihilation}
\label{m-m}
\end{figure}

The two-species monomer-monomer (MM) model was first introduced by \cite{ZPR}.
Two monomers, called $A$ and $B$, adsorb at the vacant sites of a 
one-dimensional lattice with probabilities $p$ and $q$, respectively, 
where $p+q=1$.  The adsorption of a monomer at a vacant site is affected 
by monomers present on neighboring sites.  If either neighboring site is 
occupied by the same species as that trying to adsorb, the adsorption 
probability is reduced by a factor $r<1$, mimicking the effect of a 
nearest-neighbor repulsive interaction.
Unlike monomers on adjacent sites react immediately and leave the
lattice, leading to a process limited by adsorption only.
The basic reactions are
\begin{equation}
\emptyset \to A, \ \ \ \emptyset \to B, \ \ \ AB \to \emptyset
\label{mmr}
\end{equation}
The phase diagram, displayed in Fig.~\ref{mmpd} with $p$ plotted 
{\em vs.\/} $r$, shows a reactive steady state containing vacancies
bordered by two equivalent saturated phases (labeled A and B). 
The transitions from the reactive phase to either of the saturated phases 
are continuous, while the transition between the saturated phases is 
first-order discontinuous.
The two saturated phases meet the reactive phase at a {\em bi-critical\/} 
point at a critical value of $r=r_c$. In the case of $r = 1$, the reactive 
region no longer exists and the only transition is a first-order 
discontinuous line between the saturated phases.  
Considering the density of vacancies between unlike species as the
order parameter (that can also be called a species ``C'') the model is 
the so called ``three species monomer-monomer model''. Simulations and 
cluster mean-field approximations were applied to investigate the phase 
transitions of these models \cite{Bas,BBE,BBBE,BBJ}. 
As Fig.\ref{m-m} shows if we call the extended objects filled with 
vacancies between different species as domain-walls ($C$) we can observe 
$C\to 3C$ and $2C\to\emptyset$ BARW2 processes in terms of them.
These $C$ parity conserving processes arise as the combination of the 
elementary reaction steps (\ref{mmr}). The reactions always take place 
at domain boundaries, hence the $Z_2$ symmetric $A$ and $B$ saturated 
phases are absorbing. 
 
The interacting monomer-dimer model (IMD) \cite{Park94} is a generalization 
of the simple monomer-dimer model \cite{ZGB}, in which particles of the same 
species have nearest-neighbor repulsive interactions. The IMD is parameterized 
by specifying that a monomer ($A$) can adsorb at a nearest-neighbor
site of an already-adsorbed monomer (restricted vacancy) at a rate
$r_Ak_A$ with $0 \leq r_A \leq 1$, where $k_A$ is an adsorption rate
of a monomer at a free vacant site with no adjacent monomer-occupied
sites. Similarly, a dimer ($B_2$) can adsorb at a pair of
restricted vacancies ($B$ in nearest-neighbor sites) at a rate
$r_Bk_B$ with $0 \leq r_B \leq 1$, where $k_B$ is an adsorption rate of
a dimer at a pair of free vacancies. There are no nearest-neighbor
restrictions in adsorbing particles of different species and the 
$AB\to\emptyset$ desorption reaction happens with probability 1. The case
$r_A = r_B = 1$ corresponds to the ordinary noninteracting monomer-dimer
model which exhibits a first-order phase transition between two
saturated phases in one dimension. In the other limiting case
$r_A = r_B = 0$, there exists no fully saturated phase of monomers or
dimers. However, this does not mean that this model no longer has any 
absorbing states. In fact, there are two equivalent ($Z_2$ symmetric)
absorbing states in this model. These states comprise of only the monomers 
at the odd- or even-numbered lattice sites. A pair of adjacent vacancies 
is required for a dimer to adsorb, so a state with alternating
sites occupied by monomers can be identified with an absorbing state.
The PC class phase transition of the $r_A = r_B = 0$ infinite repulsive 
case has been investigated in \cite{Park94,PKP95,Parkh,KP95,HKPP98}.
As one can see the basic reactions are similar to those of the MM model 
(eq. (\ref{mmr})) but the order parameter here is the density of dimers
($K$) that may appear between ordered domains of alternating sequences:
'0A0..A0A.' and 'A0A..0A0', where monomers are on even or odd sites only.
The recognition of an underlying BARW2 process (\ref{BARW2proc})
is not so easy in this case, 
still considering regions between odd and even filled ordered domains
one can identify domain wall random-walk, annihilation and branching 
processes through the reactions with dimers as one can see on 
the examples below.
The introduction of $Z_2$ symmetry-breaking field, that makes the system
prefer one absorbing state to the other was shown to change that transition
type from PC to DP \cite{Parkh}.
\begin{verbatim}
t     A 0 A 0 A 0 A 0 A 0 0 A 0 A 0 A        K
t+1   A 0 A 0 A 0 A 0 A B B A 0 A 0 A        K
t+2   A 0 A 0 A 0 A 0 0 0 B A 0 A 0 A    K K K
\end{verbatim}

\begin{verbatim}
t     A 0 A 0 A 0 A 0 A 0 0 0 A 0 A 0      K K
t+1   A 0 A 0 A 0 A 0 A 0 A 0 A 0 A 0           
\end{verbatim}

    \subsubsection{NEKIM with long-range correlated initial \\
    conditions} \label{nekimlong}

\begin{figure}
\epsfxsize=75mm
\epsffile{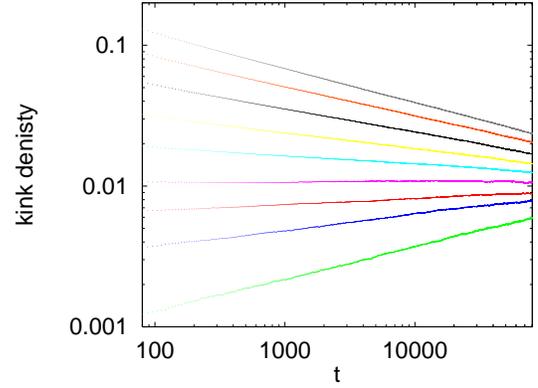}
\caption{$\log(\rho_k(t))$ versus $\log(t)$ in NEKIM simulations for 
$\sigma=0, 0.1, 0.2 ...,1$ initial conditions (from bottom to top curves)
\cite{MeOdof}.}
\label{rholri}
\end{figure}

The effect of initially long-range correlations has already been discussed
in 2d Ising models (Sect.\ref{IsinglongS}) and in 1d bond percolating 
systems (Sect.\ref{long}). In both cases continuously changing decay exponents
have been found.
In case of the NEKIM model (see Sect.~\ref{nekim}) simulations 
\cite{Geza,MeOdof} showed that the density of kinks $\rho_k(t)$ changes as
\begin{equation}
\rho_k(t)\propto t^{\kappa(\sigma)}
\end{equation}
by starting the system with two-point correlated kink distributions of the
form (\ref{TwoPointCorrelations}). The $\kappa(\sigma)$ changes linearly between
the two extremes $\beta/\nu_{||} = \pm 0.285$ shown in Fig.~\ref{rholri}.
This behavior is similar to that of the DP model case (Sect. \ref{long}),
but here one can observe a symmetry:
\begin{equation}
\sigma\leftrightarrow 1-\sigma, \qquad \kappa \leftrightarrow -\kappa
\end{equation}
which is related to the duality symmetry of the NEKIM model (\ref{duality}).

  \subsubsection{GDK with spatial boundary conditions} \label{sectspcbon}

The surface critical behavior of the PC class has been explored through the
study of the GDK model (Sect. \ref{GDKmod}) \cite{LFH98,HFL00,FHL01}. 
The basic idea is that on the surface one may include not only the usual 
BARW2 reactions (\ref{BARW2proc}) but potentially also a parity symmetry 
breaking $A\to\emptyset$ reaction. Depending on whether or not the 
$A\to\emptyset$ reaction is actually present, we may then expect different 
boundary universality classes. Since the time reversal symmetry 
(\ref{DPsymeq}) is broken for BARW2 processes two independent exponents 
($\beta_{1,\rm seed}$, $\beta_{1,\rm dens}$) characterize the
surface critical behavior.

\begin{figure}
\begin{center}
\leavevmode
\vbox{
\epsfxsize=40mm
\epsffile{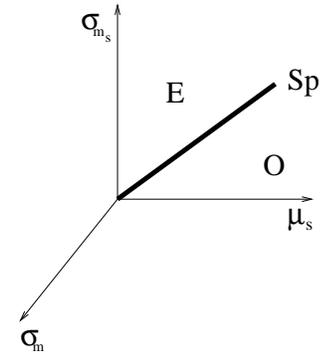}
}
\end{center}
\caption{Schematic mean field boundary phase diagram for BARW from 
\cite{FHL01}. See text for an explanation of the labeling.}
\label{psurf}
\end{figure}

The surface phase diagram for the mean field theory of BARW (valid for
$d>d_c=2$) is shown in Fig.~\ref{psurf}. Here $\sigma_m$,
$\sigma_{m_s}$ are the rates for the branching processes $A\to (m+1)A$
in the bulk and at the surface, respectively, and $\mu_s$ is the rate for 
the surface spontaneous annihilation reaction $A\to\emptyset$. Otherwise, the
labeling is the same as that for the DP phase diagram (see Figure~\ref{dpps}).
The $\mu_s>0$ corresponds to the parity symmetry breaking RBC.  

\begin{figure}
\begin{center}
\leavevmode
\vbox{
\epsfxsize=70mm
\epsffile{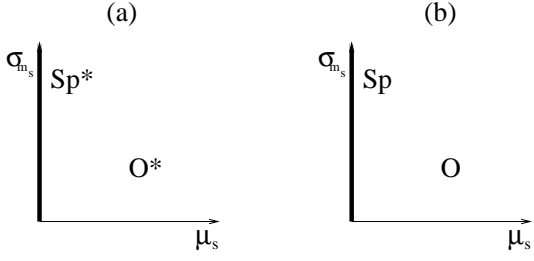}
}
\end{center}
\caption{Schematic surface phase diagrams for BARW in $d=1$ for (a)
$\sigma_m<\sigma_{m,{\rm critical}}$, and (b) $\sigma_m=\sigma_{m,{\rm
critical}}$ \cite{FHL01}. See text for an explanation of the labeling.}
\label{psurf1d}
\end{figure}

For $\mu_s=0$ (IBC) parity conserving case the surface action is of 
the form
\begin{equation}
\label{IBCactionfull}
S_{\rm s}= \int d^{d-1}x_{\parallel} \int_0^{\tau} dt ~ 
\sum_{l=1}^{m/2}\sigma_{2l_{s}}\left(1-\psi_s^{2l}
\psi_s\phi_s \right) ,
\end{equation}
where $\psi_s=\psi({\bf x_{\parallel}},x_{\perp}=0,t)$ and
$\phi_s=\phi({\bf x_{\parallel}},x_{\perp}=0,t)$.
In $d=1$ the boundary and bulk transitions inaccessible to controlled 
perturbative expansions, but scaling analysis shows that
surface branching is {\it irrelevant} leading to the Sp* and Sp
special transitions.
For the $\mu_s>0$ (RBC) parity symmetry breaking case the surface action is
 \begin{eqnarray}
S_2= \int d^{d-1}x_{\parallel} \int_0^{\tau} dt ~ [
     \sum_{l=1}^m  ~ \sigma_{l_s} (1-\psi^l_s) \psi_s\phi_s \nonumber \\
     + \mu_s(\psi_s-1) \phi_s ] .
                                                \label{RBCgenaction}
\end{eqnarray}
and the RG procedure shows that the stable fixed point corresponds to
the ordinary transition. Therefore in 1d the phase diagram looks very 
differently from the mean-field case (Fig. \ref{psurf1d}). 
One can differentiate two cases corresponding to (a) the annihilation fixed
point of the bulk and (b) the PC critical point of the bulk.
As on can see in both cases the ordinary transition ($O$,$O^*$) 
corresponds to $\mu_s>0$, RBC and the special transitions ($Sp$,$Sp^*$)
to $\mu_s=0$, IBC. The ABC condition obviously behaves as if there 
existed a surface reaction equivalent to $\emptyset\to A$, and
thus it belongs to the normal transition universality class.
By scaling considerations the following scaling relations can be 
derived:
\begin{equation}
\tau_1  = \nu_{||} - \beta_{1,\rm dens} \ \ , 
\end{equation}
\begin{equation}
        \nu_{\parallel}+d\nu_{\perp}=
        \beta_{1, \rm seed}+\beta_{\rm dens}+\gamma_{1} .
        \label{surf_hyperscaling}
\end{equation}
Ref. \cite{HFL00} showed that on the self-dual line of the
1d BARWe model (see Sect. \ref{nekim} and \cite{Mussa}) the scaling
relations between exponents of ordinary and special transitions
\begin{equation}
        \beta^{\rm O}_{1,{\rm seed}}=\beta^{\rm Sp}_{1,{\rm dens}}
                                \label{duality1}
\end{equation}
and
\begin{equation}
\beta^{\rm Sp}_{1,{\rm seed}}=\beta^{\rm O}_{1,{\rm
dens}} . \label{duality2}
\end{equation}
hold. Relying on universality they claim that they should be valid 
elsewhere close to the transition line. Numerical simulations 
support this hypothesis as shown in Table \ref{table-exp1}.
\begin{table}[htbp]
\centerline{
\begin{tabular}{|l|l|l|l|l|}
\hline
              & \makebox[16mm]{$1d$ (IBC)}
              & \makebox[16mm]{$1d$ (RBC)} 
              & \makebox[16mm]{$2d$ (O)} 
              & \makebox[16mm]{$2d$ (Sp)} \\
\hline
$\beta_{1,\rm seed}$  &  ~2.06(2)  & ~1.37(2) & 0   &  0 \\
\hline
$\beta_{1, \rm dens}$ &  ~1.34(2)  & ~2.04(2) & 3/2 &  1 \\
\hline
$\tau_1$              &  ~1.16(4)  & ~1.85(4) & 1   &  1  \\
\hline
$\gamma_1$            &  ~2.08(4)  & ~2.77(4) & 1/2 & 1/2 \\
\hline
\end{tabular}}
\caption{Critical boundary exponents of the PC class in $d=1,2$ for
ordinary and special cases.}
\label{table-exp1}
\end{table}

\subsection{Branching with $kA\to\emptyset$ annihilation} \label{kA0}

In Sect.~\ref{2A0} the $2A\to\emptyset$ annihilating random walk (ARW)
has already been introduced. 
By adding branching processes to it we defined the BARWo 
and BARWe models exhibiting continuous phase transitions 
(see Sects.~\ref{BARWo} and~\ref{BARWe}). Now we generalize this
construction to $m$-branching and $kA\to\emptyset$ annihilation type of
models (BkARW) formulating it by the field theoretical action:
\begin{eqnarray}
S = \int d^d x dt [ \psi(\partial_t - D \nabla^2)\phi
- \lambda (1-\psi^k)\phi^k \nonumber \\ 
+ \sigma(1-\psi^m) \psi \phi ]  \ .
\end{eqnarray}
These systems for $k>2$ exhibit an upper critical dimension: $d_c=2/(k-1)$ 
\cite{Lee,Cardy-Tauber2} with the mean-field exponents
\begin{equation}
\alpha = \beta = 1/(k-1) , \ \ \ Z = 2, \ \ \ \nu_{\perp}=1/2 \ .
\end{equation}
At $d_c$ (which falls below physical dimensions for $k>3$) the decay 
has logarithmic corrections:
\begin{equation}
\rho(t) = A_k (ln(t)/t)^{1/(k-1)} \ \ \ .
\end{equation}
So for the $AAA\to \emptyset$ process in one dimension this gives the 
decay behavior \cite{Lee}:
\begin{equation}
\rho(t) = (\frac{1}{4\pi\sqrt 3 D})^{\frac12} (\ln(t)/t)^{\frac12} 
+ O(t^{\frac12}) \  \ ,
\end{equation}
which is the dominant behavior of the 1d bosonic PCPD model at the
transition point (see Sect.\ref{binsp}).
Note that for $k=3$, $m=1,2$ the field theory of \cite{Cardy-Tauber2}
predicts DP class transitions in $d=1$ owing to the BARWo terms generated
by the renormalization.

\subsection{General $nA\to (n+k)A$, $mA\to (m-l)A$ processes} 
\label{pkarwsect}

For a long time reaction-diffusion models with only single parent branching
have been investigated, although there was and early forgotten
numerical result by \cite{G82Z} claiming a non-DP type of
continuous phase transition in a model where particle production can
occur by the reaction of two parents. Later it turned out than in such 
binary production lattice models, where solitary particles
follow random walk (hence they behave like a coupled system) 
different universal behavior emerges indeed (see Sect. \ref{binsp}).
In this subsection I shall discuss the mean-field classes in models
\begin{equation}
n A \stackrel{\sigma}{\to} (n+k)A, 
\qquad m A \stackrel{\lambda}{\to} (m-l) A, \label{genreactions}
\end{equation}
with $n>1$, $m>1$, $k>0$, $l>0$ and $m-l\ge 0$. In low dimensions
the site restricted and bosonic versions of these models exhibit different
behavior. The field theory of the $n=m=l=2$, bosonic model was investigated
in \cite{HT97} and concluded non-DP type of criticality with $d_c=2$
(see Sect. \ref{binsp}). 
For other cases no rigorous field theoretical treatment exist.
Numerical simulations in one and two dimensions for various $n=3$ and $n=4$ 
models resulted in somewhat contradictory results
\cite{PHK02,KC0208497,tripcikk}. There is a disagreement in the value
of the upper critical dimension, but in any case $d_c$ seems to be very
low ($d_c=1-2$) hence the number of non-mean-field classes in such models
is limited. 
On contrary there is a series of mean-field classes {\bf depending on
$n$ and $m$}.

\subsubsection{The $n=m$ symmetric case}

In such models there is a continuous phase transition at finite
production probability: $\sigma_c=\frac{l}{k+l}$ characterized by
the order parameter exponents \cite{PHK02,tripcikk}:
\begin{equation}
\beta^{MF}=1, \qquad \alpha_{MF}=1/n \ .
\end{equation}
These classes generalize the mean-field class of the DP (Sect.\ref{DPS})
and binary production models (Sect.\ref{binsp}). For referencing it
in Table \ref{MFtab} of the Summary I call it as:
PARWs (symmetric production and m-annihilating random walk class).

\subsubsection{The $n>m$ symmetric case}

In this case the mean-field solution provides a first order
transition (see \cite{tripcikk}), hence it does not
imply anything with respect to possible classes for models below
the critical dimension ($d<d_c$). Note however, that by higher order
cluster mean-field approximations, when the diffusion plays
a role the transition may turn into a continuous one
(see for example \cite{boccikk,OdSzo,meorcikk}).

\subsubsection{The $n<m$ case}

In this case the critical point is at zero production probability 
$\sigma_c=0$, where the density decays as $\alpha^{MF}=1/(m-1)$ 
as in case of the $n=1$ branching and $m=l$ annihilating models 
(BkARW classes see Sect. \ref{kA0}), but the steady state exponent
is different: $\beta^{MF}=1/(m-n)$,
defining an other series of mean-field classes (PARWa) \cite{tripcikk}.
Again cluster mean-field approximations may predict the appearance
of other $\sigma>0$ transitions with different critical behavior.

\section{Universality classes of multi-component systems} \label{multichap}

First I recall some well known results (see refs. in \cite{Priv})
for multi-component reaction-diffusion systems without particle 
creation. From the viewpoint of phase transitions these describe the 
behavior in the inactive phase or in case of some N-component BARW models 
right at the critical point. Then I show the effect of particle exclusions 
in 1d. Later universal behavior of more complex, coupled multi-component 
systems are discussed. This field is quite new and some of the results
are still under debate.

      \subsection{The $A+B\to\emptyset$ classes} \label{AB}

The simplest two-component reaction-diffusion model involves two types of
particles undergoing diffusive random walks and reacting upon contact to form
an inert particle. The action of this model is:
\begin{eqnarray}
S &=& \int d^dx dt [ \psi_A(\partial_t  - D_A \nabla^2)\phi_A
+ \psi_B(\partial_t - D_B \nabla^2)\phi_B  \nonumber \\
 &-& \lambda( 1-\psi_A \psi_B ) \phi_A\phi_B ] 
- \rho(0)(\psi_A(0)+\psi_B(0)) 
\end{eqnarray}
where $D_A$ and $D_B$ denotes the diffusion constants of species $A$ and $B$.
In $d < d_c = 4$ dimensions and for homogeneous, initially equal density of 
$A$ and $B$ particles ($\rho_0$) the density decays asymptotically as 
\cite{old,BO78,TW83,KR85,BL88,LC95}
\begin{equation} \label{ABlaw}
\rho_A(t)=\rho_B(t) \propto C\sqrt{\Delta} t^{-d/4} \ , 
\end{equation}
where $\Delta = \rho(0)-C'\rho^{d/2}(0) +...$, $C$ is a universal $C'$ is a
non-universal constant. This slow decay behavior is due to the fact that in 
the course of the reaction, local fluctuations in the initial distribution 
of reactants lead to the formation of clusters of like particles that do 
not react and they will be asymptotically segregated for $d<4$. 
The asymptotically dominant process is the diffusive decay of the
fluctuations of the initial conditions. Since this is a short ranged process
the system has a long-time memory -- appearing in the amplitude 
dependence -- for the initial density $\rho(0)$.
For $d\le 2$ controlled RG calculation is not possible, but the result
(\ref{ABlaw}) gives the leading order term in $\epsilon=2-d$ expansion.
For $D_A\ne D_B$ case a RG study \cite{LC95} found new amplitude but the
same exponents. 

The {\bf persistence} behavior in 1d with homogeneous, equal initial density 
of particles ($\rho_0 = \rho_A(0)+\rho_B(0)$) was studied by \cite{DB02}.
The probability $p(t)$, that an annihilation process has not
occurred at a given site (``type I persistence'') has the asymptotic form 
\begin{equation}
p(t) \sim {\rm const. } + t^{-\theta_l} \ .
\end{equation}
For a density of particles $\rho >> 1$, $\theta_l$ is identical to that 
governing the persistence properties of the one-dimensional diffusion 
equation, where $\theta_l \approx 0.1207$. In the case of an initially low 
density, $\rho_0 << 1$, $\theta_l \approx 1/4$ was found asymptotically. 
The probability that a site remains unvisited by any random walker 
(``type II persistence'') decays in a stretched exponential way
\begin{equation}
p(t) \sim \exp(-{\rm const. \times } \rho_0^{1/2}t^{1/4})
\end{equation}
provided $\rho_0 << 1$.

   \subsection{$AA\to\emptyset$, $BB\to\emptyset$ with hard-core repulsion}
    \label{2}

The next simplest two-component model in which particle blocking may
be effective in low dimensions was investigated first in the context
of a stochastic cellular automaton model.
At the symmetric point of the GDK model (see Sect. \ref{GDKmod})
compact domains of $I1$ and $I2$ grow separated by $A =Ac-I1$ and 
$B = Ac-I2$ kinks that cannot penetrate each other.
In particle language this system is a reaction-diffusion model of two types
$A+A\to\emptyset$, $B+B\to\emptyset$ with exclusion 
$AB\not\leftrightarrow BA$ and special {\bf pairwise initial conditions} 
(because the domains are bounded by kinks of the same type):
\begin{verbatim}
     ....A...A...B.B..B.....B..A..A..
\end{verbatim}
In the case of homogeneous, pairwise initial conditions simulations 
\cite{gdkcikk} showed a density decay of kinks $\rho\propto t^{-\alpha}$ 
characterized by a power-law with an exponent somewhat larger than 
$\alpha=0.5$.
The $\alpha=1/2$ would have been expected in case of two copies of ARW 
systems which do not exclude each other. Furthermore the deviation of 
$\alpha$ from $1/2$ showed an initial density dependence. 
Ref. \cite{gdkcikk} provided a possible explanation based on permutation
symmetry between types, according to which hard-core interactions 
cause {\bf marginal perturbation} resulting in non-universal scaling.
The situation is similar to that of the compact directed percolation
that is confined by parabolic boundary conditions 
(see Sect.\ref{sectscbon}) if we assume that $AB$ and $BA$ pairs
exert parabolic space-time confinement on coarsening domains. 
Non-universal scaling can also be observed at surface critical phenomena. 
Similarly here $AB$, $BA$ pairs produce 'multi-surfaces' in the bulk. 
However simulations and independent interval approximations 
in a similar model predict logarithmic corrections to the  
single component decay with the form
$\rho\sim t^{-1/2}/\ln (t)$ \cite{Satya}. Note that both kind of 
behavior may occur in case of marginal perturbations.
Cluster simulations \cite{gdkcikk} also showed initial density 
($\rho_{I1}(0)$) dependent survival probability of $I2$-s in the sea 
of $I1$-s:
\begin{equation}
P_{I2}(t)\propto t^{-\delta(\rho_{I1}(0))} \ .
\end{equation}

However this reaction-diffusion model with homogeneous, {\bf random initial 
distribution} of $A$-s and $B$-s exhibits a much slower density decay.
An exact duality mapping helps to understand the coarsening 
behavior. Consider the leftmost particle, which may be either $A$ or $B$, and 
arbitrarily relabel it as a particle of species $X$.  For the second particle, 
we relabel it as $Y$ if it is the same species as the initial particle; 
otherwise we relabel the second particle as $X$.  We continue to relabel 
each subsequent particle according to this prescription until all 
particles are relabeled from $\{A,B\}$ to $\{X,Y\}$. For example, the string
\begin{eqnarray*}
AABABBBA\cdots
\end{eqnarray*} 
translates to 
\begin{eqnarray*}
XYYYYXYY\cdots.
\end{eqnarray*}  
The diffusion of the original $A$ and $B$ particles at equal rates 
translates into diffusion of the $X$ and $Y$ particles.
Furthermore, the parallel single-species reactions, $A+A\to\emptyset$ 
and $B+B\to \emptyset$, translate directly to two-species annihilation 
$X+Y\to \emptyset$ (see Sect. \ref{AB}) in the dual system. 
The interesting point is that
in the $X+Y\to\emptyset$ model blockades do not exist, because
$XY$ pairs annihilate, and there is no blockade between $XX$ and $YY$ pairs. 
Therefore the density decay should be proportional to $t^{-1/4}$.
Simulations confirmed this for the $A+A\to\emptyset$, $B+B\to\emptyset$ 
\cite{barw2cikk} model, nevertheless corrections to scaling were also observed.
The pairwise initial condition transforms in the dual system to domains 
of $..XYXY..$ separated by $YY$ and $XX$ pairs, which do not allow $X$ and 
$Y$ particles to escape each other.

      \subsection{Multi-species $A_i+A_j\to\emptyset$ classes} \label{MAB}

By generalizing the diffusion-limited reactions $AB\to\emptyset$ 
(Sect. \ref{AB}) for $q>2$ species
\begin{equation}
A_i + A_j \to \emptyset
\end{equation}
in $d\ge 2$ dimensions the asymptotic density decay for such mutual 
annihilation processes with equal rates and initial densities is the same as 
for single-species pair annihilation $AA\to\emptyset$. 

In $d = 1$, however, particles of different types can not pass each other
and a segregation occurs for all $q<\infty$. 
The total density decays according to a $q$ dependent power law, 
$\rho\propto t^{-\alpha(q)}$ with
\begin{equation}
\alpha=(q-1)/2q
\end{equation}
exactly \cite{DHT02L}.
These findings were also supported through Monte Carlo simulations.
Special initial conditions such as $...ABCDABCD...$ prevent the 
segregation and lead to decay of the $2A\to\emptyset$ model (Sect.\ref{2A0}).

   \subsection{Unidirectionally coupled ARW classes} \label{uniARW}

Besides the symmetrically coupled ARW systems discussed above now I
turn to unidirectionally coupled ARW (Sect.\ref{2A0}) models:
\begin{eqnarray}
A+A &\rightarrow\emptyset \qquad & A \rightarrow A+B \nonumber \\
B+B &\rightarrow\emptyset \qquad & B \rightarrow B+C \nonumber \\
C+C &\rightarrow\emptyset \qquad & C \rightarrow C+D \nonumber \\
\rm{...}
\end{eqnarray}
introduced and analyzed with RG technique and simulations by 
\cite{Goldschmidt98}.
This kind of coupling was chosen because $A\to B$ would constitute a
spontaneous death process of $A$ particles leading to exponential
density decay. On the other hand quadratic coupling of the form
$A+A\to B+B$ leads to asymptotically decoupled systems \cite{HT97}.  
The mean-field theory is described by the rate equation for the
density $\rho_i(x,t)$ at level $i$:
\begin{equation}
\label{annmfrB}
{\partial \rho_i(x,t) \over \partial t} = D \nabla^2 \rho_i(x,t) 
- 2\lambda_i \rho_i(x,t)^2 + \sigma_{i,i-1} \rho_{i-1}(x,t) \ ,
\end{equation}
which relates the long time behavior of level $i$ to level $i-1$:
\begin{equation}
n_i(t) \propto n_{i-1}^{1/2} \ .
\end{equation}
By inserting into this the exact solution for ARW (Sect.\ref{2A0})
one gets:
\begin{equation}
\label{cpaden}
\rho_i(t) \sim \left\{ \begin{array}{ll} t^{-d/2^i} &{\rm for \ } d < 2 \
, \\ \left( t^{-1} \ln t \right)^{1/2^{i-1}} &{\rm for \ } d = d_c = 2
\ , \\ t^{-1/2^{i-1}} &{\rm for \ } d > 2 \ , \end{array} \right.
\end{equation}
The action of a two-component system with fields $a$,$\hat a$, $b$, 
$\hat b$ for equal annihilation rates ($\lambda$) 
takes the form
\begin{eqnarray}
\label{cpafth}
S &&= \int d^dx \int dt \Bigl[ {\hat a} ( \partial_t - D \nabla^2 ) a
- \lambda ( 1 - {\hat a}^2 ) a^2 +  \nonumber \\ 
&&+ {\hat b} ( \partial_t - D
\nabla^2 ) b - \lambda ( 1 - {\hat b}^2 ) b^2 + \sigma (1 - {\hat b}) 
{\hat a} a \Bigr] \ . 
\end{eqnarray}
The RG solution is plagued by IR-divergent diagrams similarly to UCDP
(see Sect.\ref{uniDP}) that can be interpreted as eventual 
{\em non-universal} crossover to the decoupled regime. Simulation results
-- exhibiting finite particle numbers and coupling strengths -- really 
show the breakdown of scaling (\ref{cpaden}), but the asymptotic behavior
could not be determined. Therefore the results (\ref{cpaden}) are valid for
an intermediate time region.

\subsection{DP coupled to frozen field classes} \label{PCPsect}

One of the first generalizations of absorbing phase transition models
were such systems, which exhibit many absorbing states, hence
the conditions of DP hypothesis \cite{DPuni,DPuni2} (Sect. \ref{genchap})
are not satisfied. 
Several variants of models with infinitely many absorbing states
containing frozen particle configurations have been introduced.
The common behavior of these models that non-diffusive (slave) 
particles are coupled to a DP-like (order parameter) process. 
In case of homogeneous, uncorrelated initial conditions
DP class exponents were found, on the other hand in cluster 
simulations -- that involve correlated initial state of the 
order parameter particles -- initial density dependent scaling 
exponents ($\eta$ and $\delta$) arise. 
These cluster exponents take the DP class values only if
the initial density of the slave particles agrees with the 
``natural density'' that occurs in the steady state.  
The first such models introduced by \cite{PCP,PCP2} 
were the so called pair contact process (PCP) (see Sect. \ref{PCPmod}) 
and dimer the reaction model. These systems seem to be single component
ones by defining rules for the pairs, but the isolated, frozen particles 
behave as a second component. 
In the threshold transfer process (TTP) \cite{TTP} the two components 
defined as: the '$2$'-s following DP process and the '1'-s which decay 
or reappear as: 
$\emptyset\stackrel{r}{\to}1\stackrel{1-r}{\longrightarrow}\emptyset$.

For the PCP model defined by the simple processes (\ref{PCPdef})
a set of coupled Langevin equations were set up \cite{MDG,MDGL}
for the fields $n_1( {\bf x}, t )$ and $n_2( {\bf x}, t )$ (order
parameter):
\begin{eqnarray}
& {\displaystyle  \partial n_2 \over\strut\displaystyle 
  \partial t} =
 [r_{2} + D_{2} \nabla^2 - u_{2}  n_2 -w_{2}  n_1]
  n_2 + \sqrt{n_2} \eta_2 \nonumber \\
 & {\displaystyle \partial n_1 \over\strut\displaystyle 
 \partial t} =
 [r_{1} + D_{1} \nabla^2 - u_{1}  n_2 -w_{1}  n_1]
  + \sqrt{n_2} \eta_1 \label{Lan}
\end{eqnarray}
where $D_i, r_i, u_i$, and $w_i$ are constants and 
$\eta_1( {\bf x}, t )$ and $\eta_2( {\bf x}, t )$ are Gaussian white noises.
Owing to the multiple absorbing states and the lack of the time reversal
symmetry (\ref{DPsymeq}) a generalized hyperscaling law (\ref{ghypers}) 
has been derived by \cite{TTP}.
As discussed in \cite{MDG} this set of equations can be simplified
by dropping the $D_1$, $u_1$, and noise terms in the $n_1$ equation,
and then solving that equation for $n_1$ in terms of $n_2$.
Substituting in the $n_2$ equation, one obtains
\begin{eqnarray}
{\displaystyle\partial n_2({\bf x},t)\over\strut\displaystyle\partial t}
&=& D_2 \nabla^2 n_2 ( {\bf x}, t )+m_2 n_2 ( {\bf x}, t )
-u_2 n_2 ^ 2( {\bf x}, t ) + \nonumber\\ 
&+& w_2 (r_1 / w_1 - n_1 ({\bf x} ,0) ) n_2 ( {\bf x}, t )e^{-w_1 \int_0^t
n_2 ({\bf x} ,s) ds } \nonumber\\
&+& \sqrt{n_2 ( {\bf x}, t )} \eta_2 ( {\bf x}, t )~,
\label{Lan2}
\end{eqnarray}
where $n_1 ({\bf x} ,0)$ is the initial condition of the $n_1$ field,
and $m_2= r_2-w_2 r_1 / w_1$. The ``natural density''\cite{PCP}, 
$n_1^{nat}$, then corresponds to the uniform density, 
$n_1 (t=0)=r_1 / w_1$, for which the coefficient of the exponential 
term vanishes, and we get back the Langevin equation
of DP (\ref{DPLangeq}). This derivation provides a simple explanation 
for the numerical observation of DP exponents in case of natural initial 
conditions. However it does not take into account the long-time memory 
and the fluctuations of passive particles (with power-law time 
and $p$ dependences \cite{OMSM98}). Therefore, some of the terms omitted in
this derivation (as for instance the term proportional to $n_2^2$ in the 
equation for $n_1$) cannot be safely eliminated \cite{Porto2} and this 
simplified theory does not generate critical fluctuations
for its background field. 
To overcome these problems a more rigorous
field theoretical analysis involving path integral representation
of a two-component variant of this class (with '$A$' activity and 
'$B$' slave):
\begin{equation}
A\to 2A, \quad A\to\emptyset, \quad A\to B
\end{equation}
has been done \cite{MSS02} and provided some evidence that the homogeneous 
state critical properties of the activity field are DP like irrespectively of 
the criticality of the slave field. For this model \cite{MSS02} 
could also prove the numerical result \cite{OMSM98} that the slave field 
decays as
\begin{equation}
\label{rhoBdec}
\rho_B(t) - \rho_B^{nat}  \propto t^{-\alpha_{DP}} \ .
\end{equation}
However this treatment has still not provided theoretical proof for the 
initial density dependent spreading exponents observed in simulations 
\cite{PCP,PCP2,TTP,OMSM98} and by the numerical integration of the
Langevin equation \cite{LM97}. Furthermore the situation is much more 
complicated when approaching criticality from the {\bf inactive phase}.
In particular, the scaling behavior of $n_A$ in this case seems to be 
unrelated to $n_B$ (this is similar to the diffusive slave field case
\cite{OSC02} Sect. \ref{apsect}). In this case it is more difficult
to analyze the field theory and dynamical percolation type of terms are 
generated that can be observed in 2d by simulations and by mean-field
analysis \cite{MDG,MDGL,WL02}. Very recently it was claimed, based on the
field theoretical analysis of the GEP model (Sect.\ref{dynperc}) -- that
exhibits similar long-time memory terms -- that the cluster variables
should follow stretched exponential decay behavior \cite{AH03}.

In {\bf two dimensions} the critical point of spreading
($p_s$) moves (as the function of initial conditions) and do not necessarily
coincide with the bulk critical point ($p_c$). The spreading behavior depends
on the coefficient of the exponential, non-Markovian term of (\ref{Lan2}). 
For positive coefficient the $p_s$ falls in the inactive
phase of the bulk and the spreading follows {\bf dynamical percolation}
(see Sect. \ref{dynperc}). For negative coefficient the $p_s$ falls in the
active phase of the bulk and spreading exponents are {\bf non-universal}
(like in 1d) but satisfy the hyper-scaling (\ref{ghypers}).

Simulations and GMF analysis \cite{Porto2,MSS02} in 
the {\bf inactive phase} in 1d showed, that the steady state density of slave 
particles approaches the natural value by a power-law
\begin{equation} 
\label{rho1scal}
|\rho_1^{nat} - \rho_1| \propto |p-p_c|^{\beta_1} \ ,
\end{equation}
with $\beta_1\sim 0.9$ for PCPD and $\beta_1=1$ for TTP transfer models.
This difference might be the consequence that in TTP models the slave
field fluctuates and relaxes to $r$ quickly, while in the PCP case it is
frozen. Due to this fact one might expect initial condition dependence
here again.

        \subsubsection{The pair contact process model} \label{PCPmod}

\begin{figure}
\epsfxsize=70mm
\epsffile{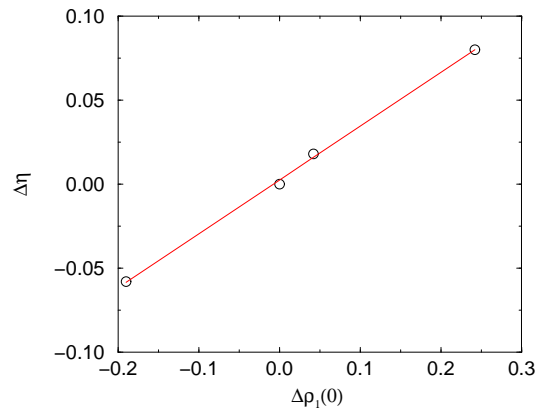}
\caption{Initial concentration dependence of the exponent $\eta$ for
PCP model \cite{OMSM98}. Linear regression gives a slope $0.320(7)$ 
between $\eta-\eta_{DP}$ and $\rho_1(0)-\rho_1^{nat}$}
\label{pcp_res}
\end{figure}

Up to now I discussed spreading processes with unary particle production. 
Now I introduce a family of systems with {\bf binary} particle production 
(i.e for a new particle production two particles are needed to collide).
The PCP model is defined on the lattice by the following processes:
\begin{equation}
\label{PCPdef}
2A\stackrel{1-p}{\longrightarrow}3A, \quad 2A\stackrel{p}{\to}\emptyset \ ,
\end{equation}
such that reactions take place at nearest-neighbor (NN) sites and we
allow single particle occupancy at most. The order parameter is the
density of NN pairs $\rho_2$.
The PCP exhibits an active phase for $p < p_c$; 
for $p \geq p_c$ the system eventually falls into an absorbing 
configuration devoid of NN pairs,
(but that contains a density $\rho_1$ of isolated particles). The best
estimate for the critical parameter {\bf in one dimension}
is $p_c \!=\! 0.077090(5)$ \cite{rdjaff}.
Static and dynamic exponents corresponding to initially uncorrelated
homogeneous state agree well with those of 1+1 d DP (\ref{DPe}). 
Spreading exponents that involve averaging over all runs, hence involving
the survival probability are non-universal (see Fig. \ref{pcp_res}) 
\cite{OMSM98}.
The anomalous critical spreading of PCP can be traced to a long memory in 
the dynamics of the order parameter $\rho_2$, arising from a coupling 
to an auxiliary field (the local particle density $\rho$), that remains 
frozen in regions where $\rho_2 \!=\! 0$. In \cite{OMSM98} a slight variation
of the spreading critical point (as the function of $\rho_1(0)$) was observed 
(similarly to the 2d case), but more detailed simulations \cite{D99} suggest 
that this can be explained by strong corrections to scaling. 
Simulations provided numerical evidence that
the $\rho_1$ exhibits anomalous scaling as eq.(\ref{rho1scal})
to $\rho_1^{nat}=0.242(1)$ with a DP exponent \cite{OMSM98} for $p<p_c$
and with $\beta_1=0.9(1)$ for $p>p_c$ \cite{Porto2}.
The {\bf DS transition} point and the DS exponents of this model 
coincide with the critical point and the critical exponents of the
PCP \cite{OMSM98}.

The effect of an {\bf external particle source} that creates isolated
particles, hence do not couple to the order parameter was investigated by
simulations and by GMF+CAM approximations \cite{pcph}.
While the critical point $p_c$ shows a singular dependence on the source 
intensity, the critical exponents appear to be unaffected by the presence of 
the source, except possibly for a small change in $\beta$. 

The properties of the {\bf two dimensional} PCP in case of homogeneous,
uncorrelated initial conditions was investigated by simulations \cite{KSD99}. 
In this case all six NN of a pair was considered for reactions (\ref{PCPdef}). 
The critical point is located at $p_c=0.2005(2)$.
By determining $\alpha$, $\beta/\nu_{\perp}$ and $Z$ exponents
and order parameter moment ratios by simulations, the universal
behavior of the 2+1 dimensional DP class (\ref{DPe}) was confirmed. 
The spreading exponents are expected to behave as described in 
Sect. \ref{PCPsect}.

\subsection{DP with coupled diffusive field classes} \label{binsp}

The next question one can pose following Section \ref{PCPsect} is
whether a diffusive field coupled to a DP process is relevant. 
Prominent representatives of such models are binary particle
production systems with explicit diffusion of solitary particles.
The critical behavior of such systems are still under investigations.
The annihilation-fission (AF) process is defined as
\begin{equation}
2A\stackrel{\lambda}{\to}\emptyset, \qquad 
2A\stackrel{\sigma}{\to}(n+2)A, \qquad 
A\emptyset\stackrel{D}{\leftrightarrow}\emptyset A \ .
\label{AFprocdef}
\end{equation}
The corresponding action for bosonic particles was derived from master 
equation by \cite{HT97}
\begin{eqnarray}
S=\int d^d x dt [ \psi(\partial_t-D\nabla^2)\phi 
- \lambda(1-\psi^2)\phi^2 \nonumber \\ 
+ \sigma(1-\psi^n) \psi^2 \phi^2 ] \ .
\end{eqnarray}
Usually bosonic field theories describe well the critical behavior of
``fermionic'' systems (i.e. with maximum one particle per site occupation). 
This is due to the fact that at absorbing phase transitions the occupation
number vanishes. In this case however the active phase of bosonic and 
fermionic models differ significantly: in the bosonic model the
particle density diverges, while in the fermionic model there is a steady 
state with finite density. As a consequence the bosonic field theory cannot 
describe the active phase and the critical behavior of the fermionic 
particle system.

As one can see this theory lacks interaction terms linear in 
the field variable $\phi$ (massless) in contrast with the DP action
(\ref{DPaction}). 
Although the field theory of {\bf bosonic AF} has turned out to be 
non-renornmali\-zible, \cite{HT97} concluded that critical 
behavior cannot be in DP class. 
In fact the upper critical behavior is $d_c=2$ that is different from 
that of DP and PC classes \cite{OSC02}. In the Langevin formulation
\begin{equation}
\frac{\partial\rho(x,t)}{\partial t} = D \nabla^2 \rho(x,t) 
+ (n\sigma-2\lambda)\rho^2(x,t) + \rho(x,t)\eta(x,t)
\label{AFLang}
\end{equation}
the noise is complex:
\begin{eqnarray}
<\eta(x,t)> & = & 0 \\
<\eta(x,t)\eta(x',t')>&=&[n(n+3)\sigma-2\lambda]\delta^d(x-x')(t-t') \nonumber
\ , 
\end{eqnarray}
that is again a new feature (it is real in case of DP and purely imaginary 
in case of CDP and PC classes). 

Eq.(\ref{AFLang}) without noise gives the {\bf MF behavior of the 
bosonic model}: for $n\sigma > 2\lambda$ the density diverges, while for 
$n\sigma < 2\lambda$ it decays with a power-law with $\alpha_b^{MF}=1$.
The MF description in the {\bf inactive phase} of the bosonic model was
found to be valid for the 2d fermionic AF system too \cite{OSC02}. 
Here the pair density decays as $\rho_2(t)\propto t^{-2}$ 
\cite{OSC02} in agreement with the MF approximation. Contrary to this 
for $\lambda\le\lambda_c$ $\rho$ and $\rho_2$ seem to be related by a 
logarithmic ratio $\rho(T)/\rho_2(t)\propto\ln(t)$. 
This behavior could not be described by the mean-field approximations.

The field theory suggests that the critical behavior of the 
{\bf 1d bosonic model} in the inactive phase is dominated by 
the $3A\to\emptyset$ process, that has an upper critical dimension 
$d_c=1$. Therefore in one dimension the particle density decays with
a power-law exponent: $\alpha=1/2$ with logarithmic corrections 
(see Sect. \ref{kA0}).
This behavior has been confirmed by simulations in case of the 1d 
annihilation-fission model \cite{OdMe02}.

The field theoretical description of the {\bf fermionic AF} process 
run into even more serious difficulties \cite{Hunp} than that of the
bosonic model and predicted an upper an critical dimension $d_c=1$ 
that contradicts simulation results \cite{OSC02}.
For the fermionic AF system {\bf mean-field} approximations 
\cite{Carlon,OSC02} give a continuous transition with exponents
\begin{equation}
\beta = 1,  \ \ \ \beta'=0, \ \ \ Z=2, \ \ \ \nu_{||}=2,
 \ \ \ \alpha=1/2, \ \ \ \eta=0 .
\label{afmftab}
\end{equation}
These MF exponents are distinct from those of other well known classes
(DP,PC,VM ...). They were confirmed in a 2d fermionic AF model, with 
logarithmic corrections, indicating $d_c=2$ \cite{OSC02}. 
An explanation for the new type of  critical behavior
based on symmetry arguments are still missing
but numerical simulations suggest \cite{Odo00,HayeDP-ARW}
that the behavior of this system can be described 
(at least for strong diffusion) by coupled sub-systems: 
single particles performing annihilating random walk 
coupled to pairs ($B$) following DP process:
$B\to 2B$, $B\to\emptyset$. The model has two non-symmetric
absorbing states: one is completely empty, in the other 
a single particle walks randomly. 
Owing to this fluctuating absorbing state this model does
not oppose the conditions of the DP hypothesis.
It was conjectured by \cite{HH} that this 
kind of phase transition appears in models where 
(i) solitary particles diffuse, (ii) particle creation requires 
two particles and (iii) particle removal requires at least two 
particles to meet. The exploration of other conditions 
that affect these classes are still under investigation.

      \subsubsection{The PCPD model} \label{pcpdsect}

\begin{figure}
\epsfxsize=70mm
\epsffile{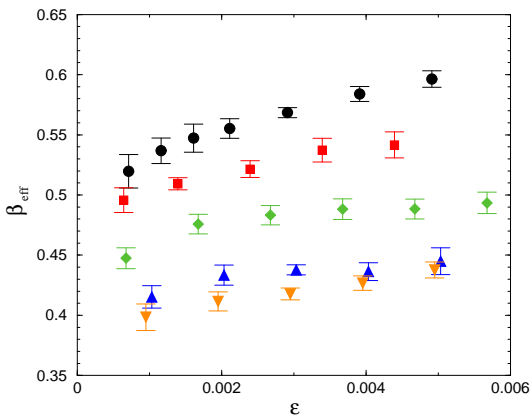}
\caption{Effective $\beta$ exponents for different diffusion rates.
The circles correspond to $D=0.05$, the squares to $D=0.1$ the diamonds
to $D=0.2$, the up-triangles to $D=0.5$ and the down-triangles to $D=0.7$.}
\label{pcpdbeta}
\end{figure}

A PCPD like model was introduced in an early work by \cite{G82Z}.
His preliminary simulations in 1d showed non-DP type transition, but the 
model has been forgotten for a long time.
The PCPD model introduced by \cite{Carlon} is controlled by 
two parameters, namely the probability of pair annihilation $p$ and the 
probability of particle diffusion $D$. The dynamical rules are
\begin{eqnarray}
AA\emptyset,\,\emptyset AA \rightarrow AAA  \qquad {\rm with \ rate}
\, & (1-p)(1-D)/2 \nonumber \\
AA \rightarrow \emptyset\emptyset \qquad  {\rm with \ rate}\,  &
p(1-D) \nonumber \\
A\emptyset \leftrightarrow \emptyset A \qquad {\rm with \ rate}\,  & D \ .
\label{DynamicRules}
\end{eqnarray}
The {\it mean-field} approximation gives a continuous transition at $p=1/3$.
For $p \le p_c(D)$ the particle and pair densities exhibit singular behavior:
\begin{equation}
\rho(\infty)\propto (p_c-p)^{\beta} \qquad
\rho_2(\infty)\propto (p_c-p)^{\beta_2}
\end{equation}
while at $p = p_c(D)$ they decay as:
\begin{equation}
\rho(t) \propto t^{-\alpha} \ , \ \ \ \rho_2(t) \propto t^{-\alpha_2} \ \ \ ,
\end{equation}
\begin{figure}
\epsfxsize=70mm
\epsffile{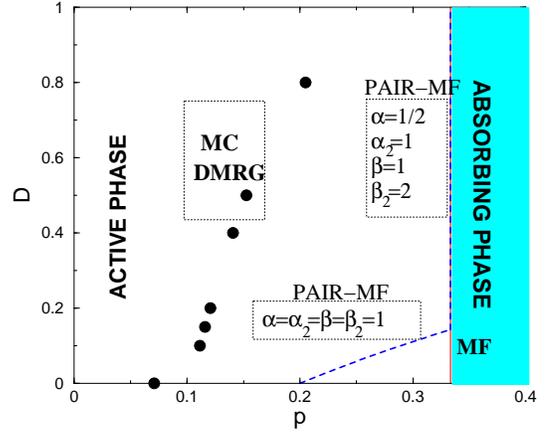}
\caption{Schematic phase diagram of the 1d PCPD model. 
Circles correspond to simulation and DMRG results, solid line
to mean-field, dashed line to pair-approximation.}
\label{pcpd_pd}
\end{figure}
with the exponents:
\begin{equation}
\alpha=1/2, \ \ \ \alpha_2=1, \ \ \ \beta=1, \ \ \ \beta_2=2 \ .
\end{equation}
According to {\it pair mean-field} approximations the phase diagram can be 
separated into two regions (see Fig.\ref{pcpd_pd}).
While for $D>1/7$ the pair approximation gives the same $p_c(D)$ and 
exponents as the simple MF, 
for $D<1/7$-s the transition line breaks and the exponents are different
\begin{equation}
\alpha=1, \ \ \ \alpha_2=1, \ \ \ \beta=1, \ \ \ \beta_2=1 \ .
\end{equation} 
In the entire inactive phase the decay is characterized by the exponents:
\begin{equation}
\alpha=1, \ \ \ \alpha_2=2 \ .
\end{equation}
The DMRG \cite{Carlon} method and simulations of the 1d PCPD model 
\cite{Hayepcpd} resulted in agreeing $p_c(D)$ values but for the critical 
exponents no clear picture was found. They could not clarify if the 
two distinct universality suggested by the pair mean-field approximations
was really observable in the 1d PCPD model. It is still a debated topic 
whether one new class, two new classes or continuously changing exponents 
occur in 1d.
Since the model has two absorbing states (besides the vacuum state there is 
another one with a single wandering particle) 
and some exponents were found to be close to those of the PC class
($Z=1.6-1.87$, $\beta/\nu_{\perp}=0.47-0.51$) \cite{Carlon} suspected that 
the transition (at least for low-$D$ values) is PC type.
However the lack of $Z_2$ symmetry, parity conservation and further
numerical data \cite{Hayepcpd,Odo00} exclude this possibility. 
Note, that the MF exponents are also different from those of the PC class.
Simulations and CAM calculations for the one dimensional $n=1$ AF model 
\cite{Odo00,pcpd2cikk} corroborated the two new universality class 
prospect (see Fig.\ref{pcpdbeta} and Table \ref{tabpcpd}).
The order parameter exponent ($\beta$) seems to be very far from 
both of the DP and PC class values \cite{Odo00,pcpd2cikk}.
\begin{table}
\begin{center}
\begin{tabular}{|l|r|r|r|r|r|}
\hline
$D$          &  0.05 & 0.1   & 0.2      &  0.5 & 0.9\\
\hline
$p_c$        &0.25078&0.24889&0.24802&0.27955&0.4324\\
$\beta_{CAM}$&  -    &0.58(6)& 0.58(2)  &0.42(4)& -   \\
$\beta$      &0.57(2)&0.58(1)& 0.58(1)  & 0.40(2)&0.39(2)\\
$\delta$     &0.273(2)&0.275(4)&0.268(2)&0.21(1)&0.20(1)\\
$\eta$       &0.10(2) &   -    & 0.14(1)  &0.23(2)&0.48(1)\\
$\delta\prime$&0.004(6)&   -    &0.004(6)  &0.008(9)&0.01(1) \\
\hline
\end{tabular}
\caption{Summary of results for 1d, $n=1$ AF model. The non-universal
critical parameter $p_c$ of the parallel model is shown here.}
\label{tabpcpd}
\end{center}
\end{table}

The two distinct class behavior may be explained on the basis of competing
diffusion strengths of particles and pairs (i.e. for large $D$-s the
explicit diffusion of lonely particles is stronger). Similar behavior 
was observed in case of 1d models with coupled (conserved) diffusive field 
(see Sect. \ref{DCF}). However a full agreement has not been achieved
in the literature with respect the precise values of the critical exponents. 
The low-$D$ $\alpha$ is supported by \cite{PK66} who considered 
the case with coagulation and annihilation rates three times the diffusion 
rate. On the other hand the high-$D$ $\alpha$ of Table \ref{tabpcpd} coincides
with that of \cite{KC0208497}, who claim a single value for $0<D<1$.
By assuming logarithmic corrections it was shown \cite{pcpd2cikk}
that a single universality class can be supported indeed with exponents
\begin{eqnarray}
\alpha=0.21(1), \quad \beta=0.40(1), \nonumber \\ \quad Z=1.75(15), \quad 
\beta/\nu_{\perp}=0.38(1) \ ,
\end{eqnarray}
however there is no strong evidence for such corrections.
Although the upper critical dimension is expected to be at $d_c=2$
\cite{OSC02} one may not exclude the possibility of a second critical
dimension ($d_c'=1$) or topological effects in 1d that may cause logarithmic
corrections to scaling.
The spreading exponent $\eta$ seems to change
continuously by varying $D$. Whether this is true asymptotically or
the effect of some huge correction to scaling is still not clear.
The simulations of \cite{Odo00} confirmed that it is irrelevant whether 
the particles production is spatially symmetric: $A\emptyset A\to AAA$ or 
spatially asymmetric: $AA\emptyset\to AAA$, $\emptyset AA\to AAA$.
Recent simulations and higher level GMF approximations suggest 
\cite{OSC02,pcpd2cikk} that the peculiarities of the pair approximation 
are not real; for $N>2$ cluster approximations the low-$D$ region scaling 
disappears. Recently two studies \cite{DM0207720,H0208345} reported 
non-universality in the dynamical behavior
of the PCPD. While the former one by Dickman and Menezes explored different
sectors (a reactive and a diffusive one) in the time evolution and gave
nontrivial exponent estimates, the latter one by Hinrichsen provided a
hypothesis that the ultimate long time behavior should be characterized by 
DP behavior. 

If we replace the annihilation process $2A\to\emptyset$ by coagulation 
$2A\to A$ in (\ref{AFprocdef}) we get the annihilation-coagulation model.
GMF approximations and simulations of this model resulted in similar phase 
diagram than that of the PCPD model albeit without any sign of two distinct 
regions.
In agreement with this CAM approximations and simulations for the 1d model
found the same kind of continuous transition independently from $D$, with 
exponents in agreement with those of the PCPD in the low-$D$ region
\cite{Odo01,PK66}.
Again the spatial symmetry of particle production was found to be irrelevant.
An exact solution was found by \cite{HH} for the 
special case in 1d, when the diffusion rate is equal to the 
coagulation rate, corresponding to the inactive phase according to which
the particle decay is like of ARW: $\rho\propto t^{-1/2}$.

        \subsubsection{Cyclically coupled spreading with pair annihilation} 
                   \label{cycsect}

In this section I show an explicit two-component realization of
the PCPD class.
A cyclically coupled two-component reaction-diffusion system was
introduced by \cite{HayeDP-ARW}
\begin{equation}
A\to 2A, \quad A\to\emptyset, \quad A\to B, \quad 2B\to A, 
\quad B\emptyset\leftrightarrow\emptyset B
\end{equation}
which mimics the PCPD model (Sect.\ref{pcpdsect}) by mapping pairs
to $A$-s and single particles to $B$-s. This model is a coupled DP+ARW 
system. Its $1+1 d$ critical space-time evolution pattern looks very 
similar to that of the PCPD model. The appearance of the evolution in 
space-time seems to be a particular feature of this class.
It is built up from compact domains with a cloud of lonely particles 
wandering and interacting with them. Furthermore this model also has two 
non-symmetric absorbing states: 
a completely empty one and another with a single wandering $B$.
By fixing the annihilation and diffusion rates of $B$-s ($r=D=1$) the model 
exhibits continuous phase transition by varying the production rate of $A$-s 
and the $A\to B$ transmutation rate. The simulations in 1d showed that
$\rho_A \propto \rho_B$ for large times and resulted in the
following critical exponent estimates
\begin{eqnarray}
\alpha = 0.21(2), \quad \beta = 0.38(6), \quad \beta'=0.27(3), \nonumber \\
Z = 1.75(5), \quad \nu_{||} = 1.8(1), 
\end{eqnarray}
satisfying the generalized hyperscaling relation (\ref{ghypers}). 
These exponents are similar to those of PCPD model in the high
diffusion region (see Table \ref{tabpcpd}) that is reasonable since
$D=1$ is fixed here.

        \subsubsection{The parity conserving annihilation-fission model} 
	\label{apsect}

As we have seen parity conservation plays an important role in unary
production systems. In case of BARW processes it changes the universality
of the transition from DP (Sect.\ref{BARWo}) to PC (Sect.\ref{BARWe}) class.
The question arises if one can see similar behavior in case of binary
production systems.
Recently \cite{binary} investigated a parity conserving 
representative ($n=2$) of the 1d AF model (\ref{AFprocdef}). By performing
simulations for low-$D$-s they found critical exponents that are in the
range of values determined for the corresponding PCPD class.
They claim that the conservation law does not affect the critical behavior
and that the binary nature of the offspring production is
sufficient condition for this class (see however Sect. \ref{cycsect},
where there is no such condition).

The two dimensional version of the parity conserving AF model was investigated
by GMF and simulation techniques \cite{OSC02}. While the $N=1,2$ GMF
approximations showed similar behavior as in case of the PCPD model
(Sect. \ref{pcpdsect}) (including the two class prediction for $N=2$) the
$N=3,4$ approximations do not show $D$ dependence of the critical 
behavior: $\beta=1$, $\beta_2=2$
was obtained for $D>0$. 
Large scale simulations of the particle density confirmed the mean-field 
scaling behavior with logarithmic corrections. 
This result can be interpreted as a numerical evidence supporting that 
the upper critical dimension in this 
model is: $d_c=2$. The pair density decays in a similar way but with an 
additional logarithmic factor to the order parameter. This kind of strongly
coupled behavior at and above criticality was observed in case of
the PCP model too (see Sect. \ref{PCPmod}). At the $D=0$ endpoint 
of the transition line 2+1 d class DP criticality (see Sect. \ref{DPS}) was 
found for $\rho_2$ and for $\rho-\rho(p_c)$. In the inactive phase for 
$\rho(t)$ we can observe the two-dimensional ARW class scaling behavior 
(see Sect.\ref{2A0}), while the pair density decays as $\rho_2\propto t^{-2}$. 
Again like in $d=1$ the parity conservation seems to be irrelevant.

    \subsection{BARWe with coupled non-diffusive field class}

Similarly to the PCP (Sect.\ref{PCPmod}) the effect of infinitely many
frozen absorbing states has been investigated in case of a BARWe model.
A parity conserving version of the 1d PCP model (Sect. \ref{PCPmod}) was 
introduced by \cite{MM99}, in which pairs follow a
BARW2 process, while lonely particles are frozen. 
Simulations showed that while the critical behavior of 
pairs in case of homogeneous, random initial distribution belongs to the 
PC class (Sect. \ref{BARWe}), the spreading exponents satisfy hyperscaling 
(\ref{ghypers}) and change continuously by varying the initial particle density. 
These results are similar to those found in the PCP model. Again long-memory 
effects are responsible for the non-universal behavior
in case of seed-like initial conditions. 
The slowly decaying memory was confirmed by studying a one-dimensional, 
interacting monomer-monomer model \cite{HSPHP01} by simulations.

     \subsection{DP with diffusive, conserved slave field classes} \label{DCF}

Motivated by {\bf model C} results (Sect.\ref{eqintro}), it is an obvious
question what happens with the phase transition to an absorbing state of a
reaction-diffusion system if a conserved secondary density is coupled to 
a non-conserved order parameter.  
One can deduce from the BARW1 spreading process (Sect.\ref{BARWo}) a
two-component, reaction-diffusion model (DCF) \cite{kss,woh} that exhibits 
total particle density conservation as follows:
\begin{equation}
\label{DCFproc}
A + B \stackrel{k}{\to} 2B, \quad B \stackrel{1/\tau}{\to} A \ .
\end{equation}
By varying the initial particle density ($\rho = \rho_A(0)+\rho_B(0)$) 
continuous phase transition occurs. General field theoretical investigation
was done by \cite{woh} (for the equal diffusion case: 
$D_A=D_B$ by \cite{kss}). The mean-field exponents that are
valid above $d_c=4$ are shown in Table \ref{DCFtab2}.
The rescaled action of this model is
\begin{eqnarray}
\label{fullaction}
S{[}\varphi,\overline{\varphi},\psi,\overline{\psi}{]} =
&\int d^dx dt\Big[\overline{\varphi}(\partial_t-\Delta)\varphi
+\overline{\psi}(\partial_t+\lambda(\sigma-\Delta))\psi \nonumber \\
&+\mu\overline{\varphi}\Delta\psi+g\psi\overline{\psi}
(\psi-\overline{\psi})+u\psi\overline{\psi}(\varphi+\overline{\varphi}) 
\nonumber \\
&+v_1\big(\psi\overline{\psi}\big)^2+v_2\psi
\overline{\psi}(\psi\overline{\varphi}-\overline{\psi}\varphi)+
v_3\varphi\overline{\varphi}\psi\overline{\psi} \nonumber \\
&-\rho_B(0)\delta(t)\;\overline{\psi}\Big]
\end{eqnarray}
where $\psi$ and $\phi$ are auxiliary fields, defined such that their average
values coincide with the average density of $B$ particles and the total 
density of particles respectively. The coupling constants are related
to the original parameters of the master equation by
\begin{equation}\begin{array}{llll}
\mu=1-D_B/D_A&g=k\sqrt{\rho}/D_A&\lambda\sigma=k(\rho_c^{\rm {mf}}-\rho)/D_A&\\
v_1=v_2=-v_3=k/D_A&u=-k\sqrt{\rho}/D_A&\lambda=
{D_B}/{D_A}\\\rho_B(0)=\rho_B(0)/\sqrt{\rho}&&&
\end{array}\end{equation}
If one omits from the action Eq.~(\ref{fullaction}) the initial time
term proportional to $\rho_B(0)$, then the remainder is, for $\mu=0$ 
(i. e. $D_A=D_B$), invariant under the time reversal symmetry
\begin{equation}\label{sym}
\begin{array}{rcl}
\psi(x,t)&\rightarrow&-\overline{\psi}(x,-t)\\
\overline{\psi}(x,t)&\rightarrow&-\psi(x,-t) \\
\varphi(x,t)&\rightarrow&\phantom{-}\overline{\varphi}(x,-t)\\
\overline{\varphi}(x,t)&\rightarrow&\phantom{-}\varphi(x,-t)
\end{array}
\end{equation}
Its epsilon expansion solution \cite{kss} and simulation results 
\cite{frei,ful} are summarized in Table \ref{DCFtab2}. Interestingly RG
predicts $Z=2$ and $\nu_{\perp}=2/d$ in all orders of perturbation theory.
\begin{table}
\begin{center}
\begin{tabular}{|c|c|c|c|}
\hline
d             &    $\beta$           &  $Z$   &  $\nu_{\perp}$ \\
\hline
1             &    0.44(1)           &   2    &  2.1(1)      \\
$4-\epsilon$  &    $1-\epsilon/8$   &   2    &   $2/d$  \\
\hline
\end{tabular}
\caption{Summary of results for DCF classes for $D_A=D_B$}
\label{DCFtab2}
\end{center}
\end{table}

The breaking of this symmetry for $\mu\neq 0$, that is, when the 
diffusion constants $D_A$ and $D_B$ are different causes different critical 
behavior for this system. For $D_A < D_B$ RG \cite{woh} predicts new classes
with $Z=2$, $\beta=1$, $\nu_{\perp}=2/d$, but simulations in 1d \cite{ful}
show different behavior (see Table \ref{DCFtab1}).
For non-poissonian initial particle density distributions the critical
initial slip exponent $\eta$ varies continuously with the width of the
distribution of the conserved density.
\begin{table}
\begin{center}
\begin{tabular}{|c|c|c|c|}
\hline
d             &    $\beta$  &  $Z$   &  $\nu_{\perp}$ \\
\hline
1             &    0.33(2)  &        &   2        \\
$4-\epsilon$  &    1        &   2    &   2/d    \\
\hline
\end{tabular}
\caption{Summary of results for DCF classes for $D_A<D_B$}
\label{DCFtab1}
\end{center}
\end{table}
The $D_A=0$ extreme case is discussed in Section \ref{NDCF}.

For $D_A > D_B$ no stable fixed point solution was found by RG hence
\cite{woh} conjectured first order transition for which signatures
were found in 2d by simulations \cite{OWLH}. However $\epsilon$ expansion
may break down in case of the occurrence of another critical dimension
$d_c'< d_c=4$ for which simulations in 1d \cite{ful} provided
numerical support (see Table \ref{DCFtab3}). 
\begin{table}
\begin{center}
\begin{tabular}{|c|c|c|c|}
\hline
d             &    $\beta$           &  $Z$   &  $\nu_{\perp}$ \\
\hline
1             &    0.67(1)           &        &      2      \\
$4-\epsilon$  &    0                 &        &              \\
\hline
\end{tabular}
\caption{Summary of results for DCF classes for $D_A > D_B$}
\label{DCFtab3}
\end{center}
\end{table}

   \subsection{DP with frozen, conserved slave field classes}
       \label{NDCF}   

If the conserved field coupled to the BARW1 process (eq. \ref{DCFproc})
is non-diffusive non-DP universality class (NDCF) behavior is
reported again \cite{rossi,pastor}. The corresponding action can be derived
from (\ref{fullaction}) in the $D_A=0$ limit:
\begin{eqnarray}
\label{NDCFaction}
S = &\int d^dx dt\Big[\overline{\varphi}(\partial_t+r-D\nabla^2)\varphi
+\overline{\psi}(\partial_t-\lambda\nabla^2))\psi \nonumber \\
& +g\psi\overline{\psi}
(\psi-\overline{\psi})+u\psi\overline{\psi}(\varphi+\overline{\varphi}) 
+v_1\big(\psi\overline{\psi}\big)^2 \nonumber \\
& +v_2\psi\overline{\psi}(\psi\overline{\varphi}-\overline{\psi}\varphi)+
v_3\varphi\overline{\varphi}\psi\overline{\psi} \Big]
\end{eqnarray}
By neglecting irrelevant terms (\ref{NDCFaction}) is invariant under the
shift transformation
\begin{equation}
\psi\to\psi+\Delta , \quad r\to r-v_2\Delta
\end{equation}
where $\Delta$ is any constant. The field theoretical analysis of this
action has run into difficulties \cite{pastor}. The main examples for 
the NDCF classes are the conserved threshold transfer process and 
the conserved reaction-diffusion model \cite{rossi,pastor}. Furthermore
the models described by the NDCF classes embrace a large group of 
stochastic sandpile models \cite{HJ98} in particular fixed-energy 
Manna models \cite{Manna,Manna2,DVZ98,M01}. 
The upper critical dimension $d_c=4$ was confirmed by simulations \cite{LubH}.

It was also shown \cite{AM01} that 
these classes describe the depinning transition of {\bf quenched 
Edwards-Wilkinson} (see Sect. \ref{qEWsect}) or linear interface models (LIM) 
\cite{HZ95,barabasi} owing to the fact that quenched disorder can 
be mapped onto long-range temporal correlations in the activity field
\cite{MM94}. However this mapping could not be done on the level of
Langevin equations of the representatives of NDCF and LIM models and in
1d this equivalence may break down \cite{AM01,DAMPVZ01,KC03L}.
The critical exponents determined by simulations 
\cite{rossi,pastor,Lubeck,Lubeck2,DTO02} 
and GMF+CAM method in 1d \cite{DCAM02} are summarized in Table \ref{NDCFtab}.
Similarly to the PCP these models exhibit infinitely 
many absorbing states, therefore non-universal spreading exponents 
are expected (in Table \ref{NDCFtab} the exponent $\eta$ corresponding
to natural initial conditions is shown).
\begin{table}
\begin{center}
\begin{tabular}{|c|c|c|c|c|c|c|c|}
\hline
d & $\alpha$ & $\beta$ & $\gamma$ & $Z$   & $\nu_{||}$ & $\sigma$ &  $\eta$ \\
\hline
1 & 0.14(1)  & 0.28(1) &         & 1.5(1) &  2.5       &        &        \\
\hline
2 & 0.50(5)  & 0.64(1) & 1.59(3)& 1.55(4)& 1.29(8) & 2.22(3) & 0.29(5) \\
\hline
3 & 0.90(3)  & 0.84(2) & 1.23(4)& 1.75(5)& 1.12(8) & 2.0(4) & 0.16(5) \\
\hline
4 &  1       &  1      &   1    &    2 &    1      &  2    &       0 \\
\hline
\end{tabular}
\caption{Summary of results for NDCF classes}
\label{NDCFtab}
\end{center}
\end{table}

     \subsection{Coupled N-component DP classes} \label{NDPS}

\begin{figure}
\epsfxsize=60mm
\epsffile{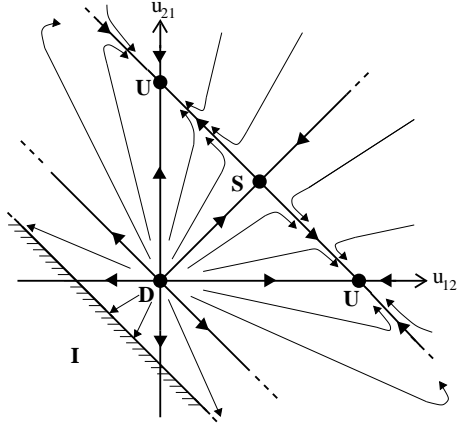}
\caption{Flow of the interspecies couplings in the two-component,
DP model under renormalization. D means decoupled, S symmetric, 
U unidirectional fixed points \cite{Jansflav}.}
\label{MDPflow}
\end{figure}

From the basic reaction-diffusion systems one can generate N-component
ones coupled by interactions symmetrically or asymmetrically.
In \cite{Ja97,Jansflav} Janssen introduced and analyzed by field 
theoretical RG method (up to two loop order) the quadratically coupled, 
N-species generalization of the DP process of the form:
\begin{eqnarray}
A_{\alpha} &\leftrightarrow & 2 A_{\alpha} \nonumber \\
A_{\alpha} &\rightarrow & \emptyset \nonumber \\
A_{\alpha} + A_{\beta} &\rightarrow & k A_{\alpha} + l A_{\beta} , \
\label{NDPeq}
\end{eqnarray}
where $k,l$ may take the values ($0,1$).
He has shown that the multi-critical behavior is always described by the 
Reggeon field theory (DP class), but this is unstable and leads to
unidirectionally coupled DP systems (see Fig.\ref{MDPflow}).

He has also shown that by this model the linearly, unidirectionally coupled
DP (UCDP) (see Sect.\ref{uniDP}) case can be described. 
The universality class behavior of UCDP is discussed in Sect.\ref{uniDP}.

In {\bf one dimension}, if BARWo type of processes are coupled 
(which alone exhibit DP class transition (see Sect. \ref{BARWo})  
{\bf hard-core interactions} can modify the phase transition universality 
(see Sect. \ref{2BARWoS}).

     \subsection{Coupled N-component BARW2 classes} \label{NBARWS}

Bosonic, $N$-component BARW systems with two offsprings (N-BARW2),
of the form
\begin{eqnarray} 
A_{\alpha} & \rightarrow & 3 A_{\alpha} \label{Ato3Aproc} \\
A_{\alpha} & \rightarrow & A_{\alpha} + 2 A_{\beta} \label{AtoABBproc} \\
2 A_{\alpha} & \rightarrow  & \emptyset 
\end{eqnarray}
were introduced and investigated by \cite{Cardy-Tauber2}
via field theoretical RG method. These models exhibit parity conservation
of each species and permutation symmetry on $N$ types (generalization
for $O(N)$ symmetry is violated by the annihilation term). The $A\to 3A$  
processes turns out to be irrelevant, because like pairs annihilate
immediately. Models with (\ref{AtoABBproc}) branching terms exhibit 
continuous phase transitions at {\bf zero branching rate}. 
The universality class is expected to be independent from $N$ 
and coincides with that of the $N\to\infty$ (N-BARW2) model that
could be solved exactly.
The critical dimension is $d_c=2$ and for $d\le 2$ the exponents are
\begin{equation} \label{NBARWe}
\beta=1, \ \ \ Z=2, \ \ \ \alpha=d/2, \ \ \ \nu_{||}=2/d, \ \ \ 
\nu_{\perp}=1/d \ ,
\end{equation}
while at $d=d_c=2$ logarithmic correction to the density decay is expected.
Simulations on a $d=2$ (fermionic) lattice model confirmed these results 
\cite{barw2cikk}. 

In one dimension it turns out that {\bf hard-core interactions} can be 
relevant and different universal behavior emerges for fermionic models 
(see Sect. \ref{2BARW2S}).
The bosonic description of fermionic models in one dimension works 
(at least for static exponents) in case of {\it pairwise initial conditions} 
(see Sect. \ref{2}) when different types of particles do not make
up blockades for each other. Such situation happens when these particles
are generated as domain walls of N+1 component systems exhibiting
$S_{N+1}$ symmetric absorbing states (see Sects. \ref{BGDP}, \ref{nekim}).

 \subsubsection{Generalized contact processes with $n>2$ \\
 absorbing states in 1d} \label{BGDP}

The generalized contact process has already been introduced in Sect.\ref{GDKmod}
with the main purpose to show an example for PC class universality class
transition in case of $Z_2$ symmetric absorbing states. 
The more general case with $n>2$ permutation symmetric absorbing states
was investigated using DMRG method by \cite{Hoy}
and turned out to exhibit N-BARW2 transition.
In the one dimensional model, where each lattice site can be occupied by 
at most one particle ($A$) or can be in any of $n$ inactive states 
($\emptyset_{1}$, $\emptyset_{2}$ \ldots$\emptyset_{n}$) the reactions are:
\begin{eqnarray}
AA   \rightarrow  A \emptyset_k,\emptyset_k A
&&
\ \ \ \ \ \ \
\mathrm{with\ rate\ \lambda/n}\label{reac_1}\\
A\emptyset_k ,\emptyset_k A  \rightarrow \emptyset_k \emptyset_k
&& 
\ \ \ \ \ \ \
\mathrm{with\ rate\ \mu_k}\label{reac_2}\\
A\emptyset_k ,\emptyset_k A  \rightarrow AA
&&
\ \ \ \ \ \ \
\mathrm{with\ rate\ 1}\label{reac_3}\\
\emptyset_k \emptyset_l \rightarrow A \emptyset_l,\,\,\emptyset_k A
&&
\ \ \ \ \ \ \
{(k\neq l)\,\,
\mathrm{with\,\,\,rate\,\,\,1}
}\label{reac_4}
\end{eqnarray}
The original contact process (Sect. \ref{CPsect}), corresponds to the 
$n=1$ case, in which the reaction (\ref{reac_4}) is obviously absent. 
The reaction (\ref{reac_4}) in the case $n \geq2$ ensures 
that configurations as $(\emptyset_{i}
\emptyset_{i} \ldots\emptyset_{i} \emptyset_{i} \emptyset_{j} \emptyset_{j}
\ldots\emptyset_{j} \emptyset_{j})$, with $i \neq j$ are not absorbing. Such
configurations do evolve in time until the different domains coarsen and one
of the $n$ absorbing states $(\emptyset_{1} \emptyset_{1} \ldots\emptyset
_{1})$, $(\emptyset_{2} \emptyset_{2} \ldots\emptyset_{2})$, \ldots
$(\emptyset_{n} \emptyset_{n} \ldots\emptyset_{n})$ is reached.
For generalized contact processes with $n=2$, simulations \cite{Hin97} 
and a DMRG study \cite{Hoy} proved that the transition falls in the PC 
class if $\mu_{1} = \mu_{2}$, while if the symmetry between the two absorbing 
states was broken ($\mu_{1} \neq\mu_{2}$) a DP transition was 
recovered. \\

The DMRG study for $n=3$ and $n=4$ showed that, the model 
is in the active phase in the whole parameter space and the critical 
point is shifted to the limit of infinite reaction rates. In this 
limit the dynamics of the model can be mapped onto the zero 
temperature $n$-state Potts model
(see also the simulation results of \cite{Lipp}).
It was conjectured by \cite{Hoy} the model is in the same N-BARW2
universality class for all $n \geq 3$. By calling a domain wall
between $\emptyset_i$ and $\emptyset_j$ as $X_{ij}$ one can follow
the dynamics of such variables. In the limit $\lambda \to \infty$, 
$X_{ij}$ coincides with the particle $A$ and in the limit $\mu \to \infty$
$X_{ij}$ coincides with the bond variable $\emptyset_i \emptyset_j$.
For finite values of these parameters one still can apply this reasoning at a
coarse-grained level. In this case $X_{ij}$ is not a sharp domain wall, but
an object with a fluctuating thickness. For $n=2$ it was shown in 
Section \ref{GDKmod} that such domain wall variables follow BARW2
dynamics (\ref{BARW2proc}). For $n>2$ one can show that besides the
N-BARW2 reactions (\ref{Ato3Aproc}),(\ref{AtoABBproc}) (involving
two types of particles maximum) reaction types involving three
 different domains 
($i\neq j$, $i \neq k$ and $j \neq k$):
\begin{eqnarray}
X_{ij} \rightarrow X_{ik} X_{kj}
\ \ \ \ \ \ \ 
X_{ik} X_{kj} \rightarrow X_{ij}
\label{barwngt2}
\end{eqnarray}
occur with increasing importance as $n\to\infty$. These reactions
break the parity conservation of the N-BARW2 process, therefore the 
numerical findings in \cite{Hoy} for $n=3,4$ indicate that they are 
probably {\bf irrelevant} or the conditions for N-BARW2 universal
behavior could be relaxed. 
Owing to the fact that the $X_{ij}$ variables are domain walls they
appear in pairwise manner, hence {\bf hard-core exclusion} effects 
are ineffective for the critical behavior i 1d. To see the effect of
pairwise initial conditions for dynamical exponents see Section 
\ref{2}.

For $n=3$ upon breaking the global $S_3$ symmetry to a lower 
one, one gets a transition either in the directed percolation
(Sect.\ref{DPS}), or in the parity conserving class (Sect.\ref{PCS}), 
depending on the choice of parameters \cite{Hoy}. Simulations indicate 
\cite{LipD} that for this model local symmetry breaking may also generate 
PC class transition).

   \subsection{Hard-core 2-BARW2 classes in 1d} \label{2BARW2S}

Besides the effects of coupling interactions in low dimensions blockades
generated by hard-core particles may also play an important role.
The effect of particle exclusion (i.e. $AB\not\leftrightarrow BA$)
in 2-BARW2 models (Sect.\ref{NBARWS}) was investigated in 
\cite{barw2cikk,Park}. 
For $d=2$ the bosonic field theoretical predictions \cite{Cardy-Tauber2} 
were confirmed \cite{barw2cikk} (mean-field class transition with
logarithmic corrections).
In one dimension however two types of phase transitions were identified 
at zero branching rate ($\sigma=0$) depending on the arrangement of 
offspring relative to the parent in process (\ref{AtoABBproc}). 
Namely if the parent separates the two offsprings (2-BARW2s):
\begin{equation} 
A\stackrel{\sigma}{\to}BAB
\end{equation}
the steady state density is higher than in the case when they are created 
on the same site (2-BARW2a):
\begin{equation}
A\stackrel{\sigma}{\to}ABB
\end{equation} 
at a given branching rate, because in the former case they are unable 
to annihilate each other. This results in different order parameter 
exponents for the symmetric (2-BARW2s) and for the asymmetric (2-BARW2a) cases
\begin{equation} 
\beta_s=1/2, \quad \beta_a=2 \ .
\end{equation}
This result is in contrast with a widespread belief that the bosonic field 
theory (where $AB\leftrightarrow BA$ is allowed) can describe these systems
(because in that case the critical behavior is different (Sect. \ref{NBARWe})). 
This observation led \cite{Park} to the conjecture that 
in one-dimensional, reaction-diffusion systems a series of new universality 
classes should appear if particle exclusion is present. 
Note however, that since the transition is at $\sigma=0$ in both cases
the on-critical exponents do not depend on how particles are created and
they can be identified with those described in Sect.(\ref{2}). 
In \cite{barw2cikk} a set of critical exponents satisfying scaling
relations have been determined for this two new classes shown in 
Table \ref{tab2}.
\begin{table}
\begin{center}
\begin{tabular}{|l|r|r|r|}
\hline
exponent   & N-BARW2 & N-BARW2s         & N-BARW2a \\
\hline
$\nu_{||}$ &  2      & 2.0(1)$|$0.915(2) & 8.0(4)$|$3.66(2) \\
\hline      
$Z$        &  2      & 4.0(2)$|$1.82(2)* & 4.0(2)$|$1.82(2)* \\
\hline
$\alpha$   &  1/2    & 0.25(1)$|$0.55(1)*& 0.25(1)$|$0.55(1)* \\
\hline  
$\beta$    &  1      & 0.50(1)$|$1.00(1) & 2.0(1)$|1.00(1)$ \\
\hline
\end{tabular}
\end{center}
\caption{Summary of critical exponents in one dimension for N-BARW2 like
models. The N-BARW2 data are quoted from \cite{Cardy-Tauber}. 
Data divided by "$|$" correspond to random vs. pairwise
initial condition cases \cite{Hoy,gdkcikk,ujcikk}.
Exponents denoted by * exhibit slight initial density dependence.}
\label{tab2}
\end{table}

    \subsubsection{Hard-core 2-BARWo models in 1d} \label{2BARWoS}

Hard-core interactions in the two-component, one-offspring production 
model (2-BARW1) were investigated in \cite{dp2cikk}. Without 
interaction between different species one would expect DP class transition. 
By introducing the $AB\not\leftrightarrow BA$ blocking to the
two-component model:
\begin{eqnarray}
A\stackrel{\sigma}{\longrightarrow} AA \ \ \ \
B\stackrel{\sigma}{\longrightarrow} BB \\
AA\stackrel{1-\sigma}{\longrightarrow}\emptyset \ \ \ \ 
BB\stackrel{1-\sigma}{\longrightarrow}\emptyset 
\end{eqnarray}
a DP class transition at $\sigma=0.81107$ was located. Note that the
effect exerted by different species on each other is irrelevant now
unlike for the case of coupled ARW (Sect.\ref{2}).
On the other hand if we couple the two sub-systems by production:
\begin{eqnarray}
A\stackrel{\sigma/2}{\longrightarrow} AB \ \ \ \ 
A\stackrel{\sigma/2}{\longrightarrow} BA \\
B\stackrel{\sigma/2}{\longrightarrow} AB \ \ \ \ 
B\stackrel{\sigma/2}{\longrightarrow} BA \\
AA\stackrel{1-\sigma}{\longrightarrow}\emptyset \ \ \ \ 
BB\stackrel{1-\sigma}{\longrightarrow}\emptyset 
\end{eqnarray}
a continuous phase transition emerges at $\sigma=0$ rate  
-- therefore the on-critical exponents are the same as those described 
in Sect. (\ref{2}) -- and the order parameter exponent was found to 
be $\beta=1/2$. Therefore this transition belongs to the same 
class as the 2-BARW2s model (see Sect.\ref{2BARW2S}). 
The parity conservation law, which is relevant in case of one component 
BARW systems (PC versus DP class) turns out to be irrelevant here. 
This finding reduces the expectations suggested by \cite{Park} for a 
whole new series of universality classes in 1d systems with exclusions. 
In fact the blockades introduced by exclusions generate robust classes. 
In \cite{dp2cikk} a hypothesis was set up that {\bf in coupled branching 
and annihilating random walk systems of N-types of excluding particles 
for continuous transitions at $\sigma=0$ two universality classes exist, 
those of 2-BARW2s and 2-BARW2a models}, depending on whether the reactants 
can immediately annihilate (i.e. when similar particles are not 
separated by other type(s) 
of particle(s)) or not. Recent investigations in similar
models \cite{HSPHP01,Lip} are in agreement with this hypothesis.

   \subsubsection{Coupled binary spreading processes}

Two-component versions of the PCPD model (Sect. \ref{pcpdsect}) with particle
exclusion in 1d were introduced and investigated by simulations in 
\cite{parwcikk} with the aim to test if the hypothesis for N-component 
BARW systems set up in \cite{dp2cikk} (Sect.\ref{2BARWoS}) can be 
applied for such models.
The following models with the same diffusion and annihilation terms 
($AA\to\emptyset$, $BB\to\emptyset$) as in (Sect.\ref{2}) and different 
production processes were investigated.

1) Production and annihilation random walk model (2-PARW):
\begin{eqnarray}
AA\stackrel{\sigma/2}{\longrightarrow}AAB, \ \ \ 
AA\stackrel{\sigma/2}{\longrightarrow}BAA, \\
BB\stackrel{\sigma/2}{\longrightarrow}BBA, \ \ \
BB\stackrel{\sigma/2}{\longrightarrow}ABB \ .
\end{eqnarray}

2) Symmetric production and annihilation random walk model 
(2-PARWS):
\begin{eqnarray}
AA\stackrel{\sigma}{\longrightarrow}AAA, \\ 
BB\stackrel{\sigma}{\longrightarrow}BBB \ .
\end{eqnarray}
These two models exhibit active steady states for $\sigma>0$ with a 
continuous phase transition at $\sigma_c=0$. Therefore the exponents at the 
critical point are those of the 2-component ARW model with
exclusion (Sect. \ref{2}). Together with the exponent $\beta=2$ result for 
both cases this indicates that they belong to the N-BARW2a class. 
This also means that the hypothesis set up for N-BARW2 systems \cite{dp2cikk} 
(Sect.\ref{2BARWoS}) can be extended. \\
\begin{figure}
\epsfxsize=70mm
\epsfysize=60mm
\epsffile{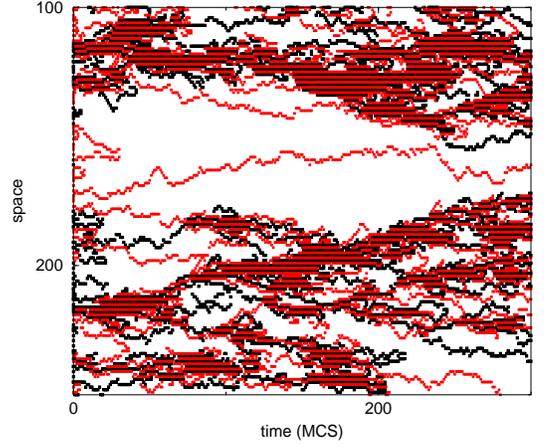}
\caption{Space-time evolution of the 2-PARWAS model at the critical point
\cite{parwcikk}. Black dots correspond to $A$ particles, others to $B$-s.}
\label{parwrajz}
\end{figure}

3) Asymmetric  production and annihilation random walk model
(2-PARWA):
\begin{eqnarray}
AB\stackrel{\sigma/2}{\longrightarrow}ABB, \ \ \
AB\stackrel{\sigma/2}{\longrightarrow}AAB, \\ 
BA\stackrel{\sigma/2}{\longrightarrow}BAA, \ \ \
BA\stackrel{\sigma/2}{\longrightarrow}BBA \ .
\end{eqnarray}
This model does not have an active steady state. The $AA$ and $BB$
pairs annihilate themselves on contact, while if an $A$ and $B$ particle 
meet an $AB\to ABB\to A$ process eliminates blockades and the 
densities decay with the $\rho\propto t^{-1/2}$ law for $\sigma > 0$.
For $\sigma=0$ the blockades persist, and in case of random initial state 
$\rho\propto t^{-1/4}$ decay (see Sect. \ref{2}) can be observed here. \\

4) Asymmetric production and annihilation random walk model
with spatially symmetric creation (2-PARWAS):
\begin{eqnarray}
AB\stackrel{\sigma/2}{\longrightarrow}ABA, \ \ \
AB\stackrel{\sigma/2}{\longrightarrow}BAB, \\ 
BA\stackrel{\sigma/2}{\longrightarrow}BAB, \ \ \
BA\stackrel{\sigma/2}{\longrightarrow}ABA \ .
\end{eqnarray}
In this case  $AB$ blockades proliferate by production events. 
As a consequence of this an active steady state 
appears for $\sigma > 0.3253(1)$ with a continuous phase transition.
The space-time evolution from random initial state shows 
(Fig.\ref{parwrajz}) that compact domains of alternating 
$..ABAB..$ sequences separated by lonely wandering particles are formed.
This pattern is very similar to what was seen in case of one-component binary 
spreading processes \cite{HayeDP-ARW}: compact domains within a cloud of
lonely random walkers, except that now domains are built up from 
alternating sequences only. This means that the $..AAAA...$ and $...BBBB...$ 
domains decay by this annihilation rate and the particle blocking is 
responsible for the formation of compact clusters. 
In the language of coupled DP + ARW model 
\cite{HayeDP-ARW} the pairs following DP process are the $AB$ pairs 
now, which cannot decay spontaneously but through an annihilation process: 
$AB + BA\to\emptyset$. They interact with two types of particles 
executing annihilating random walk with exclusion. The simulations 
resulted in the critical exponent estimates: $\beta=0.37(2)$, $\alpha=0.19(1)$
and $Z=1.81(2)$ which agree fairly well with those of the PCPD model in the 
high diffusion rate region \cite{Odo00}.

\section{Interface growth classes} \label{growth}

Interface growth classes are strongly related to the basic universality
classes discussed so far and can be observed in experiments more easily.
For example one of the few experimental realizations of the robust DP
class (\ref{DPS}) is related to a depinning transition in inhomogeneous 
porous media \cite{BBCHSV92} (see Sect.\ref{CGsect}). 
The interface models can either be defined by continuum equations or by
lattice models of solid-on-solid (SOS) or restricted solid-on-solid 
(RSOS) types.
In the latter case the height variables ${h_i}$ of adjacent sites are
restricted
\begin{equation}
\vert h_i - h_{i+1} \vert \le 1 \ \ \ .
\end{equation}
The morphology of a growing interface is usually characterized
by its width
\begin{equation}
W(L,t) =
\Bigl[
\frac{1}{L} \, \sum_i \,h^2_i(t)  -
\Bigl(\frac{1}{L} \, \sum_i \,h_i(t) \Bigr)^2
\Bigr]^{1/2}
\,.
\end{equation}
In the absence of any characteristic length, growth processes are expected to
show power-law behavior of the correlation functions in space and height and 
the {\em Family-Vicsek} scaling~\cite{family} form:
\begin{equation}
\label{FV-forf}
W(L,t) = t^{\tilde\alpha/Z} f(L / \xi_{||}(t)),
\end{equation}
describes the surface, with the scaling function $f(u)$ 
\begin{equation}
\label{FV-fu}
f(u) \sim
\left\{ \begin{array}{lcl}
     u^{\tilde\alpha}     & {\rm if} & u \ll 1 \\
     {\rm const.} & {\rm if} & u \gg 1
\end{array}
\right. .
\end{equation}
Here $\tilde\alpha$ is the roughness exponent and characterizes the stationary
regime in which correlation length $\xi_{||}(t)$ has reached a value larger 
than the system size $L$. 
The ratio $\tilde\beta= \tilde\alpha/Z$ is called as the growth exponent and
characterizes the short time behavior of the surface. Similarly to equilibrium 
critical phenomena, these exponents do not depend on the microscopic 
details of the system under investigation. Using these exponents it is 
possible to divide growth processes into universality classes 
\cite{barabasi,krug-rev}.
The (\ref{FV-forf}) scaling form of $W^2$ is invariant under $\Lambda$ the 
rescaling
\begin{equation}
x\to\Lambda x, \ \ \ \  t\to\Lambda^Z t, \ \ \ \ 
h(x,t)\to\Lambda^{-\tilde\alpha} h(x,t)
\end{equation}
Recently {\bf anomalous roughening} has been observed in many growth models 
and experiments. In these cases the measurable $\tilde\alpha_{loc}$ roughness 
exponent is different from $\tilde\alpha$ and may satisfy different scaling 
law \cite{krug-rev,schro,lopez96,das1,das2,lopez97c,O92,yang,jef,fracture1,
fracture2,bru}. 

\begin{center}
\begin{figure}
\epsfxsize=75mm
\epsffile{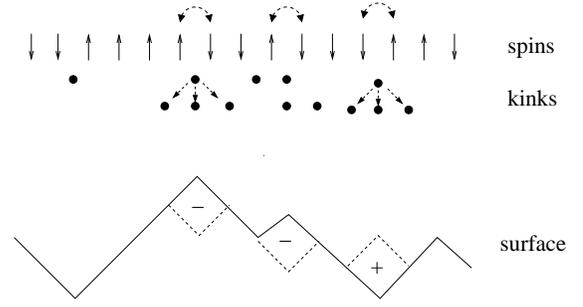}
\caption{Mapping between spins, kinks and surfaces}
\label{surfmap}
\end{figure}
\end{center}

Surfaces in $d+1$ dimensional systems can be {\bf mapped} onto a 
time step of a $d$ dimensional particle reaction-diffusion or spin models. 
For example the 1d Kawasaki spin model corresponding to the
$K\leftrightarrow 3K$ process with random walk of kinks 
the mapping onto the 1d surface is shown in Fig. \ref{surfmap}.
This means that to a unique $\{s_j\}$ spin configuration at time $t$ 
corresponds a spatial profile $\{h_i(t)\}$ by accumulating the spin values
\begin{equation}
h_i(t) = \sum_{j=1}^i s_j
\end{equation} 
In 1d the surface can also be considered as a random walker with 
fluctuation
\begin{equation}
\Delta x \propto t^{1/Z_w}
\end{equation}
hence the roughness exponent is related to the dynamical exponent
$Z_w$ as
\begin{equation}
\tilde\alpha=1/Z_w \ .
\end{equation}
The $\tilde\alpha=1/2$ corresponds to uncorrelated 
(or finite correlation length) random walks. If $\tilde\alpha > 1/2$ the 
surface exhibits correlations, while if $\tilde\alpha < 1/2$ the 
displacements in the profile are anti-correlated. Since the surfaces may 
exhibit drifts the fluctuations around the mean is measured defining the 
local roughness (Hurst) exponent.
Using this surface mapping \cite{Sales} have characterized
the different classes of Wolfram's 1d cellular automata \cite{Wolfram}.

One can show that by coarse graining the 1d Kawasaki dynamics
\begin{equation}
w_i = \frac{1}{4\tau} \left[ 1 - \sigma_i\sigma_{i+1} 
+ \lambda(\sigma_{i+1} - \sigma_i) \right]
\end{equation}
a mapping can be done onto the KPZ equation (\ref{KPZ-e}), and the surface 
dynamics for $\lambda\not = 0$ (corresponding to anisotropic case) 
is in the KPZ class (Sect.\ref{KPZsect}) while for $\lambda=0$ it is in 
the Edward Wilkinson class (Sect. \ref{EWsect}).
While these classes are related to simple random walk with 
$Z_w=1/\tilde\alpha=2$ the question arises what surfaces
are related to other kind of random walks (example Levy-flights or 
correlated random walks etc ...).
Recently it was shown that globally constrained random walks (i.e. when
a walker needs to visit each site an even number of times) can be mapped
onto surfaces with dimer-type of dynamics \cite{NPKN01} with
$Z_w = 3 = 1/\tilde\alpha$. 

By studying correspondence between lattice models with 
absorbing states and models of pinned interfaces in random media 
\cite{DM00} established the scaling relation
\begin{equation}
\tilde\beta = 1 - \beta/\nu_{||}
\end{equation}
that was confirmed numerically for $d=1,2,3,4$ contact processes 
(Sect. \ref{CPsect}). The local roughness exponent was found to be
smaller than the global value indicating anomalous surface growth
in DP class models.

In interface models different types of transitions may take place.
{\bf Roughening transitions} may occur between smooth phase characterized by 
finite width $W$ (in an infinite system) and rough phase when the width and
diverges in an infinite system (but saturates in finite ones) by varying some
control parameters ($\epsilon$). Near the transition point the spatial
($\xi_{||}$) and growth direction correlations ($\xi_{\perp}$) diverge as
\begin{eqnarray}
\xi_{||} \propto \epsilon^{\nu_{||}} \\
\xi_{\perp} \propto \epsilon^{\nu_{\perp}}
\end{eqnarray}
(note that in RD systems $\xi_{||}$ denote temporal correlation length).
In the smooth phase the heights $h_i(t)$ are correlated below $\xi_{\perp}$.
While in equilibrium models roughening transitions exist in $d>1$ dimensions
only in nonequilibrium models this may occur in $d=1$ as well.

An other surface transition is the {\bf depinning transition}, 
when as the consequence of changing some control parameter 
(usually an external force $F$) the surface starts propagating with 
speed $v$ and evolves in a rough state. Close to the transition $v$ is 
expected to scale as 
\begin{equation}
v \propto (F-F_c)^{\tilde\theta}
\end{equation}
with the $\tilde\theta$ velocity exponent and the correlation
exponents diverge. Known depinning transitions (in random media) 
are related to absorbing phase transitions with conserved quantities
(see Sects. \ref{qEWsect}, \ref{qKPZsect}).

In the rough phase a so-called {\bf faceting phase transition} 
may also take place when up-down symmetrical facets appear. In this case the 
surface scaling behavior changes (see Section \ref{dimer}).

   \subsection{The random deposition class} \label{RDsect}

The random deposition is the simplest surface growth process that involves
uncorrelated adsorption of particles on top of each other. Therefore columns
grow independently, linearly without bounds. The roughness exponent 
$\tilde\alpha$ (and correspondingly $Z$) is not defined here. 
The width of the surface grows as $W\propto t^{1/2}$ hence 
$\tilde\beta=1/2$ in all dimensions. An example for such behavior is shown
in a dimer growth model in Sect. \ref{DAT}.

   \subsection{Edwards-Wilkinson (EW) classes} \label{EWsect}

As it was mentioned in Sect.\ref{growth} growth models of 
this class can easily mapped onto spins with symmetric Kawasaki
dynamics or to particles with ARW (Sect.\ref{2A0}).
If we postulate the following translation and reflection symmetries
\begin{equation} \label{EW-s}
{\bf x} \to {\bf x+\Delta x} \ \ \
t \to t+\Delta t \ \ \
h \to h+\Delta h \ \ \
{\bf x} \to {\bf -x} \ \ \
h \to -h
\end{equation}
we are led to the Edwards-Wilkinson (EW) equation \cite{EWc}
\begin{equation} \label{EW-e}
\partial_t h({\bf x},t) = v + \sigma\nabla^2 h({\bf x},t) + \zeta({\bf x},t) ,
\end{equation}
which is the simplest stochastic differential equation that describes a 
surface growth with these symmetries.
Here $v$ denotes the mean growth velocity, $\sigma$ the surface tension 
and $\zeta$ the zero-average Gaussian noise field with variance
\begin{equation}
\label{gaussnoise}
<\zeta({\bf x},t)\zeta({\bf x'},t')> = 2 D \delta^{d-1} ({\bf x-x'})(t-t')
\end{equation}
This equation is linear and exactly solvable. The critical exponents 
of EW classes are 
\begin{equation}
{\tilde\beta} = ({1\over 2} - {d\over 4}), \ \ Z=2
\end{equation}

    \subsection{Quenched EW classes} \label{qEWsect}

In random media linear interface growth is described by the so called 
quenched Edwards-Wilkinson equation
\begin{equation} \label{qEW-e}
\partial_t h({\bf x},t) = \sigma\nabla^2 h({\bf x},t) + F 
+ \eta({\bf x},h({\bf x},t)) ,
\end{equation}
where $F$ is a constant, external driving term and $\eta({\bf x},h({\bf x},t))$
is the quenched noise. The corresponding linear interface models (LIM) 
exhibit a depinning transition at $F_c$. 
The universal behavior of these models were investigated in 
\cite{Nat,Nar,Hyun} and it was shown to be equivalent with NDCF 
classes (Sect. \ref{NDCF}).
Analytical studies \cite{Nar} predict $\tilde\alpha=(4-d)/3$ and 
$Z=2-(2/9)(4-d)$.

    \subsection{Kardar-Parisi-Zhang (KPZ) classes} \label{KPZsect}

If we drop the $h\to -h$ symmetry from (\ref{EW-s}) we can add a term
to the (\ref{EW-e}) equation that is the most relevant term in renormalization
sense breaking the up-down symmetry:
\begin{equation}
\label{KPZ-e}
\partial_t h({\bf x},t) = v + \sigma\nabla^2 h({\bf x},t) + 
\lambda(\nabla h({\bf x},t))^2 + \zeta({\bf x},t)
\end{equation}
that is the Kardar-Parisi-Zhang (KPZ) equation \cite{KPZeq}. 
Here again $v$ denotes the mean growth velocity, $\sigma$ the surface tension 
and $\zeta$ the zero-average Gaussian noise field with variance
\begin{equation}
<\zeta({\bf x},t)\zeta({\bf x'},t')> = 2 D \delta^{d-1} ({\bf x-x'})(t-t')
\end{equation}
This is non-linear, but exhibits a tilting symmetry as the result of
the Galilean invariance of (\ref{KPZ-e}):
\begin{equation}
h \to h' +\epsilon{\bf x}, \qquad {\bf x} \to {\bf x'} - \lambda\epsilon t,
\qquad t \to t'
\end{equation}
where $\epsilon$ is an infinitesimal angle. As a consequence the scaling
relation 
\begin{equation}
\tilde\alpha + Z = 2
\end{equation}
holds in any dimensions. In one dimension the critical exponents are known
exactly whereas for $d>1$ dimensions numerical estimates exist \cite{barabasi}
(see Table \ref{KPZtab}).
\begin{table}
\begin{center}
\begin{tabular}{|c|c|c|c|}
\hline
$d$ & $\tilde\alpha$ & $\tilde\beta$ & $Z$ \\
\hline
1   &  1/2           &  1/3          & 3/2 \\
\hline
2   &  0.38          &  0.24         & 1.58 \\
\hline
3   &  0.30          &  0.18         & 1.66 \\
\hline
\end{tabular}
\caption{Scaling exponents of KPZ from \cite{barabasi}}
\label{KPZtab}
\end{center}
\end{table}
The upper critical dimension of this model is debated. Mode coupling
theories and various phenomenological field theoretical schemes 
\cite{H90,L95,L97} settle to $d_c=4$. Contrary to analytical approaches
numerical solution of the KPZ equation \cite{MWK91}, simulations
\cite{Kim,WK87,TFW92,AHK93,MPP}, 
and the results of real-space renormalization group 
calculations \cite{Piet,Piet2,Piet3} provide no evidence for a finite $d_c$.
Furthermore, the only numerical study \cite{Tu} of the mode-coupling 
equations gives no indication for the existence of a finite $d_c$ either.
Recently simulations of the restricted solid-on-solid growth models 
were used to build the width distributions of $d = 2-5$ dimensional 
KPZ interfaces. The universal scaling function associated with the steady-state width distribution was found to change smoothly as $d$ is increased, thus 
strongly suggesting that $d = 4$ is not an upper critical dimension for the 
KPZ equation. The dimensional trends observed in the scaling functions indicate
that the upper critical dimension is at infinity  \cite{MPPR02}.

      \subsubsection{Multiplicative noise systems} \label{MNsect}

In the hope of classifying nonequilibrium phase transition classes
according to their noise terms \cite{GMTMN,GMTMN2} 
introduced and studied systems via the Langevin equation
\begin{equation}
        \partial_t n(x,t) = D \, \nabla^2 n(x,t) - r \, n(x,t) 
                        - u \, n(x,t)^2 + n(x,t) \, \eta(x,t) \ ,
\label{GE}
\end{equation}
exhibiting real multiplicative noise (MN) :
\begin{equation}
        \langle \eta(x,t) \rangle = 0 \ , \quad
        \langle \eta(x,t) \eta(x',t') \rangle = 2 \nu \,
                \delta^d(x-x') \delta(t-t') \ ,
\label{GEnoise}
\end{equation}
that is proportional to the field ($n$).
Since for DP class the noise is square root proportional to the field;
for ARW and PC systems the noise is imaginary one may expect distinct 
universal behavior for such systems. Ref. \cite{HT97} on 
the other hand argued on field theoretical basis that such 'naive' 
Langevin equations fail to accurately describe systems controlled by 
particle pair reaction processes, where the noise is in fact `imaginary'
and one should derive a proper Langevin equation by starting from master
equation of particles.
Therefore they investigated the simplest RD systems where both real and 
imaginary noise is present and compete:
(a)  $2A \to \emptyset$, $2A \to 2B$, $2B \to 2A$, and 
$2B \to \emptyset$ 
(b)  $2A \to \emptyset$ and $2A \to (n+2) A$  (see Sect. \ref{binsp})
by setting up the action first -- derived from the master equation
of particles. In neither case were they able to recover the MN critical
behavior reported in \cite{GMTMN,GMTMN2}. Therefore they suspected that there
might not be real RD system possessing the MN behavior.

On the other hand \cite{GMTMN,GMTMN2} have established connection of MN systems
via the Cole-Hopf transformation: $n(x,t) = e^{h(x,t)}$ to the KPZ theory.
They have shown in 1d that the phase diagram and the critical exponents
$Z$, $\nu_{\perp}$ and $\beta$ of the two systems agree within numerical 
accuracy. They have found diverging susceptibility (with continuously
changing exponent as the function of $r$) for the entire range of $r$.

    \subsection{Quenched KPZ classes} \label{qKPZsect}

In random media nonlinear interface growth is described by the so called 
{\bf quenched KPZ equation} \cite{barabasi},
\begin{equation}\label{qKPZ-e}
\partial_t h({\bf x},t) = \sigma\nabla^2 h({\bf x},t) + 
\lambda(\nabla h({\bf x},t))^2 + F + \eta({\bf x},h({\bf x},t))
\end{equation}
where $F$ is a constant, external driving term and 
$\eta({\bf x},h({\bf x},t))$ is the quenched noise (do not fluctuate in
time). Its universal behavior was investigated in \cite{Les,Bul} and 
predicted $\tilde\alpha\simeq 0.63$ in one dimension, 
$\tilde\alpha\simeq 0.48$ in two dimensions and $\tilde\alpha\simeq 0.38$ 
in three dimensions. 
It was shown numerically that in 1d this class is described by 1+1 d
directed percolation depinning \cite{TL92}. In higher dimensions 
however it is related to percolating directed surfaces \cite{BGM96}.

   \subsection{Other continuum growth classes} \label{CGsect}

For continuum growth models exhibiting the symmetries
\begin{equation} \label{CG-s}
{\bf x} \to {\bf x+\Delta x} \ \ \
t \to t+\Delta t \ \ \
h \to h+\Delta h \ \ \
{\bf x} \to {\bf -x} \ \ \
\end{equation}
the possible general Langevin equations with relevant terms were
classified as follows \cite{barabasi}.
\begin{itemize}
\item The deterministic part describes conservative or nonconservative
process (i.e. the integral over the entire system is zero or not).
Conservative terms are $\nabla^2h$, $\nabla^4h$ and $\nabla^2(\nabla h)^2$.
The only relevant nonconservative terms is the $(\nabla h)^2$.
\item The system is linear or not.
\item The noise term is conservative 
(i.e. the result of some surface diffusion) with correlator
\begin{equation} 
\langle\zeta_d({\bf x},t)\zeta_d({\bf x'},t')\rangle
=(-2D_d\nabla^2+D'_d\nabla^4) \delta({\bf x-x'})\delta(t-t')
\end{equation}
or nonconservative like in eq.(\ref{gaussnoise}) (as the result of
adsorption, desorption mechanisms).
\end{itemize}
Analyzing the surface growth properties of such systems besides the 
EW and KPZ classes five other universality classes were identified 
(see Table \ref{CGtab}).
\begin{table}
\begin{center}
\begin{tabular}{|c|c|c|c|}
\hline
Langevin equation                 &  $\tilde\alpha$ & $\tilde\beta$  & Z \\
\hline 
$\partial_t h = -K\nabla^4 h + \zeta$  &  $\frac{4-d}{2}$& $\frac{4-d}{8}$ 
& 4 \\
\hline
$\partial_t h = \nu\nabla^2h + \zeta_d$&  $\frac{-d}{2}$ & $\frac{-d}{4}$ 
& 2 \\
\hline
$\partial_t h =-K\nabla^4 h + \zeta_d$ &  $\frac{2-d}{2}$& $\frac{2-d}{8}$ 
& 4 \\
\hline
$\partial_t h=-K\nabla^4h+\lambda_1\nabla^2(\nabla h)^2+\zeta$&$\frac{4-d}{3}$&
                                 $\frac{4-d}{8+d}$ & $\frac{8+d}{3}$  \\
\hline
                                 &      \multicolumn{3}{c|}{$d\le 1$}   \\
$\partial_t h=-K\nabla^4h+\lambda_1\nabla^2(\nabla h)^2 +\zeta_d$ &  
                                 $\frac{2-d}{3}$& 
                                 $\frac{2-d}{10+d}$ & $\frac{10+d}{3}$  \\
                                 &      \multicolumn{3}{c|}{$d > 1$}   \\
                                 & $\frac{2-d}{2}$  & $\frac{2-d}{8}$ & 4 \\
\hline
\end{tabular}
\caption{Summary of continuum growth classes discussed in this section
following ref. \cite{barabasi}.}
\label{CGtab}
\end{center}
\end{table}

   \subsection{Classes of mass adsorption-desorption, aggregation \\
    and chipping models} \label{agrsect}

Based on the interest in self-organized critical systems in which different
physical quantities exhibit power law distributions in the steady state over 
a wide region of the parameter space \cite{BTW} a family of lattice models,
in which masses diffuse, aggregate on contact, and also chip off a single
unit mass was introduced by \cite{MKBL98,MKBJ00}. 
Self-organized criticality has been studied in a variety of model systems
ranging from sandpiles to earthquakes.
A particularly simple lattice model due to Takayasu, with mass diffusion,
aggregation upon contact and adsorption of unit masses from outside at a 
constant rate, exhibits self-organized criticality \cite{Takayasu,TakaB}:
the steady-state mass distribution has a nontrivial power law decay for 
large masses in all dimensions \cite{Takayasu}.
These mass adsorption-desorption models in one dimension are defined as 
follows. 
A site $i$ is chosen randomly and then one of the following events can occur: 
\begin{enumerate}
\item Adsorption:
With rate $q$, a single particle is adsorbed at site $i$; 
thus $m_i \rightarrow m_{i}+1$.
\item Desorption: 
With rate $p$, a single particle is desorbed from site $i$; 
thus $m_i \rightarrow m_{i}-1 $ provided  $m_i \ge 1$.
\item Diffusion and Aggregation:
With rate $1$, the mass $m_i$ at site $i$ moves to a
nearest neighbor site [either $(i-1)$ or $(i+1)$] chosen at random.
If it moves to a site which already has some 
particles, then the total mass just adds up; thus $m_i \rightarrow 0$ and
$m_{i\pm1}\rightarrow m_{i\pm1}+m_i$.
\item Chipping: With rate $w$ a bit of mass at the site ``chips'' off,
e. provided $m_i\ge 1$ a single particle leaves site $i$ 
\end{enumerate}
While the Takayasu model ($p=0$, $w=0$) does not have a phase 
transition in the steady state, by introducing a nonzero desorption 
rate $p$ induces a critical line $p_c(q)$ in the $p-q$ plane. 
For fixed $q$, if one increases $p$ from $0$, one finds that for all 
$p<p_c(q)$, the steady state mass distribution has the same large $m$ 
behavior as in the Takayasu case, i.e., 
\begin{equation}
P(m)\propto m^{-\tau}
\end{equation}
where the exponent $\tau$ is the Takayasu exponent and is independent of $q$. 
For $p=p_c(q)$, we find the steady state mass distribution still decays 
algebraically for large $m$, but with a new critical exponent $\tau_c$ 
which is bigger than the Takayasu exponent $\tau_t$. 
For $p>p_c(q)$, we find that
\begin{equation}
P(m)\sim \exp (-m/m^{*})
\end{equation}
for large $m$ where $m^{*}$ is a characteristic mass that diverges if one 
approaches $p_c(q)$ from the $p>p_c(q)$ side. The critical exponent 
$\tau_c$ is the same everywhere on the critical line $p_c(q)$. 
This phase transition occurs in all spatial dimensions including $d=1$.
The $\tau$ exponent was determined for the mean-field and one dimensional
cases (\cite{Takayasu,MKBE00})
\begin{equation}
\tau_t^{MF} = 3/2, \quad \tau_c^{MF}=5/2, \quad \tau_t^{1d}=4/3, 
\quad \tau_c^{1d}=1.833 \ ,
\end{equation}
although the location of $d_c$ is not known.

This model can also be mapped onto an interface dynamics, if we interpret 
the configuration of masses as an interface profile regarding $m_i$ as a local 
height variable. The phase transition of the model can be qualitatively
interpreted as a nonequilibrium {\it wetting transition} of the interface.
The corresponding surface growth exponents are \cite{MKBE00}
\begin{equation}
\tilde\beta^{MF}=1/6, \quad Z=2, \quad \tilde\beta^{1d}=0.358, \quad Z=2 \ .
\end{equation}

In the $p=q=0$ conserved model (CM) a nonequilibrium phase transition occurs by varying the chipping rate or the average mass per site $\rho$.
There is a critical line $\rho_c(w)$ in the $\rho$-$w$ plane that separates 
two types of asymptotic behaviors of $P(m)$.  
For fixed $w$, as $\rho$ is varied across the critical value
$\rho_c(w)$, the large $m$ behavior of $P(m)$ was found to be,
\begin{equation}
  P(m)\sim \cases
          {e^{-m/m^{*}} &$\rho<\rho_c(w)$,\cr
	  m^{-\tau} &$\rho=\rho_c(w)$,\cr
	  m^{-\tau}+ {\rm infinite\,\,\, aggregate} &$\rho>\rho_c(w)$.\cr}
	 \label{pm}
\end{equation}
As one increases $\rho$ beyond $\rho_c$, this asymptotic algebraic
part of the critical distribution remains unchanged but in addition an
infinite aggregate forms. This means that all the additional mass
$(\rho-\rho_c)V$ (where $V$ is the volume of the system) condenses onto a
single site and does not disturb the background critical distribution.
This is analogous, in spirit, to the condensation of a macroscopic number
of bosons onto the single $k=0$ mode in an ideal Bose gas as the
temperature goes below a certain critical value.
Ref. \cite{RME01} proved analytically that the mean field phase
boundary, $\rho_c(w)=\sqrt {w+1}-1$, is {\it exact} and independent
of the spatial dimension $d$. They also provided unambiguous numerical evidence
that the exponent $\tau=5/2$ is also independent of $d$.
The corresponding growth exponents are: $Z=2$, $\tilde\alpha=2/3$
\cite{MKBJ00}. Even though the single site distribution $P(m)$ may be given 
exactly by the mean field solution, that does not prove that mean field theory 
or product measure is the exact stationary state in all dimensions.

The left-right asymmetric version of the CM model was also studied
in \cite{MKBJ00}. This has a qualitatively similar phase
transition in the steady state as the CM model but exhibits a different 
class phase transition owing to the mass current in this system. 
Simulations in one dimension predict: $Z=1.67$, $\tilde\alpha=0.67$.

   \subsection{Unidirectionally coupled DP classes} \label{uniDP}

As it was mentioned in Section \ref{NDPS} in case of coupled, multi-species
DP processes field theoretical RG analysis \cite{Ja97} predicts DP 
criticality with an unstable, symmetrical fixed point, such that 
sub-systems with unidirectionally coupled DP behavior emerge.
This was shown to be valid for linearly coupled N-component,
contact processes too. 
Unidirectionally coupled DP systems of the form (UCDP) 
\begin{eqnarray}
A &\leftrightarrow 2A \qquad & A \rightarrow A+B \nonumber \\
B &\leftrightarrow 2B \qquad & B \rightarrow B+C \nonumber \\
C &\leftrightarrow 2C \qquad & C \rightarrow C+D \nonumber \\
\rm{...}
\end{eqnarray}
were investigated by \cite{THH98,GHHT99} with the motivation 
that such models can describe interface growth models, where 
adsorption-desorption are allowed at terraces and edges (see Sect.\ref{MAT}).
The simplest set of Langevin equations for such systems was set up 
by \cite{AEHM98}:
\begin{eqnarray}
\label{CoupledLangevin}
\partial_t{\phi_k}(\mbox{\bf x},t) \;&=&\; \nonumber
\sigma \phi_k(\mbox{\bf x},t) -
\lambda \phi_k^2(\mbox{\bf x},t)  +
D \nabla^2 \phi_k(\mbox{\bf x},t) + \\
&+& \mu \phi_{k-1}(\mbox{\bf x},t) +
\zeta_k(\mbox{\bf x},t) \ ,
\end{eqnarray}
where $\zeta_k$ are independent multiplicative noise fields for level
$k$ with correlations
\begin{eqnarray}
\label{CoupledNoise}
\langle \zeta_k(\mbox{\bf x},t) \rangle \;&=&\; 0 \,, \\ \nonumber
\langle \zeta_k(\mbox{\bf x},t) \zeta_l(\mbox{\bf x}',t')  \rangle \;&=&\;
2 \Gamma \, \phi_k(\mbox{\bf x},t) \, \delta_{k,l} \,
\delta^d(\mbox{\bf x}-\mbox{\bf x}') \, \delta(t-t')
\ ,
\end{eqnarray}
for $k>0$, while for the lowest level ($k=0$) $\phi_{-1} \equiv 0$ is fixed.
The parameter $\sigma$ controls the offspring production, $\mu$ is the coupling
and $\lambda$ is the coagulation rate. As one can see the $k=0$ equation 
is just the Langevin equation of DP (\ref{DPLangeq}). 
The {\bf mean-field} solution of these equations that is valid above
$d_c=4$ results in critical exponents for the level $k$:
\begin{equation}
\beta_{MF}^{(k)} = 2^{-k} \ .
\end{equation}
and  $\nu_{\perp}^{MF}=1/2$ and $\nu_{||}^{MF}=1$ independently of $k$.
For $d<d_c$ field theoretical RG analysis of the action for $k<K$ levels 
\begin{eqnarray}
\label{CoupledDP}
S = & \sum_{k=0}^{K-1} \int d^dx\,dt \,\Biggl\{
  \psi_k \,
  \Bigl(\tau \partial_t - D \nabla^2  - \sigma \Bigr)\,\phi_k \nonumber \\
  & -\mu \psi_k \phi_{k-1} 
  + \frac{\Gamma}{2}\,\psi_k \Bigl(\phi_k - \psi_k \Bigr) \phi_k
   \Biggr\} \ ,
   \end{eqnarray}
was performed in \cite{THH98,GHHT99}. The RG treatment run into several 
difficulties. 
Infrared-divergent diagrams were encountered~\cite{Goldschmidt98} and
the coupling constant $\mu$ was shown to be a relevant quantity
(that means it diverges under RG transformations). 
Ref. \cite{GHHT99} argued that this is the reason why 
scaling seems to break down in simulations for large times  
(in lattice realization $\mu$ is limited). 
The exponents of the one-loop calculations for the first few levels, 
corresponding to the interactive fixed line as well as results of
lattice simulations are shown in Table \ref{uniDPexp}. 
\begin{table}
\begin{center}
\begin{tabular}{|c||c|c|c|c|} \hline
                & $d=1$         & $d=2$         & $d=3$      & $d=4-\epsilon$
\\\hline
$\beta_1$       & $0.280(5)$    & $0.57(2)$     & $0.80(4)$
&$1-\epsilon/6+O(\epsilon^2)$\\
$\beta_2$       & $0.132(15)$   & $0.32(3)$     & $0.40(3)$
&$1/2-\epsilon/8+O(\epsilon^2)$\\
$\beta_3$       & $0.045(10)$   & $0.15(3)$     & $0.17(2)$
&$1/4-O(\epsilon)$ \\\hline
$\delta_1$      & $0.157(4)$    & $0.46(2)$     & $0.73(5)$
&$1-\epsilon/4+O(\epsilon^2)$\\
$\delta_2$      & $0.075(10)$   & $0.26(3)$     & $0.35(5)$
&$1/2-\epsilon/6+O(\epsilon^2)$\\
$\delta_3$      & $0.03(1)$     & $0.13(3)$     & $0.15(3)$
&$1/4 - O(\epsilon)$\\\hline
$\eta_1$        & $0.312(6)$    & $0.20(2)$     & $0.10(3)$
&$\epsilon/12+O(\epsilon^2)$\\
$\eta_2$        & $0.39(2)$     & $0.39(3)$     & $0.43(5)$
&$1/2+O(\epsilon^2)$\\
$\eta_3$        & $0.47(2)$     & $0.56(4)$     & $0.75(10)$
&$3/4-O(\epsilon)$\\\hline
$2/Z_1$         & $1.26(1)$     & $1.10(2)$     & $1.03(2)$     &\\
$2/Z_2$         & $1.25(3)$     & $1.12(3)$     & $1.04(2)$
&$1+\epsilon/24+O(\epsilon^2)$\\
$2/Z_3$         & $1.23(3)$     & $1.10(3)$     & $1.03(2)$     & \\ 
\hline
$\nu_{\perp,1}$         & $1.12(4)$     & $0.70(4)$     & $0.57(4)$
&\\
$\nu_{\perp,2}$         & $1.11(15)$    & $0.69(15)$    &
$0.59(8)$
&$1/2+\epsilon/16+O(\epsilon^2)$\\
$\nu_{\perp,3}$         & $0.95(25)$    & $0.65(15)$    & $0.62(9)$
&\\\hline
$\nu_{\parallel,1}$     & $1.78(6)$     & $1.24(6)$     & $1.10(8)$
&\\
$\nu_{\parallel,2}$     & $1.76(25)$    & $1.23(17)$    &
$1.14(15)$
&$1+\epsilon/12+O(\epsilon^2)$\\
$\nu_{\parallel,3}$     & $1.50(40)$    & $1.15(30)$    & $1.21(15)$
&\\\hline
\end{tabular}
\caption{
\label{uniDPexp}
Critical exponents of UCDP \cite{GHHT99}.}
\end{center}
\end{table}
These scaling exponents can be observed for intermediate times but it is 
not clear if in the asymptotically long time they drift to the decoupled 
values or not.

The main representatives of these classes are certain monomer 
adsorption-desorption models (Sect.\ref{MAT}) and polynuclear growth 
models (PNG) with depinning transitions
\cite{KerteszWolf89,LRWK90,Toom94,Toom94b}. 
The latter type of systems are defined by {\bf parallel} update dynamic 
rules and coupled DP processes emerge in a {\em co-moving} frame.

   \subsubsection{Monomer adsorption-desorption at terraces}\label{MAT}

Here I show an example how a coupled particle system is related to an
interface growth model.
Refs. \cite{AEHM98,AEHM96} defined SOS and RSOS models that can be 
mapped onto UCDP (Sect.\ref{uniDP}). In this models adsorption an desorption 
processes may take place at terraces and edges. For each update a site~$i$
is chosen at random and an atom is adsorbed
\begin{equation}
\label{adsorption}
h_i \rightarrow h_{i}+1 \; \; {\rm with \ probability \ } q
\end{equation}
or desorbed at the edge of a plateau
\begin{equation}
\begin{array}{ll}
\label{desorption}
&h_i \rightarrow {\rm min}(h_{i},h_{i+1}) \; \;
{\rm with \ probability \ } (1-q)/2 \ , \\
&h_i \rightarrow {\rm min}(h_{i},h_{i-1}) \; \;
{\rm with \ probability \ } (1-q)/2 \ .
\end{array}
\end{equation}
Identifying empty sites at a given layer as $A$ particles, the adsorption 
process can be interpreted as the decay of $A$ particles ($A\to\emptyset$), 
while the desorption process corresponds to $A$ particle production
($A\to 2A$). These processes generate reactions on subsequent layers,
hence they are coupled. The simulations in 1d have shown that this coupling
is relevant in the upward direction only hence the model is equivalent to
the UCDP process. Defining the order parameters on the $k-th$ layer as
\begin{equation}
\label{OrderSOS}
n_k=\frac{1}{N}\,\sum_i \, \sum_{j=0}^{k}\,\delta_{h_i,j} \ ,
\end{equation}
where $h_i$ is the height at site $i$, they are expected to scale as
\begin{equation}
\label{CoupledDensityScaling}
n_k \sim (q_c-q)^{\beta^{(k)}} \ . \qquad \qquad k=1,2,3,\ldots
\end{equation}
By varying the growth rate ($q$) these models exhibit a roughening transition 
at $q_c=0.189$ (for RSOS) and at $q_c=0.233(1)$ (for SOS) from a non-moving, 
smooth phase to a moving, rough phase in one spatial dimension.
The $\beta^k$ (and other exponents) take those of the 1d UCDP class
(see Table \ref{uniDPexp}).

The scaling behavior of the interface width is characterized 
by many different length scales. At criticality it increases as 
\begin{equation}
W^2(t) \propto \tau \ln t
\end{equation}
with $\tau \simeq 0.102(3)$ for RSOS \cite{H02mono}.

   \subsection{Unidirectionally coupled PC classes} \label{dimer}

Similarly to the UCDP case (see Sect.\ref{uniDP}) surface growth processes
of dimers stimulated the introduction of unidirectionally coupled BARW2 
(Sect.\ref{BARWe}) models \cite{dimerlcikk,HiOd99}:
\begin{eqnarray}
\label{ReactionScheme}
&&A \rightarrow 3A  \hspace{5.5mm} \qquad B \rightarrow 3B
\hspace{5.5mm}\qquad C \rightarrow 3C
\, \nonumber\\
&&2A \rightarrow \O \hspace{5.5mm} \qquad 2B \rightarrow \O
\hspace{5.5mm}\qquad 2C \rightarrow \O
\, \\
&&A \rightarrow A+B \qquad B \rightarrow B+C \qquad C \rightarrow C+D
\,, \ldots \nonumber
\end{eqnarray}
generalizing the concept of UCDP. The mean field approximation 
of the reaction scheme~(\ref{ReactionScheme}) looks as
\begin{eqnarray}
\partial_t n_A &=& \sigma n_A - \lambda n_A^2           \,, \nonumber \\
\partial_t n_B &=& \sigma n_B - \lambda n_B^2 + \mu n_A \,, \\
\partial_t n_C &=& \sigma n_C - \lambda n_C^2 + \mu n_B \,, \,\ldots\nonumber
\end{eqnarray}
where $n_A,n_B,n_C$ correspond to the densities $n_0,n_1,n_2$ in the 
growth models. $\sigma$ and $\lambda$ are  the rates for offspring production
and pair annihilation respectively. The coefficient $\mu$ is an effective 
coupling constant between different particle species.
Since these equations are coupled in only one direction,
they can be solved by iteration. Obviously, the mean-field
critical point is $\sigma_c=0$. For small values of $\sigma$
the stationary particle densities in the active state are given by
\begin{equation}
n_A =      \frac{\sigma}{\lambda} \,, \,\,\,
n_B \simeq \frac{\mu}{\lambda} \left( \frac{\sigma}{\mu} \right)^{1/2} , \,\,\,
n_C \simeq \frac{\mu}{\lambda} \left( \frac{\sigma}{\mu} \right)^{1/4} \,,
\end{equation}
corresponding to the mean field critical exponents
\begin{equation}
\label{MeanFieldBeta}
\beta^{MF}_A=1 ,\,\,\,\,\beta^{MF}_B=1/2 ,\,\,\,\,\, \beta^{MF}_C=1/4
,\ldots \,.
\end{equation}
These exponents should be valid for $d>d_c=2$. Solving the asymptotic 
temporal behavior one finds $\nu_\parallel=1$, implying that 
$\delta^{MF}_k=2^{-k}$. 

The effective action of unidirectionally coupled BARW2's 
should be given by
\begin{eqnarray}
\label{EffectiveAction}
&&S[\psi_0,\psi_1,\psi_2,\ldots,\bar{\psi}_0,
\bar{\psi}_1,\bar{\psi}_2,\ldots]= \nonumber \\
&&\int d^dx\,dt\,\sum_{k=0}^\infty \Bigl\{
\bar{\psi}_k(\partial_t-D\nabla^2)\psi_k - \lambda(1-\bar{\psi}_k^2)\psi_k^2 +
\nonumber\\
&&\hspace{4mm}
+ \sigma(1-\bar{\psi}_k^2)\bar{\psi}_k\psi_k 
+ \mu (1-\bar{\psi}_k) \bar{\psi}_{k-1} \psi_{k-1}  \Bigr\}
\end{eqnarray}
where $\psi_{-1}=\bar{\psi}_{-1} \equiv 0$. Here the fields $\psi_k$ and 
$\bar{\psi_k}$ represent the configurations of the system at level~$k$.
Since even the RG analysis of the one component BARW2 model suffered serious 
problems \cite{Cardy-Tauber} the solution of the theory of 
(\ref{EffectiveAction}) seems to be hopeless. Furthermore one expects
similar IR diagram problems and diverging coupling strengths as in case
of UCDP that might be responsible for violations of scaling in the long 
time limit.

Simulations of a 3-component model in 1d coupled by instantaneous particle 
production of the form (\ref{ReactionScheme}) resulted in decay exponents 
for the order parameter defined as (\ref{OrderSOS}): 
\begin{equation}
\delta_A = 0.280(5)
\ , \
\delta_B = 0.190(7)
\ , \
\delta_C = 0.120(10)
\ ,
\end{equation}
For further critical exponents see Sect.\ref{DAT}.

It would be interesting to investigate parity-conserving growth processes 
in higher dimensions. Since the upper critical dimension $d^\prime_c$ is 
less than 2, one expects the roughening transition -- if still existing -- 
to be described by mean-field exponents. In higher dimensions, $n$-mers 
might appear in different shapes and orientations.

The main representatives of these classes are certain dimer
adsorption-desorption models (Sect.\ref{MAT}) and polynuclear growth
models (PNG) with depinning transitions~\cite{HiOd99}.
The latter type of systems are defined by {\bf parallel} update dynamic
rules and coupled DP processes emerge in a {\em co-moving} frame.

      \subsubsection{Dimer adsorption-desorption at terraces}\label{DAT}

\begin{figure}
\epsfxsize=70mm
\epsffile{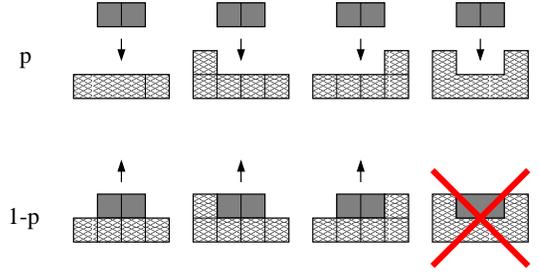}
\caption{
\label{dimerrule} Dimers are adsorbed with probability $p$
and desorbed at the edges of terraces with probability $1-p$. 
Evaporation from the middle of plateaus is not allowed \cite{HiOd99}.
}
\end{figure}

\begin{figure}
\epsfxsize=80mm
\epsffile{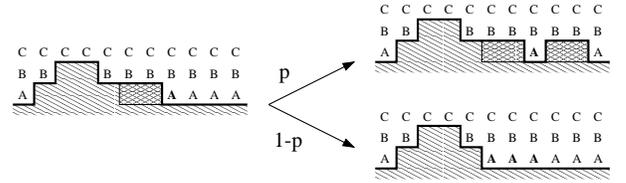}
\caption{
\label{dimermap}
Extended particle interpretation. Dimers are adsorbed ($2A \rightarrow \O$)
and desorbed ($A \rightarrow 3A$) at the bottom layer. 
Similar processes take place at higher levels \cite{HiOd99}.
}
\end{figure}

Similarly to the monomer case (Sect.\ref{MAT}) dimer adsorption-desorption 
models were defined \cite{dimerlcikk,HiOd99,NPN00}. With the restriction that 
desorption may only take place at the edges of a plateau the models can be
mapped onto the unidirectionally coupled BARW2 (Sect.\ref{dimer}). 
The dynamical rules in $d=1$ are defined in Fig. \ref{dimerrule}.
The mapping onto unidirectionally coupled BARW2 can be seen on 
Fig.\ref{dimermap}. \\

Such dimer models can be defined in arbitrary spatial dimensions.
In \cite{HiOd99} four one dimensional variants were investigated: \\

\noindent 1) Variant A is a restricted solid-on-solid
(RSOS) model evolving by random sequential updates. \\

\noindent 2) Variant B is a solid-on-solid (SOS) model evolving by 
random sequential updates. \\

\noindent 3) Variant C is a restricted solid-on-solid
(RSOS) model evolving by parallel updates. \\

\noindent 4) Variant D is a solid-on-solid (SOS) model evolving by 
parallel updates. 

\begin{figure}
\epsfxsize=80mm
\epsffile{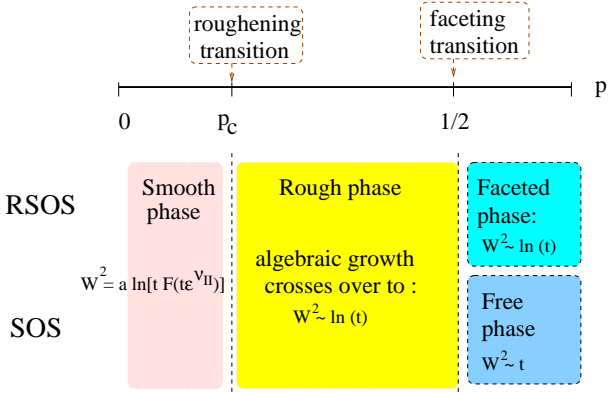}
\caption{Phase diagram of 1d dimer models.}
\label{dimerpd}
\end{figure}

Variants A and B exhibit transitions in contrast
with PNG models, 
in which only parallel update rules permit roughening transitions.
By varying the adsorption rate $p$ the phase diagram shown in 
Fig.\ref{dimerpd} emerges for RSOS and SOS cases.
If $p$ is very small, only a few dimers are adsorbed at the surface, 
staying there for a short time before they evaporate back into the
gas phase. Thus, the interface is anchored to the actual bottom layer 
and does not propagate. In this smooth phase the interface with growths
logarithmically until it saturates to a finite value (even for $L\to\infty$).

As $p$ increases, a growing number of dimers covers the surface 
and large islands of several layers stacked on top of each other are formed.
Approaching a certain critical threshold $p_c$ the mean size of the islands 
diverges and the interface evolves into a rough state 
with the the finite-size scaling form
\begin{equation}
\label{WidthFSScaling}
W^2(L,t) \simeq a \, \ln \Bigl[ t \, G(t/L^Z) \Bigr]\, .
\end{equation}
The order parameter defined on the $k$-th layer as (\ref{OrderSOS}) 
exhibits unidirectionally coupled BARWe critical behavior. 
The transition rates and exponents are summarized in Table \ref{dimertab}. 
\begin{table}
\begin{center}
\begin{tabular}{||c||c|c|c|c||}
\hline
variant         & A & B & C & D \\ \hline

restriction     & yes & no & yes & no  \\
updates         & random & random & parallel & parallel \\ \hline
$p_c$           & $0.3167(2)$ & $0.292(1)$ & $0.3407(1)$ & $0.302(1)$ \\
$a$             & $0.172(5)$ & $0.23(1)$ & $0.162(4)$ & $0.19(1)$ \\
$Z$             & $1.75(5)$ & $1.75(5)$ & $1.74(3)$ & $1.77(5)$ \\
$\delta_0$      & $0.28(2)$ & $0.29(2)$ & $0.275(10)$ & $0.29(2)$ \\
$\delta_1$      & $0.22(2)$ & $0.21(2)$ & $0.205(15)$ & $0.21(2)$ \\
$\delta_2$      & $0.14(2)$ & $0.14(3)$ & $0.13(2)$ & $0.14(2)$ \\
\hline
$\tilde{\alpha}$& $1.2(1)$ & undefined & 1.25(5) & undefined \\
$\tilde{\beta}$ & $0.34(1)$ & $0.50(1)$ & $0.330(5)$ & $0.49(1)$ \\
\hline
\end{tabular}
\end{center}
\caption{ \label{dimertab} 
Numerical estimates for the four variants of the dimer model
at the roughening transition $p=p_c$ (upper part) and at the 
transition $p=0.5$ (lower part).
}
\end{table}
Above $p_c$ one may expect the interface to detach from the bottom layer 
in the same way as the interface of monomer models starts to propagate 
in the supercritical phase. However, since dimers are adsorbed at neighboring
lattice sites, solitary unoccupied sites may emerge. These pinning centers 
prevent the interface from moving and lead to the formation of `droplets'.
Due to interface fluctuations, the pinning centers can slowly diffuse to 
the left and to the right. When two of them meet at the same place, they 
annihilate and a larger droplet is formed. Thus, although the interface 
remains pinned, its roughness increases continuously. The width initially 
increases algebraically until it slowly crosses over to a
logarithmic increase $W(t) \sim \sqrt{a \ln t}$. 

The restricted as well as the unrestricted variants
undergo a second phase transition at $p=0.5$ \cite{NPN00} where the 
width increases {\em algebraically} with time as $W \sim t^{\tilde{\beta}}$.
In the RSOS case ordinary {\em Family-Vicsek} scaling (\ref{FV-forf} 
occurs with the exponents given in Table \ref{dimertab}.
The dynamic exponent is $Z = \tilde{\alpha}/\tilde{\beta} \simeq 3$.
This value stays the same if one allows dimer digging at the faceting
transition but other surface exponents $\tilde\alpha\simeq 0.29(4)$ and 
$\tilde\beta\simeq 0.111(2)$ will be different \cite{NPN00}. 
An explanation of the latter exponents is given in \cite{NPKN01} based on 
mapping to globally constrained random walks. 
In the SOS cases (variants B,D) large spikes are formed, 
the surface roughens much faster with a growth exponent of 
$\tilde{\beta}\simeq 0.5$. The interface evolves into configurations with
large columns of dimers separated by pinning centers. These spikes can grow 
or shrink almost independently. As the columns are spatially 
decoupled, the width does not saturate in finite systems, i.e., the dynamic
exponents $\tilde{\alpha}$ and $Z$ have no physical meaning.

For $p>0.5$ the restricted models A and C evolve into faceted 
configurations. The width first increases algebraically until the pinning 
centers become relevant and the system crosses over to a logarithmic
increase of the width. Therefore, the faceted phase may be considered as a 
rough phase. The unrestricted models B and D, however, evolve into spiky 
interface configurations. The spikes are separated and grow independently 
by deposition of dimers. Therefore, $W^2$ increases {\em linearly} with time, 
defining the {\em free} phase of the unrestricted models.

In the simulations mentioned by now the interface was grown from flat
initial conditions. It turns out that starting with {\bf random initial 
conditions} $h_i=0,1$ the densities $n_k$ turn out to decay much slower.
For restricted variants an algebraic decay of $n_0$ with an exponent 
\begin{equation}
\delta_0 \simeq 0.13
\end{equation}
was observed. Similarly, the critical properties of the faceting transition 
at $p=0.5$ are affected by random initial conditions.
The non-universal behavior for random initial conditions is
related to an {\em additional} parity conservation law. 
The dynamic rules not only conserve parity of the particle number 
but also conserve the parity of droplet sizes. Starting with a flat 
interface the lateral size of droplets is always even, allowing them 
to evaporate entirely.
However, for a random  initial configuration,  droplets of odd size
may be formed which have to recombine in pairs before they can evaporate,
slowing down the dynamics of the system.
In the language of BARW2 processes the additional parity conservation law
is due to the absence of nearest-neighbor diffusion.
Particles can only move by a combination of offspring production
and annihilation, i.e., by steps of {\em two} lattice sites. Therefore,
particles at even and odd lattice sites have to be distinguished.
Only particles of different parity can annihilate. Starting with
a fully occupied lattice all particles have alternating parity
throughout the whole temporal evolution, leading to the usual
critical behavior at the PC transition. For random initial conditions,
however, particles of equal parity cannot annihilate, slowing
down the decay of the particle density.
Similar sector decomposition has been observed in diffusion of
$k$-mer models \cite{Barma,Barma2}.

\section{Summary}

In summary dynamical extensions of classical equilibrium classes were
introduced in the first part of this review. New exponents, concepts,
sub-classes, mixing dynamics and some unresolved problems were discussed.
The common behavior of these models was the strongly fluctuating ordered 
state. In the second part genuine nonequilibrium dynamical classes of 
reaction-diffusion systems and interface growth models were overviewed.
\begin{table}
\begin{tabular}{|c|c|c|}
\hline
CLASS ID    &  features             &   Section\\
\hline\hline
DP          & time reversal symmetry   & \ref{DPS} \\
DyP         &  long memory          & \ref{dynperc} \\       
VM          &  $Z_2$ symmetry       & \ref{glau}  \\
PCP         &  coupled frozen field & \ref{PCPsect} \\
NDCF        &  global conservation  & \ref{NDCF} \\
\hline
PC          &  $Z_2$ symmetry, BARW2 conservation  & \ref{BARWe} \\
BP          &  DP coupled to ARW   &  \ref{binsp} \\
DCF         &  coupled diffusive, conserved field &  \ref{DCF} \\
N-BARW2     &  N-component BARW2  conservation & \ref{NBARWS} \\
N-BARW2s    &  symmetric NBARW2 + exclusion &  \ref{2BARW2S} \\
N-BARW2a    &  asymmetric NBARW2 + exclusion &  \ref{2BARW2S} \\
\hline          
\end{tabular}
\caption{Summary of known absorbing state universality classes in homogeneous
isotropic systems.}
\label{tab}
\end{table}
These were related to phase transitions to absorbing states 
of weakly fluctuating ordered states.
The class behavior is usually determined by the spatial dimension,
symmetries, boundary conditions and inhomogeneities like in case of 
equilibrium models but in low dimensions hard-core exclusion was
found to be a relevant factor too, splitting up criticality in
fermionic and bosonic models. 
The symmetries are not so evident as in case of equilibrium models, 
they are expressed in terms of the relations of fields and response 
fields most precisely.
Furthermore in recently discovered coupled systems with binary,
triplet or quadruplet particle production no special symmetry seems to 
be responsible for a novel type of critical behavior.
Perhaps a proper field theoretical analysis of the coupled DP+ARW system 
could shed some light on this mystery.
The parity conservation in hard-core and in binary spreading models seems 
to be irrelevant. 
In Table \ref{tab} I summarized the most well known families of 
absorbing phase transition classes of homogeneous, spatially isotropic
systems. Those which are below the horizontal line exhibit 
{\it fluctuating absorbing states}. The necessary and sufficient 
conditions for these classes are usually unknown.
The mean-field classes can also give a guide to distinguish
classes below $d_c$. In Table \ref{MFtab} I collected the mean-field 
exponents and upper critical dimensions of the known
absorbing-state model classes. Note that in the general
$nA \to (n+k) A$, $mA\to(m-l)A$ type of RD systems the values of
$n$ and $m$ determine the critical behavior.
\begin{table}
\begin{center}
\begin{tabular}{|l|l|l|l|l|l|l|l|l|}
\hline
CLASS &
$\beta$& $\beta'$& $Z$ &$\nu_{||}$& $\alpha$& $\delta$ &$\eta$ &$d_c$\\
\hline
DP  &  1  &   1    &  2   &  1        &  1      &  1     &  0  & 4   \\
\hline
DyP &  1  &   1    &  2   &  1        &  1      &  1     &  0  & 6   \\
\hline
VM  &  0  &   1    &  2   &  1        &  0      &  1     &  0  & 2   \\
\hline
PC  &  1  &   0    &  2   &  1        &  1      &  0     & -1/2 & 2  \\
\hline
BkARW & $\frac{1}{k-1}$ & 0  & 2  &  1        & $\frac{1}{k-1}$ &  0      & 0    & $\frac{2}{k-1}$\\
\hline
PARWs  & 1    & 0     & 2  &  n      & 1/n     &  0      & 0    &  \\
\hline
PARWa  & $\frac{1}{m-n}$ &    & 2  & $\frac{m-1}{m-n}$  & $\frac{1}{m-1}$ &        &     &  \\
\hline
NDCF&  1      & 1  & 2    & 1         & 1       &  1      & 0    & 4 \\
\hline
NBARW2& 1     & 0  & 2    & 1         & 1       &  0      & 0    & 2 \\
\hline
\end{tabular}
\caption{Mean-field classes of known, homogeneous absorbing-state transitions}
\label{MFtab}
\end{center}
\end{table}

In $d>1$ dimension the mapping of spin-systems onto RD systems of particles 
is not straightforward, instead one should also deal with the theory of
branching interfaces \cite{Cardy98}. Preliminary simulations found interesting
critical phenomena by generalized Potts models, exhibiting absorbing states
\cite{Lipp}.

Further research should also explore the universality classes of
such nonequilibrium phase transitions that occur by external 
current driven systems or by other models exhibiting fluctuating ordered 
states \cite{E00,EKKM98}. Nonequilibrium phase transitions in quantum 
systems \cite{Racznewrev} or by irregular graph or network based systems are
also of current interest of research.
Finally, having settled the problems arisen by fundamental nonequilibrium
models one should turn towards the study of more application motivated 
systems.

{\section*{Acknowledgements}

I thank H. Hinrichsen, N. Menyh\'ard and M. A. Mu\~noz for their
comments to the manuscript. Support from Hungarian research funds OTKA
(No.25286) and Bolyai (BO/00142/99) is acknowledged.

\begin{widetext}

\section*{Abbreviations}

\begin{tabular}{lll}
$\hspace{15mm}$     & \\

ABC                 & active boundary condition &  (\ref{sectsbon}) \\
ARW             & annihilating random walk ($AA\to\emptyset$) & (\ref{2A0}) \\
AF                  & annihilation-fission process & (\ref{binsp}) \\
BARW                & branching and annihilation random walk & (\ref{DPS}) \\
BARWe & even-offspring branching and annihilation random walk &(\ref{BARWe}) \\ 
BARWo & odd-offspring branching and annihilation random walk & (\ref{BARWo}) \\ 
BARW2 & two-offspring branching and annihilation random walk & (\ref{BARWe}) \\
BkARW & branching and k particle annihilating random walk & (\ref{kA0}) \\ 
BP                  & binary production & (\ref{binsp}) \\
CAM                 & coherent anomaly method & (\ref{DPS}) \\
CDP                 & compact directed percolation & (\ref{sectsca}) \\
DCF                 & diffusive conserved field & (\ref{DCF}) \\
DK                  & Domany-Kinzel cellular automaton & (\ref{sectsca}) \\
DP                  & directed percolation & (\ref{DPS}) \\
DS                  & damage spreading & (\ref{DSsect}) \\
DyP                 & dynamical percolation & (\ref{dynperc}) \\
EW                  & Edwards-Wilkinson & (\ref{EWsect})  \\
DMRG                & density matrix renormalization group & (\ref{genchap}) \\
GEP                 & generalized epidemic process & (\ref{dynperc}) \\
GDK & generalized Domany-Kinzel cellular automaton & (\ref{GDKmod}) \\
GMF                 & generalized mean-field approximation & (\ref{1storder}) \\
IBC                 & inactive boundary condition & (\ref{sectsbon}) \\
KPZ                 & Kardar-Parisi-Zhang & (\ref{KPZsect}) \\
LIM                 & linear interface model & (\ref{qEWsect}) \\
MF                  & mean-field approximation & (\ref{fieldthint}) \\
N-BARW2 & even-offspring, N-component branching and 
annihilation random walk &(\ref{NBARWS})\\ 
NDCF                & non-diffusive conserved field & (\ref{NDCF}) \\
NEKIM               & nonequilibrium Ising model & (\ref{nekim}) \\
PC                  & parity conserving & (\ref{PCS}) \\
PCP                 & pair contact process & (\ref{PCPsect}) \\
PCPD        & pair contact process with particle diffusion & (\ref{pcpdsect})\\
PARWs      & symmetric production and m-particle annihilating random walk &
               (\ref{pkarwsect}) \\
PARWa      & asymmetric production and m-particle annihilating random walk &
               (\ref{pkarwsect}) \\
PNG        & polinuclear growth models & (\ref{uniDP}) \\	       
RBC                 & reflecting boundary condition & (\ref{sectsbon}) \\
RD                  & reaction-diffusion & (\ref{fieldthint})\\
RG                  & renormalization group & (\ref{fieldthint}) \\
RSOS                & restricted solid on solid model & (\ref{growth}) \\
SCA                 & stochastic cellular automaton & (\ref{sectsca}) \\
SOS                 & solid on solid model & (\ref{growth}) \\
TTP                 & threshold transfer process & (\ref{PCPsect}) \\
UCDP         & unidirectionally coupled directed percolation & (\ref{uniDP}) \\
VM                  & voter model & (\ref{glau}) \\
\end{tabular}

\end{widetext}

\bibliographystyle{apsrmp}
\bibliography{rmpb}

\end{document}